\documentclass[sn-nature]{sn-jnl}
\usepackage{graphicx}
\makeatletter\providecommand{\headerps@out}[1]{}\makeatother
\usepackage{amsmath,amssymb,amsfonts,mathtools}
\usepackage{array,booktabs,longtable,calc}
\usepackage{caption}
\usepackage{microtype}
\usepackage{textcomp}
\usepackage{etoolbox}
\usepackage{xcolor}
\hypersetup{colorlinks=true,linkcolor=black,citecolor=black,urlcolor=blue,
    pdftitle={Charge-state determination unifies electrostatic generator behaviour},
    pdfauthor={Hongfa Zhao, Baiqiao Wang, Hanlin Zhou, Wenbo Ding}}
\begin{document}
\title[Charge-state determination]{Charge-state determination unifies electrostatic generator behaviour}
\author[1]{\fnm{Hongfa} \sur{Zhao}}
\equalcont{These authors contributed equally to this work.}
\author[1]{\fnm{Baiqiao} \sur{Wang}}
\equalcont{These authors contributed equally to this work.}
\author[1]{\fnm{Hanlin} \sur{Zhou}}
\equalcont{These authors contributed equally to this work.}
\author*[1]{\fnm{Wenbo} \sur{Ding}}\email{dingwenbo@sz.tsinghua.edu.cn}
\affil*[1]{\orgdiv{Shenzhen International Graduate School}, \orgname{Tsinghua University}, \orgaddress{\city{Shenzhen}, \postcode{518055}, \country{China}}}
\abstract{Electrostatic generators operate through distinct mechanisms of charge generation, polarization and transport, yet their diverse electrical behaviours have largely been described through separate theoretical models. Establishing how these different electrical sources determine conductor-charge states and connect seemingly distinct observations remains a fundamental challenge. Here we establish a source-to-state-to-observation framework in which material-established electrical sources, geometry, electrostatic constraints, electrical boundaries and charge history self-consistently determine conductor charge and its measurable consequences. We show that established triboelectric, electret and piezoelectric formulations emerge from a common charge-state determination process, which also connects ionoelectric transport and liquid-solid interfacial sources to their electrical responses. We resolve this route using a contact-separation triboelectric generator, whose multiple source, compensation and induced charges expose the underlying state constraints. This perspective explains why voltage can change under unchanged electrical sources and mechanical motion when the measurement boundary or charge history changes. It also connects nonlinear charge-generated harmonics with their load-constrained evolution, and shows how capacitive measurement and resistive loading approach the same charge-frozen reference state under matched conditions. Together, these results provide a coherent physical basis for interpreting distinct electrostatic-generator behaviours and the relationships among them.}
\maketitle

Electrostatic generators convert mechanical and interfacial processes into electrical energy and signals, supporting energy harvesting and self-powered sensing. Across triboelectric, electret, piezoelectric and related systems, extensive research has revealed diverse electrical behaviours and established powerful models for predicting their outputs\cite{mref1,mref2,mref3,mref4,mref5,mref6}. Yet understanding why these behaviours arise, and how they relate to one another, remains a deeper challenge. As the field expands across increasingly diverse materials, devices and applications, establishing a shared physical understanding becomes essential for connecting experimental findings and guiding further development. Under unchanged electrical sources and mechanical motion, the voltage observed from a generator can still depend on its measurement boundary\cite{mref7}. Such observations raise a fundamental question: what determines the physical state underlying an electrical output, and can the diverse behaviours of electrostatic generators be understood through a common physical route?

Device-specific models\cite{mref8,mref9,mref10,mref11}, studies of interfacial charging and electromechanical transduction\cite{mref12,mref13,mref14,mref15,mref16}, and analyses of generator architectures, circuits and performance\cite{mref17,mref18,mref19,mref20,mref21} have established essential foundations for understanding electrostatic-generator operation. Source-resolved electrostatic theories\cite{mref22,mref23} have further clarified how initially established electrical charges constrain conductor responses. Together, these advances have enabled quantitative predictions of electrical outputs and revealed important relationships among device geometry, electrical conditions and generator performance. Yet predicting individual behaviours does not necessarily establish how they arise from a common physical state or why seemingly distinct phenomena are connected. This challenge extends beyond electrostatic generation: in many physical systems, the ability to describe individual observations does not necessarily reveal the common physical processes connecting them\cite{mref24,mref25}. Material-established electrical sources, induced conductor charges and terminal observables represent different but interdependent physical levels. Their relationships are determined by electrostatic constraints, geometry, electrical boundaries and preceding charge history. Making these connections explicit is essential for moving from separate descriptions of generator behaviour towards a coherent physical understanding across devices and operating conditions.

Here we establish a source-to-state-to-observation framework that connects electrostatic generators through the self-consistent determination of conductor-charge states. We examine its generality by recovering established fixed-charge and piezoelectric descriptions and connecting ionic-transport and liquid-solid interfacial descriptions to electrode-charge responses (Supplementary Notes 1-7). We then use a contact-separation triboelectric generator to resolve the source, compensation and induced charges in detail, revealing how different electrical observations arise from the determination and evolution of the same physical charge state. Three controlled changes expose these connections. Measurement capacitance selects how much charge can move, the preceding equilibration determines the charge reference, and load resistance constrains its subsequent evolution. The resulting states account for measurement-dependent voltage, persistent baselines, nonlinear electrical harmonics and load relaxation. In particular, capacitive and resistive boundaries approach a common charge-frozen limit under matched conditions, while the load reshapes the charge trajectory that generates the observed spectrum. Experiments across generator classes place these linked behaviours within the wider physics of electrostatic generation.

\section{A common electrostatic problem across distinct electrical sources}

Different microscopic mechanisms establish the electrical sources of electrostatic generators, but their conductor-charge responses are governed by a common physical determination problem. We denote the upstream source state by \(\mathcal{S}(t)\). Depending on the material system, \(\mathcal{S}(t)\) may contain prescribed volume or surface charge \((\rho_{s},\sigma_{s})\), polarization \(\mathbf{P}_{s}\), mobile ionic populations, interfacial states or other material-specific electrical variables. Their common feature is that they are established upstream of conductor-response charge.

For a specified source state \(\mathcal{S}(t)\), geometry \(\Gamma(t)\), electrical boundary \(\mathcal{B}(t)\) and preceding electrical history \(\mathcal{H}_{0}\), the conductor-charge state is constrained by electrostatic boundary conditions, conductor equipotentiality and free-charge conservation. Here \(\mathcal{H}_{0}\) denotes the preceding electrical history establishing the reference conductor-charge state. The resulting hierarchy is

\begin{equation*}
[\mathcal{S}(t),\,\Gamma(t),\,\mathcal{B}(t),\,\mathcal{H}_{0}]\rightarrow\rho(\mathbf{r},t)\rightarrow[\mathbf{E}(\mathbf{r},t),\phi(\mathbf{r},t)]\rightarrow Q(t)\rightarrow[I(t),V(t)]\tag{1}\label{eq:1}
\end{equation*}

Here \(\rho(\mathbf{r},t)\) contains both source and conductor-response charge. For a defined transfer direction, \(Q(t)\) denotes transferred conductor charge relative to the specified reference state and \(I = dQ/dt\).

Equation (1) makes material-established sources, conductor-response charges and measurable outputs successive, physically connected levels. Across generator classes, specified sources, geometry, electrical boundary and preceding charge history enter a common charge-conserving electrostatic problem. In spatially uniform limits, the solved state yields analytical terminal relations; in finite geometries, it resolves spatial charge distributions (Supplementary Notes 7-8). Material-specific laws establish fixed charge, deformation-induced polarization and evolving interfacial or ionic sources, without changing the downstream state-determination route.

The shared mathematical procedure is to determine the electric field and the unknown free-charge distribution on each conductor from the material-established sources, the instantaneous geometry and the electrical constraints. Each connected metal must remain equipotential, while its total free charge is selected by the circuit boundary and preceding charge history. The source mechanisms differ, but they enter this same state-determination problem through their respective material variables:

\[\begin{matrix} \text{TENG}:\quad & \mathcal{S}=\{\rho_{src}(\mathbf{r}),\,\sigma_{trap}(\mathbf{r}),\,\mathbf{P}_{rem}(\mathbf{r})\}, \\ \text{Electret}:\quad & \mathcal{S}=\{\rho_{trap},\,\sigma_{trap},\,\mathbf{P}_{rem}\}, \\ \text{Piezoelectric}:\quad & \mathbf{P}_{s} = \mathbf{e}:\mathbf{\varepsilon}_{mech}, \\ \text{Ionoelectric}:\quad & \mathcal{S}=\{c_{i},\,\mathbf{J}_{i},\,\rho_{fix}\},\ \rho_{s} = \sum_{i}^{}z_{i}ec_{i} + \rho_{fix}, \\ \text{Liquid-solid}:\quad & \sigma_{int} = \beta_{EDL}\sigma_{CE},\ \Gamma = \Gamma_{wet}(t). \end{matrix}\]

These source descriptions are not mutually exclusive. A triboelectric generator, for example, may contain surface and volume charges as well as dipolar polarization, all of which can be incorporated into the same electrostatic determination problem. The large-area analytical model considered below adopts a reduced surface-charge representation, whereas the finite-geometry formulation accommodates spatially distributed sources and polarization contributions (Supplementary Note 8).

For ionoelectric generators, the source state also retains species-resolved concentrations and fluxes; for liquid-solid generators, the changing wetted region enters the geometry. These are different ways of establishing the electrical source, not separate rules for determining conductor charge. Once the corresponding electrostatic state has been solved, the familiar terminal charge-voltage, capacitance and equivalent-circuit descriptions can be recovered as reduced expressions (Supplementary Notes 3-6).

Generator classes differ primarily in how \(\mathcal{S}\) is established and evolves. TENGs and electrets can approximate fixed-source regimes after source establishment\cite{mref26,mref27,mref28}; piezoelectric generators possess deformation-dependent constitutive sources\cite{mref29,mref30,mref31}; piezoelectrets combine persistent and deformation-dependent components\cite{mref8,mref32}; and ionoelectric or solid-liquid systems may require transport or electrochemical source dynamics\cite{mref33,mref34,mref35}. These different source mechanisms therefore share a downstream problem: determining how electrical sources constrain conductor-charge states and measurable behaviour (Fig. 1).

\begin{figure}[!htbp]
\centering
\includegraphics[width=0.99\linewidth,height=0.69\textheight,keepaspectratio]{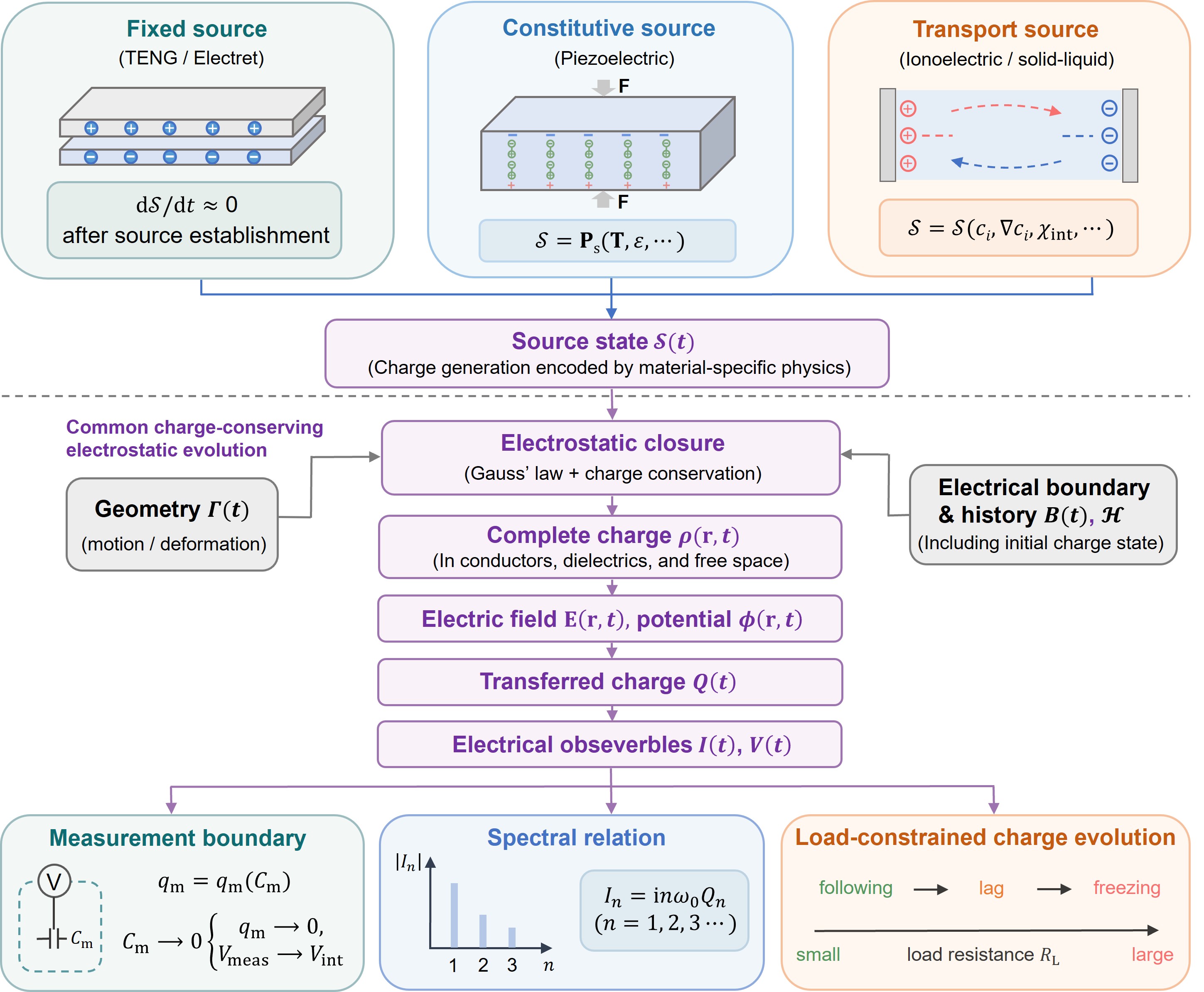}
\caption{A common physical route connects diverse sources to electrostatic-generator behaviour. Different material mechanisms establish persistent, deformation-dependent, transport-controlled or interfacial electrical source states (top). The centre of the hierarchy is the conductor-charge state, selected by geometry, electrostatic equilibrium, conservation laws, electrical boundary and preceding charge history. The source and solved conductor response determine the field and potential; their constrained evolution yields transferred charge, current and measured voltage. The lower panels illustrate measurement-conditioned voltage, nonlinear spectral response and load-constrained charge evolution as related consequences of this state determination, rather than independent output descriptions. Classical generator models arise as source- and geometry-specific reductions of this common physical route.}
\label{fig:1}
\end{figure}

The common determination route can be made explicit by tracing differently formulated generator theories back to their material sources and conductor-charge constraints. Stored charges and persistent polarization lead to the bias-source-variable-capacitance description and rotating electret model (Supplementary Note 3). Deformation-induced polarization, together with its mechanical coupling, yields the piezoelectric bimorph\textquotesingle s electromechanical port relations (Supplementary Note 4). Nernst-Planck transport and pressure-driven flow establish ionic source states whose electrode response recovers characteristic piezoionic potential and load relations (Supplementary Note 5). Contact electrification and electric-double-layer screening establish an interfacial source whose geometry-dependent induced charge determines a liquid-solid generator\textquotesingle s response (Supplementary Note 6). The contact-separation TENG follows the same source-to-conductor-state sequence (Supplementary Note 7). Together, these classical recoveries establish a shared state-determination route across distinct electrical sources. We next use the TENG to reveal how that route connects the physical charge states underlying multiple electrical behaviours.

\section{Self-consistent charge-state solutions across geometrical scales}

We first demonstrate quantitative charge-state determination in a contact-separation TENG. A dielectric of thickness \(d_{0}\) and relative permittivity \(\varepsilon_{r}\) is backed by a metallic electrode and separated from a counter-electrode by an air gap \(z\). Contact-generated\cite{mref36,mref37} and pre-existing surface charges\cite{mref22,mref38} are combined into the prescribed source state, \(\sigma_{s} = \sigma_{\text{te}} + \sigma_{\text{tp}}.\) Subsequent motion changes the geometrical coupling, whereas the conductor free charges remain response variables to be determined (Fig. 2a-b).

The TENG charge inventory includes contact-generated and pre-existing source charges, compensating charges associated with source establishment, and conductor charges induced by electrostatic equilibrium. In the large-area limit, material-established charges are prescribed while conductor-response charges remain unknown. Vanishing fields inside metals, equipotentiality of connected electrodes and free-charge conservation then determine the complete conductor state before transferred charge or terminal voltage is evaluated (Supplementary Note 7).

Consider a generator first equilibrated under short circuit at separation \(z_{0}\). Independently solving the conductor-charge state at \(z_{0}\) and at a subsequent separation \(z\) gives

\begin{equation*}
Q_{\text{sc}}(z;z_{0}) = \frac{\sigma_{s}\varepsilon_{r}Sd_{0}\left( z - z_{0} \right)}{\left( \varepsilon_{r}z + d_{0} \right)\left( \varepsilon_{r}z_{0} + d_{0} \right)}\tag{2}\label{eq:2}
\end{equation*}

Here \(S\) is the active area. \(Q_{\text{sc}}\) is therefore the difference between two closed conductor states rather than an independently assigned source quantity. The compensation charge associated with source establishment remains in the absolute conductor-charge bookkeeping but cancels from the geometry-dependent redistribution and transferred charge (Supplementary Note 7).

Equation (2) gives \(Q_{\text{sc}} \propto S\) in the laterally uniform limit and an intrinsically nonlinear dependence on separation. For a prescribed trajectory \(z(t)\), \(I_{\text{sc}}(t) = \frac{dQ_{\text{sc}}}{dt} = \frac{\partial Q_{\text{sc}}}{\partial z}\dot{z}.\) Mechanical velocity is therefore weighted by a position-dependent charge sensitivity, so single-frequency motion need not generate a sinusoidal electrical trajectory.

The CS-TENG charge-transfer law thus emerges from conductor-state determination rather than an assigned output relation. Analogous reductions follow for fixed-charge electrets with equivalent electrostatic geometries and for deformation-induced polarization in piezoelectric systems (Supplementary Notes 3-4). In each case, the electrical output is evaluated only after solving the conductor state.

The same electrostatic closure extends to finite geometry without prescribing a laterally uniform electrode response. The unknowns are spatially varying conductor surface-charge densities coupled nonlocally through their electrostatic potentials. For each conductor, the surface must remain equipotential even though its free-charge density is generally nonuniform; the corresponding total-charge constraints depend on the electrical boundary and preceding history (Fig. 2c; Supplementary Note 8). Integrated charge, transferred charge, potential and electric field follow only after this self-consistent spatial state has been determined.

The analytical and finite-geometry solutions express the same electrostatic constraints at different levels of spatial resolution. At small normalized separation, the finite-plate solution approaches the nearly uniform analytical state in the interior; as separation becomes comparable to lateral dimensions, edge redistribution and fringing become important (Fig. 2d-f). Their systematic differences delimit where the parallel-plate description is physically justified, while showing how a terminal-scale relation emerges from a spatially resolved state. The same state constraints also underlie the measurement-boundary analysis in Fig. 3.

\begin{figure}[!htbp]
\centering
\includegraphics[width=0.99\linewidth,height=0.69\textheight,keepaspectratio]{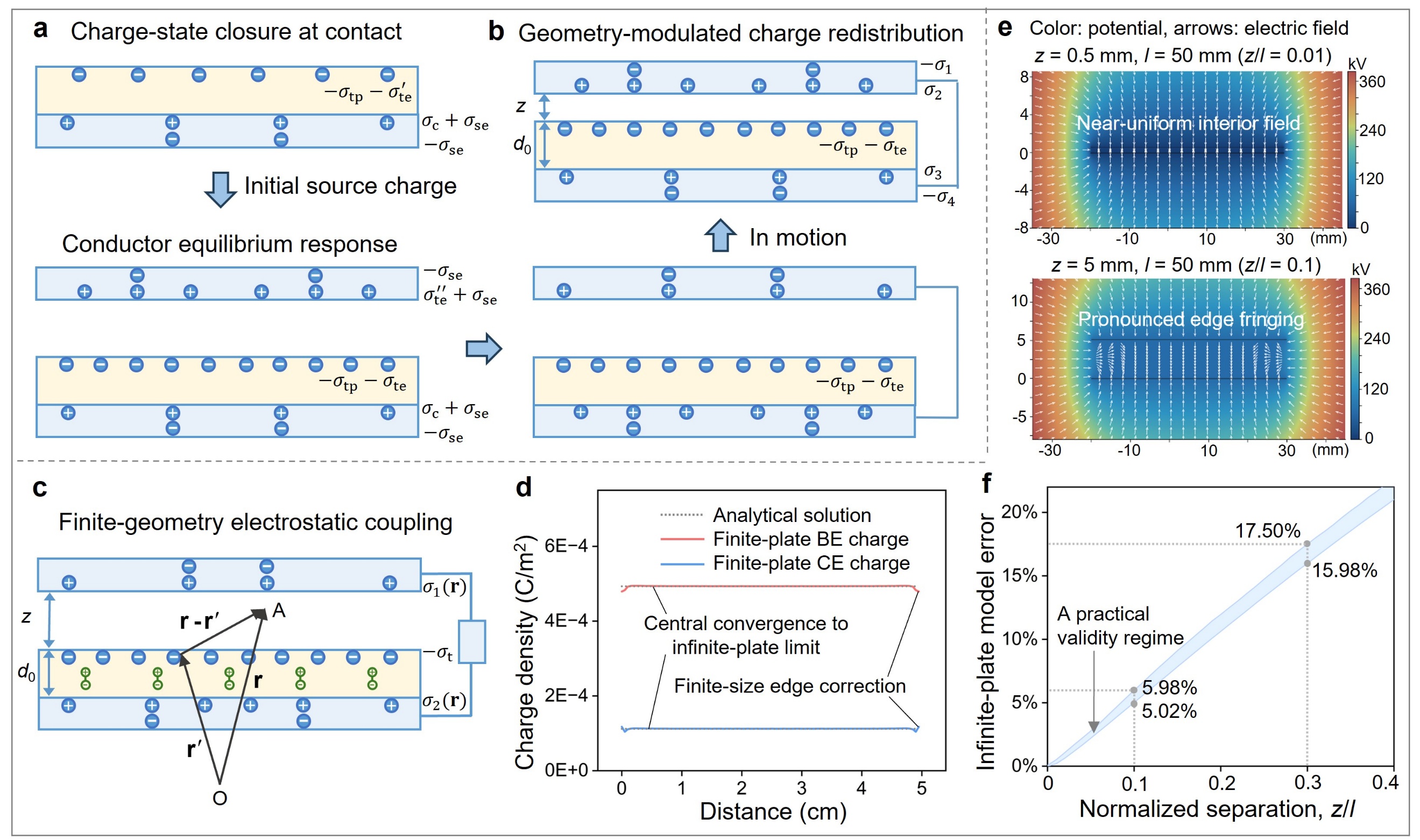}
\caption{Charge-state closure and finite-geometry convergence in a contact-separation TENG. \textbf{a}, Contact-generated and pre-existing surface charges establish the source state; conductor free charge subsequently redistributes to satisfy electrostatic equilibrium and charge conservation. \textbf{b}, Separation changes the electrostatic coupling and hence the equilibrium conductor-charge distribution. \textbf{c}, In finite geometry, surface elements interact nonlocally and the complete conductor-charge distribution is solved self-consistently. \textbf{d}, Comparison between the large-area analytical and finite-geometry charge densities on the base and counter electrodes. The finite solution approaches the uniform analytical state in the central region while edge corrections emerge near the boundaries. \textbf{e}, Representative potential distributions and electric-field vectors for a 5 \(\times\) 5 cm\textsuperscript{2} TENG. At \(z = 0.5\) mm (\(z/L = 0.01\)), the interior field is nearly uniform, whereas at \(z = 5\) mm (\(z/L = 0.1\)), pronounced edge fringing develops. \textbf{f}, Relative deviation of the large-area analytical approximation from finite-geometry calculations over representative dimensions. The deviation increases systematically with \(z/L\) and remains below approximately 6\% at \(z/L = 0.1\) over the examined parameter range.}
\label{fig:2}
\end{figure}

\section{Voltage reveals boundary-selected electrostatic states}

Voltage measurement first reveals why the distinction between electrical source and conductor state matters. Across electrostatic generators, the same source and motion can produce different observed voltages when the measuring circuit selects different conductor-charge states. A contact-separation TENG makes this boundary selection analytically explicit; other generator classes test its shared physical meaning. The zero-transfer, charge-frozen state identified here will also provide an independent reference for the resistive-load behaviour examined below.

A striking prediction emerges when the electrical source and mechanical motion remain unchanged but the measurement boundary is varied. The voltage need not be a unique, source-defined output: different electrical boundaries can select different conductor-charge states. For generators of different electrode areas, the theory predicts a common intrinsic-voltage limit under charge isolation and an area-dependent measured-voltage ratio when charge exchange is permitted. Consider a TENG initially equilibrated under short circuit at separation \(z_{0}\). The conductor groups are then isolated, fixing their total free charges. When the separation changes to \(z_{1}\), local surface charge can redistribute, but no charge can pass between conductor groups. Solving the electrostatic state under these charge-frozen constraints gives

\begin{equation*}
V_{\text{int}}(z_{1};z_{0}) = \frac{\sigma_{s}d_{0}\left( z_{1} - z_{0} \right)}{\varepsilon_{0}\left( \varepsilon_{r}z_{0} + d_{0} \right)}\tag{3a}\label{eq:3a}
\end{equation*}

We define \(V_{\text{int}}\) as the intrinsic voltage: the zero-transfer potential difference for a specified source state, geometry and preceding charge history. It is therefore not a history-independent material or terminal constant (Fig. 3a).

A finite measurement capacitance \(C_{m}\) permits charge \(q_{m}\) to transfer between conductor groups\cite{mref39}. This transferred charge is an additional unknown of the coupled electrostatic problem. Charge conservation, conductor equilibrium and the capacitor boundary are solved simultaneously, generating a family of \(C_{m}\)-selected conductor-charge states (Fig. 3b, Supplementary Note 9). For motion from \(z_{0}\) to \(z_{1}\),

\begin{equation*}
q_{m} = \frac{\sigma_{s}\varepsilon_{r}Sd_{0}\left( z_{1} - z_{0} \right)C_{m}}{\left( \varepsilon_{r}z_{0} + d_{0} \right)\left\lbrack C_{m}\left( \varepsilon_{r}z_{1} + d_{0} \right) + \varepsilon_{0}\varepsilon_{r}S \right\rbrack},\tag{3b}\label{eq:3b}
\end{equation*}

with \(V_{\text{meas}} = \frac{q_{m}}{C_{m}}.\) The dependence \(q_{m} = q_{m}\left( C_{m} \right)\) shows that measurement capacitance changes inter-conductor charge exchange and hence the state whose potential difference is observed. In the limit \(C_{m} \rightarrow 0\), \(q_{m} \rightarrow 0\) and \(V_{\text{meas}} \rightarrow V_{\text{int}}\); increasing \(C_{m}\) progressively moves the system towards its short-circuit response.

The physical conclusion extends across electrostatic generators: the measurement circuit is a state-selecting electrical boundary. Equations (3a) and (3b) make that effect explicit in the representative CS-TENG, while electret, piezoelectric and other source mechanisms enter through their own charge-potential response. In every case the observed voltage is read from the conductor state established under the actual measurement boundary, rather than from the material source alone.

After the relevant electrostatic state family has been solved, we define the local charge-potential response coefficient

\begin{equation*}
\chi_{Q}\equiv-\left(\frac{\partial V}{\partial Q}\right)^{-1}_{\mathcal{S},\Gamma,\mathcal{H}_{0}}\tag{4}\label{eq:4}
\end{equation*}

For the linear one-dimensional CS-TENG, \(\chi_{Q}\left( z_{1} \right) = \frac{Q_{\text{sc}}(z_{1};z_{0})}{V_{\text{int}}(z_{1};z_{0})} = \frac{\varepsilon_{0}\varepsilon_{r}S}{\varepsilon_{r}z_{1} + d_{0}} = \frac{\varepsilon_{0}S}{z_{1} + d_{0}/\varepsilon_{r}}.\)

In this linear one-dimensional limit, the independently determined charge-potential response coefficient coincides with the geometrical capacitance of the parallel-plate system. Conventional voltage-charge relations can therefore reproduce the terminal response exactly under these conditions. The state-based derivation establishes the physical origin and scope of this reduction; finite-geometry capacitance-matrix descriptions similarly retain integrated terminal information\cite{mref40}, whereas spatial charge redistribution and fringing fields follow from the full electrostatic solution. The finite-measurement solution becomes \(\frac{q_{m}}{Q_{\text{sc}}} = \frac{C_{m}}{C_{m} + \chi_{Q}},\ \frac{V_{\text{meas}}}{V_{\text{int}}} = \frac{\chi_{Q}}{C_{m} + \chi_{Q}}.\) Defining \(\Pi_{m} = \frac{C_{m}}{\chi_{Q}},\) the measurement crossover is controlled by competition between the external measurement boundary and the intrinsic charge-potential response of the solved generator state. Because \(\chi_{Q} \propto S\) for the analytical CS-TENG, \(C_{m}^{*} \propto S\).

Equation (4) illustrates how familiar terminal descriptions emerge from solved charge states rather than replace them. Electret, piezoelectric, ionic and liquid-solid generators likewise yield their own source- and boundary-dependent terminal charge-response relations (Supplementary Notes 3-6), linking the full physical state to measurable terminal behaviour.

The voltage crossover makes the difference between electrical source and measured state experimentally accessible. With source and mechanical excitation held fixed, two TENG areas approach the same intrinsic-voltage limit at small measurement capacitance, whereas increasing capacitance reveals the area dependence of the boundary-selected state (Fig. 3c). Piezoelectric generators show a corresponding crossover at their characteristic capacitance scale (Fig. 3d); electret, piezoelectret, ionoelectric-gel and solid-liquid devices provide further tests of measurement-boundary sensitivity (Extended Data Figs. 1 and 2). Changing the initial short-circuit equilibration separation also leaves a persistent voltage offset under otherwise identical motion (Fig. 3e; Extended Data Fig. 1b). Thus different measurement conditions and different preceding charge histories can select different conductor states without requiring a different material source. The charge-frozen limiting state provides a physical reference against which the resistive-load response can later be compared.

\begin{figure}[!htbp]
\centering
\includegraphics[width=0.99\linewidth,height=0.69\textheight,keepaspectratio]{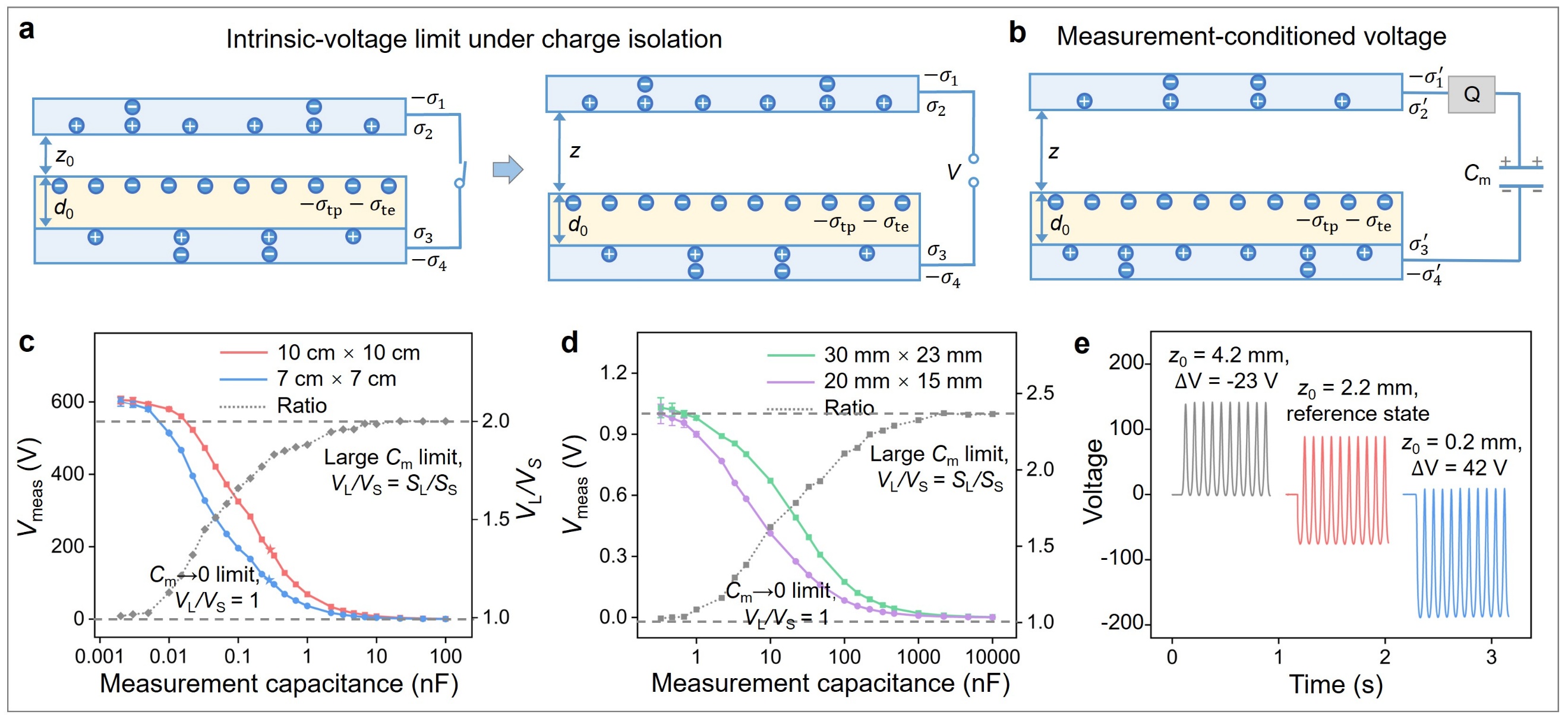}
\caption{Predicted and measured voltage crossover reveals boundary-selected charge states and charge-history dependence. \textbf{a}, Intrinsic-voltage state established by short-circuit equilibration at \(z_{0}\), followed by charge isolation and motion to \(z_{1}\). \textbf{b}, Finite measurement capacitance \(C_{m}\) permits charge exchange and selects a different conductor-charge state, so both \(q_{m}\) and \(V_{\text{meas}}\) depend on the measurement boundary. \textbf{c}, Measured voltage of 10 \(\times\) 10 cm\textsuperscript{2} and 7 \(\times\) 7 cm\textsuperscript{2} TENGs versus \(C_{m}\). As \(C_{m} \rightarrow 0\), charge transfer vanishes and \(V_{L}/V_{S} \rightarrow\)1; increasing \(C_{m}\) reveals area scaling and \(V_{L}/V_{S} \rightarrow S_{L}/S_{S}\). Error bars denote the standard deviation. \textbf{d}, Corresponding response of piezoelectric generators with electrode areas of 30 \(\times\) 23 mm\textsuperscript{2} and 20 \(\times\) 15 mm\textsuperscript{2}. Despite the distinct upstream polarization source, the response crosses from an approximately area-independent small-\(C_{m}\) intrinsic-voltage limit towards an area-scaled large-\(C_{m}\) state. \textbf{e}, TENG voltage waveforms following different initial equilibration separations \(z_{0}\). The subsequent mechanical trajectory is unchanged, and the persistent voltage-state shifts demonstrate the role of preceding charge history.}
\label{fig:3}
\end{figure}

\section{Nonlinear charge trajectories determine electrical harmonics}

Electrical harmonics probe the time dependence of the determined charge state rather than an independently assigned output spectrum. The contact-separation TENG provides an explicit geometry-to-charge calculation; other generator classes enter through their distinct source dynamics. External loading can subsequently constrain this charge trajectory.

The charge-state solution also predicts the spectral content of electrical output from prescribed mechanical motion. Even a single-frequency drive can generate higher electrical harmonics through the nonlinear dependence of transferred charge on geometry. For a specified electrical boundary and preceding charge history, the transferred charge follows the evolving source and geometry, \(Q(t) = Q\text{ }[\mathcal{S}(t),\Gamma(t)\mathcal{\mid B}\mathcal{H}_{0}\rbrack.\) The time dependence can arise from source evolution, geometrical modulation of an established source, or both. The resulting nonlinear mapping establishes which additional frequencies appear in electrical output (Fig. 4a).

For the analytical CS-TENG, Eq. (2) gives the nonlinear mapping \(Q_{\text{sc}}(z)\). Substituting the prescribed periodic trajectory \(z(t)\) generates the charge waveform, which is then decomposed spectrally. The exact sinusoidal-motion series and local nonlinear expansion are derived in Supplementary Note 10.

Smaller internal separation and larger mechanical excursion sample stronger curvature of \(Q_{\text{sc}}(z)\) and therefore enhance normalized higher-order spectral components. In contrast, active area primarily rescales absolute charge and current in the large-area limit while leaving the normalized harmonic distribution unchanged.

Once the nonlinear charge trajectory is determined, the electrical current spectrum follows without a separate source model. If \(Q_{n}\) denotes the Fourier component at \(\omega_{n} = n\omega_{0}\), \(I_{n} = in\omega_{0}Q_{n},\  \mid I_{n} \mid = n\omega_{0} \mid Q_{n} \mid .\) Higher harmonics therefore originate in the nonlinear source/geometry-to-charge mapping; differentiation introduces no new frequencies and only determines how the generated charge harmonics appear in current amplitude and phase (Supplementary Note 11).

TENG measurements also show that current harmonics, when normalized by their corresponding angular frequencies, coincide with the transferred-charge harmonics: \(\mid I_{n} \mid /n\omega_{0}\) matches \(\mid Q_{n} \mid\) over the resolved orders (Fig. 4b,c). This correspondence confirms that the measured current inherits the spectral structure of the solved charge trajectory. The strongest spectral tests concern how geometry shapes the harmonic distribution: increasing excursion or reducing mean separation enhances normalized higher-order components, whereas active area mainly rescales absolute amplitudes (Fig. 4d,e; Extended Data Fig. 3).

Piezoelectric, electret, piezoelectret, ionoelectric-gel and solid-liquid generators also satisfy the harmonic-resolved charge-current correspondence under the examined conditions. This cross-generator result tests the common downstream identity \(I = dQ/dt\) and extends the charge-state description across distinct generator classes (Fig. 4f, Extended Data Figs. 4 and 5).

Across generator classes, electrical frequencies are carried by the evolving transferred-charge state, whose time dependence follows source establishment or evolution, geometry and electrical boundary. The TENG makes nonlinear geometrical coupling explicit; other generators involve deformation-induced polarization, ionic transport, or changing contact and induction geometry at liquid-solid interfaces (Supplementary Notes 4-6). Distinct mechanisms therefore generate different charge waveforms within the same state-to-observation route. The load experiments below show how restricted charge exchange reshapes this spectral structure.

\begin{figure}[!htbp]
\centering
\includegraphics[width=0.99\linewidth,height=0.69\textheight,keepaspectratio]{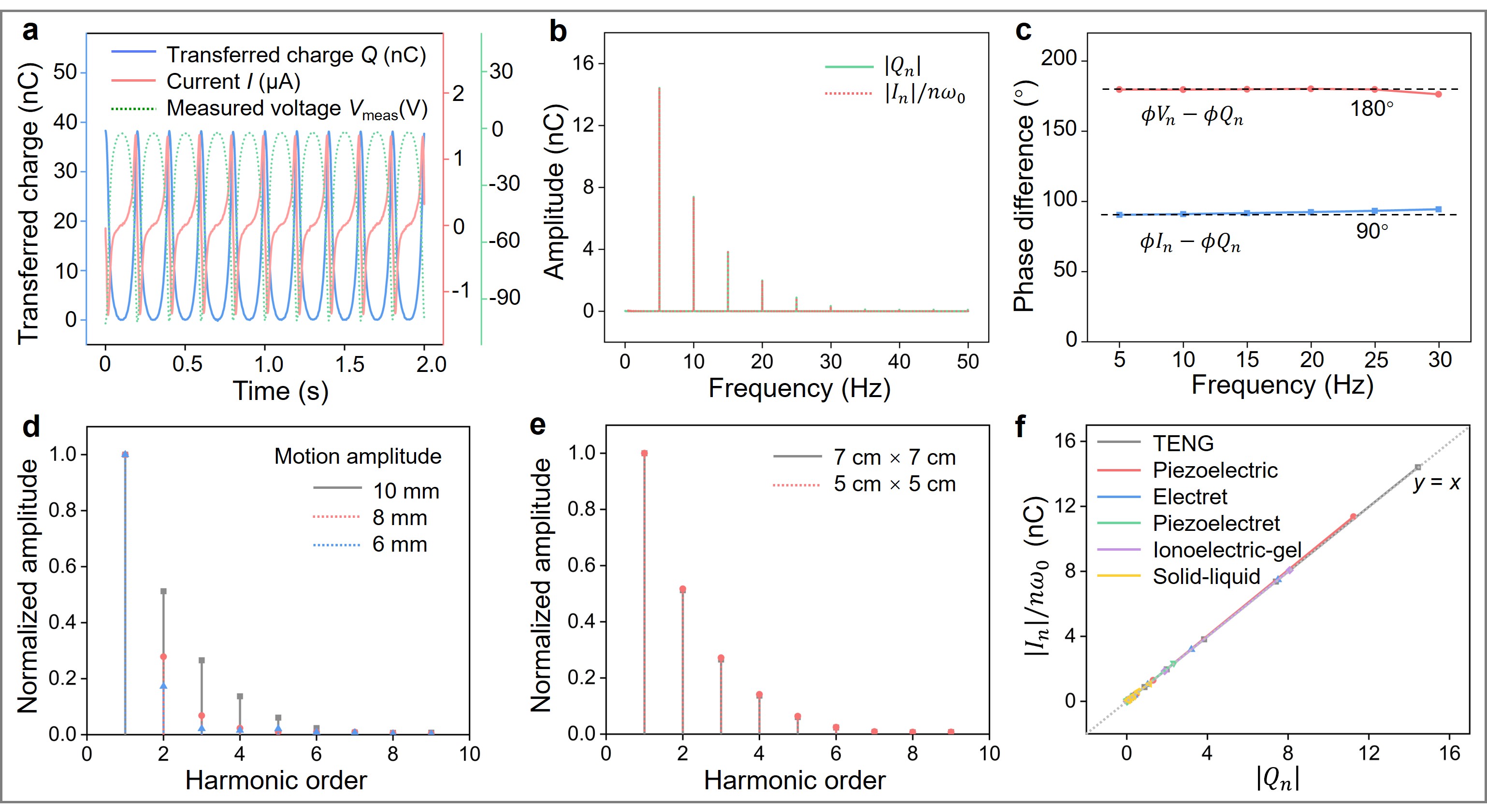}
\caption{Geometry-predicted electrical harmonics and charge-determined spectral response across generators. \textbf{a}, Representative transferred-charge, current and measured-voltage waveforms of a contact-separation TENG under periodic mechanical excitation. The non-sinusoidal electrical response arises from the nonlinear geometry-to-charge mapping. \textbf{b}, Harmonic amplitudes \(\mid Q_{n} \mid\) and frequency-normalized current \(\mid I_{n} \mid /n\omega_{0}\), showing harmonic-by-harmonic correspondence. \textbf{c}, Harmonic-resolved phase relations among current, transferred charge and measured voltage. Current exhibits the quadrature relation arising from differentiation; the apparent voltage-charge inversion reflects the adopted polarity convention. \textbf{d}, Normalized harmonic spectra at different mechanical amplitudes; larger excursions enhance higher-order components by sampling a broader nonlinear region of the charge-geometry relation. \textbf{e}, Normalized spectra for different TENG areas under otherwise identical excitation; area primarily rescales absolute output while leaving the normalized spectral distribution largely unchanged. \textbf{f}, Cross-generator comparison of \(\mid I_{n} \mid /n\omega_{0}\) and \(\mid Q_{n} \mid\) for triboelectric, piezoelectric, electret, piezoelectret, ionoelectric-gel and solid-liquid generators. The correspondence reflects the common downstream identity \(I = dQ/dt\), not a shared microscopic source law.}
\label{fig:4}
\end{figure}

\section{Charge relaxation predicts load-dependent behaviour}

Electrical loading reveals how the same charge trajectory is reshaped by a different electrical boundary. An external resistance constrains charge exchange while the material source or mechanical geometry evolves; the conductor state can therefore follow, lag behind or remain near its charge-frozen reference. This explains load-dependent transitions in different generator classes, with the contact-separation TENG providing an explicit analytical example. The load problem also connects the preceding observations in two ways. In the high-resistance limit it approaches the zero-transfer state encountered in the capacitive voltage measurements (Fig. 3); at finite resistance it reshapes the nonlinear charge trajectory and preferentially suppresses higher spectral components (Fig. 4). Thus voltage measurement, harmonic generation and electrical loading are different probes of the same state-determination process.

Under a resistance \(R_{L}\), the accumulated transferred charge fixes the total charge of the conductor groups at each instant. Together with the instantaneous source state and geometry, these totals define the electrostatic state and its terminal potential \(V_{p}(t)\); the resistor then determines the subsequent charge transfer. Updating the conductor totals and re-solving the electrostatic state gives the general load-constrained evolution (Supplementary Note 12): \([\mathcal{S}(t),\Gamma(t),\mathcal{H}_{0},Q(t)\rbrack \rightarrow \rho(\mathbf{r},t) \rightarrow V_{p}(t),\) with \(V_{p} = R_{L}I,\ I = \frac{dQ}{dt}\), and therefore \(R_{L}\frac{dQ}{dt} = V_{p}[\mathcal{S}(t),\Gamma(t),\mathcal{H}_{0},Q(t)\rbrack\). In finite geometry, each charge increment changes the conductor totals and the complete spatial conductor-charge distribution is re-solved at the updated geometry.

For the analytical one-dimensional CS-TENG, eliminating the explicit conductor-surface charges gives an exact scalar reduction. Let \(Q_{\text{eq}}(t)\) denote the transferred charge of the independently closed instantaneous short-circuit state referenced to the same initial condition as \(Q(t)\). The actual terminal potential is \(V_{p}(t) = \frac{Q_{\text{eq}}(t) - Q(t)}{\chi_{Q}(t)}\), with \(\chi_{Q}(t) = \frac{\varepsilon_{0}\varepsilon_{r}S}{\varepsilon_{r}z(t) + d_{0}}\). Hence

\begin{equation*}
\frac{dQ}{dt} = \frac{Q_{\text{eq}}(t) - Q(t)}{R_{L}\chi_{Q}(t)}\tag{5}\label{eq:5}
\end{equation*}

Equation (5) is an exact reduced consequence of the one-dimensional electrostatic charge solution, not an assumed RC-equivalent starting law. The difference \(Q_{\text{eq}}(t) - Q(t)\) measures the departure of the actual conductor state from its moving short-circuit equilibrium and drives subsequent charge transfer.

Equation (5) makes the charge-relaxation picture explicit for a linear two-terminal electrostatic response. Each generator supplies its own instantaneous reference charge and charge-potential response through the source-driven state solution. The electret description yields a variable-capacitance charge equation; the piezoelectric description couples that equation to mechanical evolution; and ionoelectric and liquid-solid systems connect their transport-controlled or interfacial source terms to electrode-charge and external-load constraints (Supplementary Notes 3-6). The shared organizing principle is that electrical charge transfer and the evolving source jointly determine the conductor state from which load-dependent observations arise.

The instantaneous electrical relaxation time is \(\tau_{L}(t) = R_{L}\chi_{Q}(t).\) For comparison of complete periodic trajectories, we choose a fixed reference geometry with charge-response coefficient \(\chi_{\star}\) and define \(\Pi_{L} = \omega R_{L}\chi_{\star}.\) For \(\Pi_{L} \ll\)1, conductor charge closely follows the moving electrostatic equilibrium; for \(\Pi_{L} \sim\)1, electrical relaxation competes with mechanical driving; and for \(\Pi_{L} \gg\)1, inter-conductor transfer becomes progressively frozen and the terminal potential approaches the intrinsic-voltage state.

For a fixed dielectric stack and separation trajectory in the large-area limit, nondimensionalization reduces the nonlinear load response to \(\Pi_{L}\), giving

\begin{equation*}
R_{c},\text{\:\,}R_{L,\text{opt}} \propto \frac{1}{\omega\chi_{\star}} \propto \frac{1}{\omega S}\tag{6}\label{eq:6}
\end{equation*}

The numerical coefficients depend on the operational definitions of crossover and optimum, whereas the inverse area-frequency scaling follows when the charge-response coefficient scales with electrode area. Using the field-standard peak-to-peak convention, \(P_{\text{pp}} = I_{\text{pp}}^{2}R_{L}.\) The nonlinear optimum and its reduced single-frequency limit are derived in Supplementary Note 12.

The inverse area-frequency dependence in Eq. (6) provides a concrete example of how a characteristic load scale emerges from the determined charge-response coefficient and the driving timescale. Across generator types, the same physical competition between source or geometry evolution and constrained conductor-charge exchange explains the appearance of load transitions, while each device\textquotesingle s response coefficient establishes its own scale.

The measured current-voltage transition moves to lower resistance with increasing area or frequency (Fig. 5a), and responses obtained under different conditions approximately collapse when resistance is normalized by the theoretical characteristic scale (Fig. 5b). Time-domain charge and current waveforms reveal progressively restricted transfer (Fig. 5c), while higher-order harmonics are preferentially attenuated (Fig. 5d), directly linking load-induced trajectory changes to the nonlinear spectra established in Fig. 4. The predicted inverse area-frequency scaling is borne out by the TENG load measurements. For a 10 \ensuremath{\times} 10 cm\textsuperscript{2} device, increasing the driving frequency from 8 to 12 and 16 Hz shifts the peak-to-peak power optimum from approximately 90 to 60 and 40 M\ensuremath{\Omega}, respectively. At 12 Hz, increasing the active area from 5 \ensuremath{\times} 5 to 7 \ensuremath{\times} 7 and 10 \ensuremath{\times} 10 cm\textsuperscript{2} shifts the optimum from approximately 200 to 100 and 60 M\ensuremath{\Omega} (Fig. 5e).

Piezoelectric generators show corresponding area-dependent load transitions, normalized responses and optimum-resistance shifts at their own characteristic scales (Fig. 5f-h), with electret generators providing further cross-mechanism evidence (Extended Data Fig. 6). Crucially, the load experiments return to the charge-frozen state independently identified by the voltage measurements. Reducing measurement capacitance limits the charge exchanged with the measurement branch, whereas increasing load resistance restricts charge transfer over the observation interval. Under matched source and initial-charge conditions, both routes approach the intrinsic-voltage state established by the charge-state solution (Fig. 3; Fig. 5a,f; Extended Data Fig. 7). Thus measurement-conditioned voltage and load-dependent dynamics are not merely analogous responses: they access the same underlying physical state through different electrical constraints.

\begin{figure}[!htbp]
\centering
\includegraphics[width=0.99\linewidth,height=0.69\textheight,keepaspectratio]{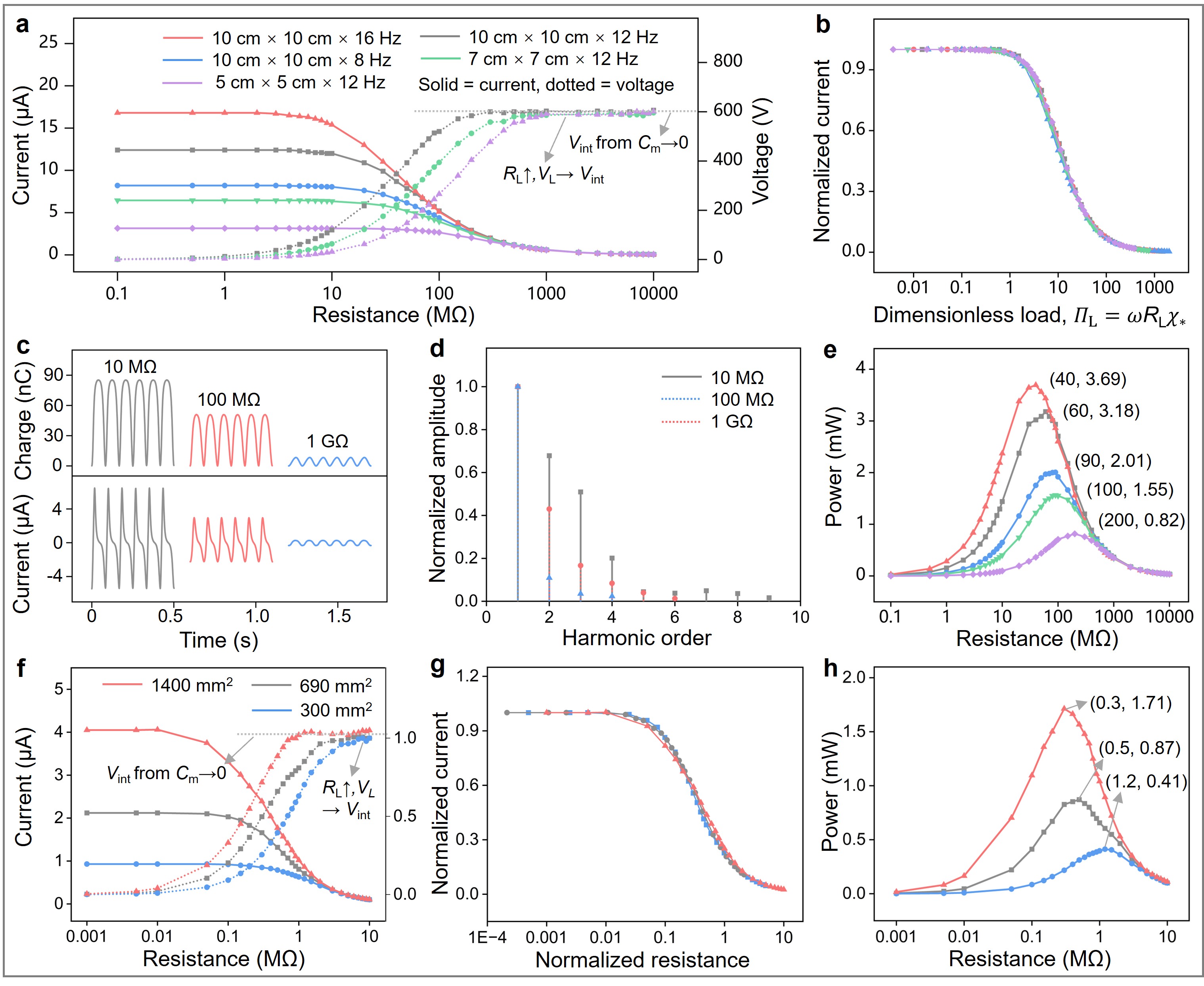}
\caption{Predicted load scaling and charge-state convergence across active electrostatic generators. \textbf{a}, Load-dependent current and voltage of contact-separation TENGs at different active areas and operating frequencies. Increasing \(R_{L}\) suppresses current while the terminal voltage approaches the independently determined intrinsic-voltage limit. For \(R_{L}\) \(>\) 500 M\ensuremath{\Omega}, the current standard deviation is below 0.006 \ensuremath{\mu}A. \textbf{b}, Normalized load response plotted against \(\Pi_{L} = \omega R_{L}\chi_{\star}\). Measurements obtained at different device areas and frequencies approximately collapse, showing that the transition is governed by competition between electrical charge relaxation and mechanical driving. \textbf{c}, Representative transferred-charge and current waveforms under increasing \(R_{L}\). Increasing resistance progressively prevents the actual conductor-charge trajectory from following the instantaneous equilibrium trajectory. \textbf{d}, Normalized current harmonic spectra under different load resistances. Higher-order harmonics are preferentially suppressed because each spectral component experiences an effective load parameter \(n\omega_{0}R_{L}\chi_{Q}\). \textbf{e}, Peak-to-peak power metric \(P_{\text{pp}} = I_{\text{pp}}^{2}R_{L}\) as a function of resistance for different device areas and frequencies. The optimal resistance shifts towards lower values with increasing frequency and active area, consistent with \(R_{L,\text{opt}} \propto (\omega S)^{- 1}\). \textbf{f}, Piezoelectric current and voltage as functions of resistance for different active areas. For \(R_{L}\) \(>\) 0.5 M\ensuremath{\Omega}, the current standard deviation is below 0.026 \ensuremath{\mu}A. The load transition and characteristic resistance shift with electrode area despite the distinct upstream polarization mechanism. \textbf{g}, Normalized piezoelectric load curves plotted against normalized resistance. \textbf{h}, Area dependence of the optimal piezoelectric load resistance.}
\label{fig:5}
\end{figure}

\section{Discussion}

The central result of this work is a common physical determination route connecting the sources of electrostatic generators to the charge states responsible for their measured behaviour. Surface and trapped charges, deformation-induced polarization, mobile ionic populations and liquid-solid interfacial sources arise from different material processes. Yet the conductor charges induced by these sources are selected by geometry, electrostatic constraints, electrical boundaries and preceding charge history. The contact-separation TENG, with its coexisting source, compensating and induced charges, makes that route explicit: analytical and finite-geometry solutions resolve the physical charge state before obtaining the electrical observations. This shifts the organizing question from how to describe a particular output to which charge state gives rise to it.

A first line of evidence is the recovery of differently formulated classical generator theories within this source-to-state sequence. The charge relations of the TENG, the electret bias-source-variable-capacitance model and the piezoelectric electromechanical port model emerge from their respective source states and conductor constraints. The same organization connects Nernst-Planck-Darcy ion transport to electrode charge and characteristic piezoionic response, and contact electrification with electric-double-layer screening to the induced-charge response of a liquid-solid generator (Supplementary Notes 3-7). Together, these recoveries establish a physically explicit charge-state determination across distinct source mechanisms, providing a common basis for interpreting different electrical measurements of the same system.

The second line of evidence is the way independently varied electrical conditions expose relationships between behaviours usually examined separately. At unchanged material source and mechanical excitation, a changed measurement capacitance selects a different conductor-charge state, while changing the preceding equilibration establishes a different reference state. Nonlinear source-geometry coupling gives a charge trajectory with higher harmonics; a resistive load then constrains that very trajectory and preferentially suppresses its faster components. The two independent approaches to charge isolation are particularly revealing: reducing measurement capacitance and increasing load resistance both approach the charge-frozen state when source, geometry and charge history are matched (Figs. 3 and 5). The spectral and load observations likewise describe successive stages of one charge-evolution process (Figs. 4 and 5). Together, these cross-boundary and cross-spectral connections reveal how distinct electrical measurements interrogate a shared, evolving physical charge state.

The resulting interpretive framework changes the question asked of an electrical measurement. A voltage change need not indicate new material charging: it can reveal a changed boundary-selected or history-selected conductor state. This distinction is essential for interpreting electrical measurements and comparing generator behaviour across instruments, devices and operating conditions. An electrical spectrum reflects the trajectory of charge established by source physics and motion, whereas a load changes the trajectory being measured rather than merely assigning a voltage to an otherwise fixed current. The same distinction can be carried across generator classes because their material-specific constitutive, transport and interfacial laws supply different sources to a common electrostatic determination of conductor response. More broadly, this work illustrates how a field can advance from separately describing observed behaviours to understanding their relationships through a common, physically constrained state-determination route. The resulting quantitative connections allow independent measurements to test a shared physical explanation, making unification predictive and experimentally testable rather than merely descriptive.

\section{Methods}

\subsection*{Generator preparation and mechanical excitation}

The primary experimental platform was a basic contact-separation TENG selected for its structural simplicity and ease of reproduction. Square devices with active areas of 5 \(\times \ \)5, 7 \(\times \ \)7 and 10 \(\times \ \)10 cm\textsuperscript{2} were used. A 100-\ensuremath{\mu}m FEP film served as the dielectric layer, with a conductive-ink electrode printed on its back surface. The independent electrode was a 100-\ensuremath{\mu}m aluminium film. Both films were fixed to acrylic plates and mounted on a programmable LinMot linear motor (model PS01-37x120-C).

Before measurement, each TENG was repeatedly operated in contact-separation mode until the output became stable after the initial charge-accumulation process. The contact position was established by moving the two surfaces to just-contact and taking this position as the mechanical reference. The motor was then retracted by a prescribed distance to set the internal separation. The programmed motor displacement therefore defined \(z(t)\), including the initial gap, excursion and driving frequency.

For sinusoidal-motion experiments, the sinusoidal trajectory available in the LinMot control software was used. Because the complete mechanical assembly can cause small deviations from the programmed trajectory, the controller compensation parameters were adjusted when necessary so that the forward and return velocities at corresponding positions, particularly near the TENG current peaks, were comparable.

A commercial PTFE electret sheet of 10 \(\times \ \)10 cm\textsuperscript{2} area and 100-\ensuremath{\mu}m thickness was used as received, without an additional charging treatment. A finite gap was maintained between the PTFE and the independent electrode so that the two surfaces did not contact during operation.

For piezoelectric and piezoelectret measurements, a force sensor (model SBT730) was mounted on the linear-motor stage and coupled to an acrylic loading plate used to compress the sample. The piezoelectric device was a commercial PZT ceramic with a thickness of 1 mm and a supplier-reported \(d_{33} =\) 650 pC N\textsuperscript{-}\textsuperscript{1}. The piezoelectret was a commercial 6 \(\times \ \)6 cm\textsuperscript{2} sheet with a thickness of 100 \ensuremath{\mu}m and a supplier-reported effective piezoelectric coefficient of 200 pC N\textsuperscript{-}\textsuperscript{1}.

The solid-liquid extension device used a droplet configuration. An FEP film covered two aluminium electrodes, each 4 \(\times \ \)1 cm\textsuperscript{2}, separated by 4 cm. Water droplets were supplied from a reservoir/dropper assembly onto the FEP surface.

The ionoelectric-gel extension sample was prepared from EMIMTFSI and PDMS at a mixing ratio of 1:10, thermally cured at 80 \textdegree{}C for 12 h and subsequently UV-cured for 1 h. The ionoelectric-gel and solid-liquid devices provide complementary experimental evidence of boundary-conditioned electrical responses across distinct source mechanisms; the source-to-electrode-charge route is demonstrated with representative literature models in Supplementary Notes 5 and 6.

\subsection*{Electrical acquisition}

Charge, current and voltage were measured with a Keithley 6514 electrometer. The electrometer output was acquired through an NI-9223 data-acquisition module and recorded in LabVIEW. The acquisition program displayed the time-domain waveform and could directly report peak-to-peak amplitudes.

For finite-capacitance measurements, the generator, an external high-voltage capacitor and the electrometer were connected in series, with the electrometer operated in charge mode. During generator motion, the transferred charge admitted by the measurement boundary was recorded in LabVIEW. The corresponding measurement-boundary voltage was obtained from \(V_{\text{meas}} = \frac{q_{m}}{C_{m}}.\) The capacitance was varied across the boundary transition of each generator class while the device structure and mechanical condition were otherwise maintained.

For TENG charge-history measurements, the device was first positioned at a prescribed internal separation \(z_{0}\) and the two electrodes were electrically connected to establish the corresponding equilibrium conductor-charge state. The connection was then interrupted and the selected measurement boundary was applied. The subsequent mechanical trajectory was kept unchanged while \(z_{0}\) was varied between measurements.

Harmonic measurements used 2-min time records sampled at 1000 Hz. Fourier analysis was performed on the recorded transferred-charge, current and voltage waveforms to extract the mechanical fundamental and resolved harmonics. Harmonic frequencies were defined by \(\omega_{n} = n\omega_{0}\), and charge-current comparisons used the relation \(I_{n} = in\omega_{0}Q_{n}.\)

\subsection*{Resistive-load measurements}

Commercial resistors spanning the range required by the different generator classes were inserted in the generator current path; the complete available range was approximately 1 k\(\Omega\) to 10 G\(\Omega\). The piezoelectric load transition occurred mainly in the k\(\Omega\) regime, whereas the other generator classes were characterized mainly in the M\(\Omega\) regime. Current was reported using the peak-to-peak amplitude \(I_{\text{pp}}\). Following the peak-to-peak convention commonly used for load curves in this field, the corresponding voltage and power metric were calculated as \(V_{\text{pp}} = I_{\text{pp}}R\) and \(P_{\text{pp}} = I_{\text{pp}}^{2}R\). The same convention was used throughout each resistance series. The high-resistance value of \(V_{\text{pp}}\) was compared with the voltage approached in the small-capacitance measurement limit as an internal consistency check of the charge-frozen state. This peak-to-peak metric is used for consistent comparison of resistance-dependent output; it is distinct from the cycle-averaged electrical power.

\subsection*{Finite-geometry electrostatic calculations}

Finite-plate calculations implemented the charge-distribution solution described in Supplementary Note 8 in Python. The model used the experimental TENG geometry with a 5 \(\times \ \)5 cm\textsuperscript{2} active area, FEP relative permittivity \(\varepsilon_{r} =\) 2.2, dielectric thickness \(d_{0} = \ \)100 \ensuremath{\mu}m and source-charge density \(\sigma_{s} = \ \)1.8 \(\times \ \)10\textsuperscript{-5} C m\textsuperscript{-}\textsuperscript{2}. Representative internal separations of 0.5 mm and 5 mm were calculated.

The conductor surfaces were discretized into square elements, and the unknown free-charge distribution was solved from electrostatic interaction, conductor equipotentiality and total-charge constraints. The self-interaction term was treated separately from the off-diagonal element interactions. After the charge distribution was obtained, potential, electric field and integrated electrode charge were calculated from the same solution.

\section*{Data availability}

The data that support the findings of this study are available from the corresponding author upon reasonable request.

\section*{Code availability}

The custom Python code used for the finite-geometry electrostatic calculations is available from the corresponding author upon reasonable request.

\section*{References}
\begingroup\small\raggedright
\renewcommand{\refname}{}

\endgroup

\section*{Funding}

This work was supported by National Key R\&D Program of China (Grant No. 2024YFB3816000), Shenzhen Science and Technology Program (Grant Nos. KJZD20240903100905008 and JCYJ20240813111910014), Guangdong Innovative and Entrepreneurial Research Team Program (Grant No. 2021ZT09L197), China Postdoctoral Science Foundation (Grant No. 2025M780094), and the Postdoctoral Fellowship Program of CPSF (Grant No. GZB20260152).

\section*{Competing interests}

The authors declare no competing interests.

\clearpage
\section*{Extended Data Figures and Tables}

\begin{figure}[!htbp]
\centering
\includegraphics[width=0.99\linewidth,height=0.69\textheight,keepaspectratio]{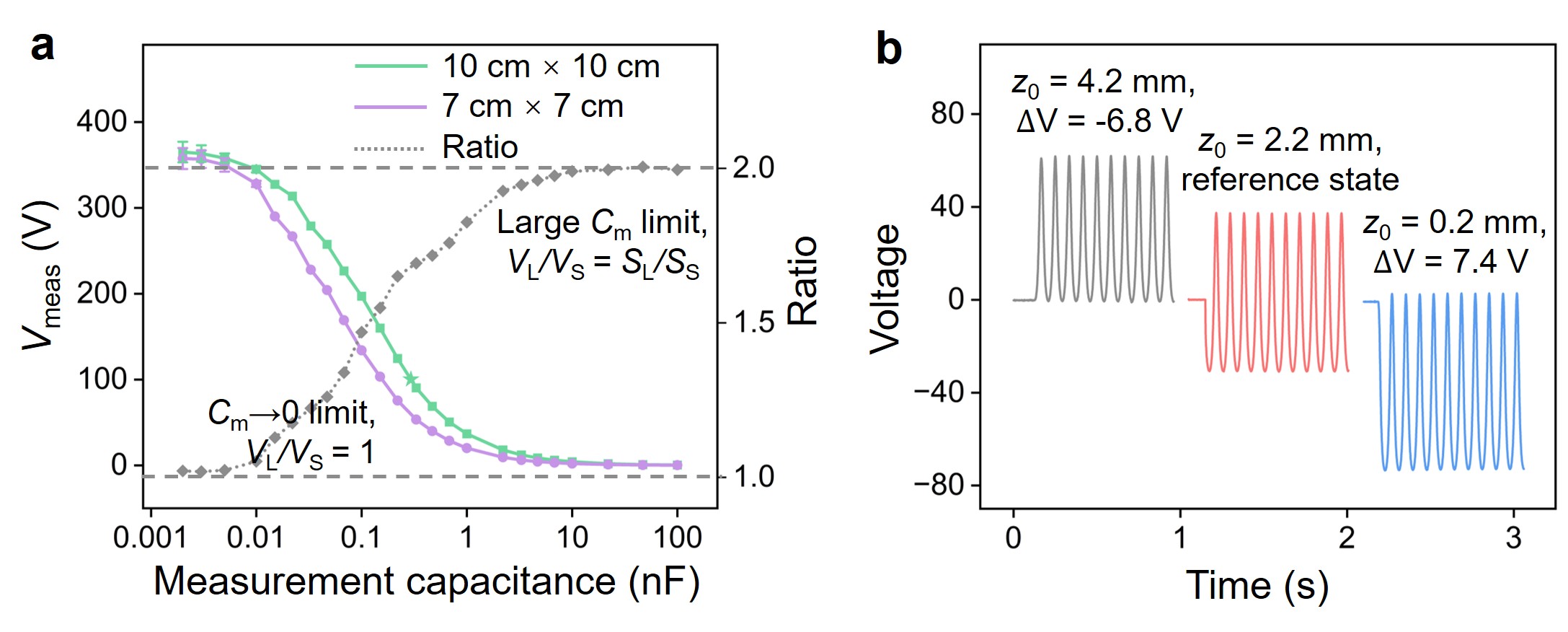}
\caption*{\textbf{Extended Data Fig. 1 | Electret boundary-selected voltage and charge-history dependence.} a, Measurement-conditioned voltage of 10 \(\times \ \)10 cm\textsuperscript{2} and 7 \(\times \ \)7 cm\textsuperscript{2} electret generators as a function of measurement capacitance \(C_{m}\), together with the voltage ratio \(V_{L}/V_{S}\). As \(C_{m}\) decreases, charge exchange through the measurement branch is progressively suppressed and the two devices approach an approximately area-independent small-capacitance voltage limit; with increasing \(C_{m}\), the area dependence becomes apparent and \(V_{L}/V_{S}\) approaches \(S_{L}/S_{S}\), reproducing the boundary-selected crossover observed for the TENG in Fig. 3c. Error bars denote the standard deviation. b, Electret voltage waveforms after establishing different initial charge-equilibration states at \(z_{0}\) \(=\) 4.2, 2.2 and 0.2 mm. The subsequent mechanical trajectory is unchanged. The resulting persistent voltage-state shifts retain information about the preceding charge-equilibration history, analogous to the TENG response in Fig. 3e.}

\end{figure}

\begin{figure}[!htbp]
\centering
\includegraphics[width=0.99\linewidth,height=0.69\textheight,keepaspectratio]{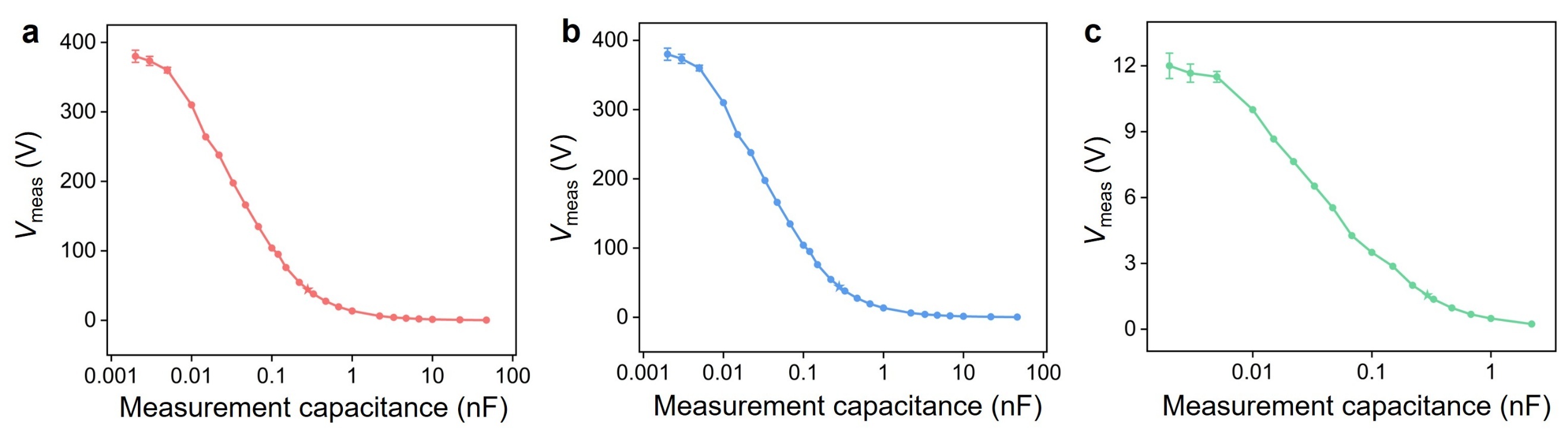}
\caption*{\textbf{Extended Data Fig. 2 | Boundary-selected voltage across piezoelectret, ionoelectric-gel and solid-liquid generators.} a-c, Measured voltage as a function of measurement capacitance \(C_{m}\) for piezoelectret (a), ionoelectric-gel (b) and solid-liquid (c) generators under otherwise fixed device and mechanical conditions. In each case, decreasing \(C_{m}\) suppresses charge exchange through the measurement branch and the measured voltage approaches a finite small-capacitance limit, whereas increasing \(C_{m}\) permits progressively greater charge transfer and reduces the measured voltage. Error bars denote the standard deviation. These measurements extend the boundary-selected voltage response across distinct source mechanisms without implying a common microscopic source law.}

\end{figure}

\begin{figure}[!htbp]
\centering
\includegraphics[width=0.99\linewidth,height=0.69\textheight,keepaspectratio]{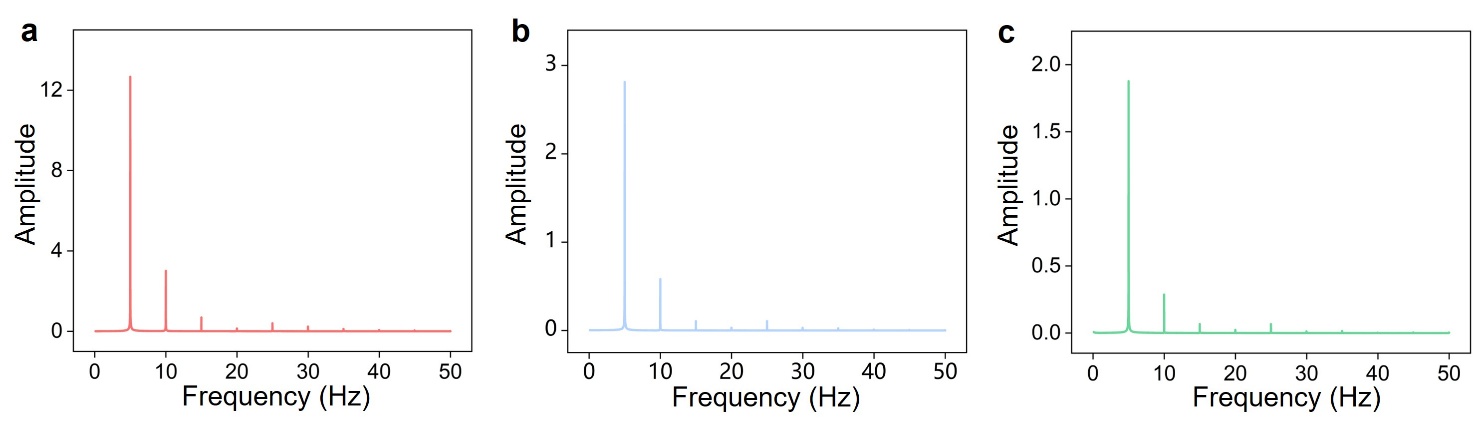}
\caption*{\textbf{Extended Data Fig. 3 | Internal-separation dependence of TENG harmonic generation.} a-c, Representative output spectra for mean internal separations of 1 mm (a), 2 mm (b) and 4 mm (c), with the mechanical amplitude fixed at 2 mm and the remaining operating conditions unchanged. Reducing the mean internal separation increases the higher-order spectral content, consistent with stronger sampling of the nonlinear charge-geometry relation at smaller separation.}

\end{figure}

\begin{figure}[!htbp]
\centering
\includegraphics[width=0.99\linewidth,height=0.69\textheight,keepaspectratio]{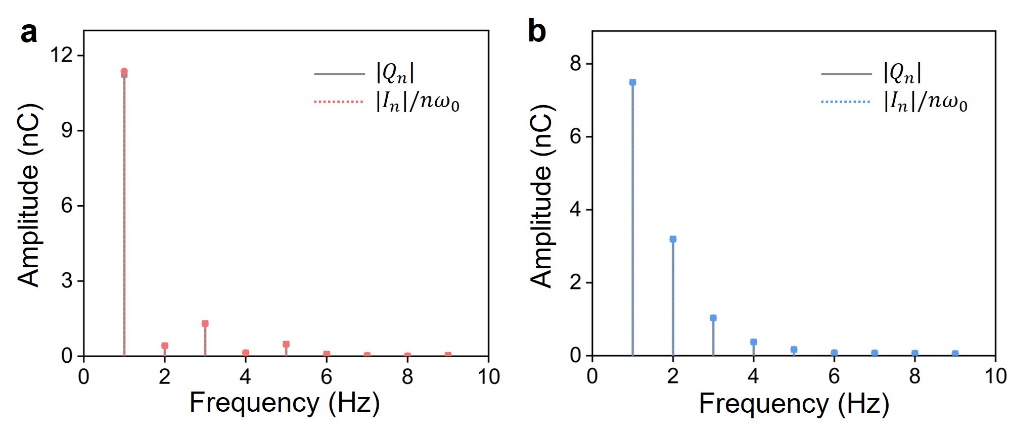}
\caption*{\textbf{Extended Data Fig. 4 | Harmonic-resolved charge-current correspondence in electret and piezoelectric generators.} a,b, Harmonic amplitudes of transferred charge \(\left| Q_{n} \right|\) and current normalized by the corresponding harmonic angular frequency, \(\left| I_{n} \right|/n\omega_{0}\), for electret (a) and piezoelectric (b) generators. The two spectra coincide harmonic by harmonic within experimental resolution, consistent with the exact downstream relation \(I_{n} = in\omega_{0}Q_{n}\) and showing that the resolved current harmonics originate from the corresponding transferred-charge trajectory.}

\end{figure}

\begin{figure}[!htbp]
\centering
\includegraphics[width=0.99\linewidth,height=0.69\textheight,keepaspectratio]{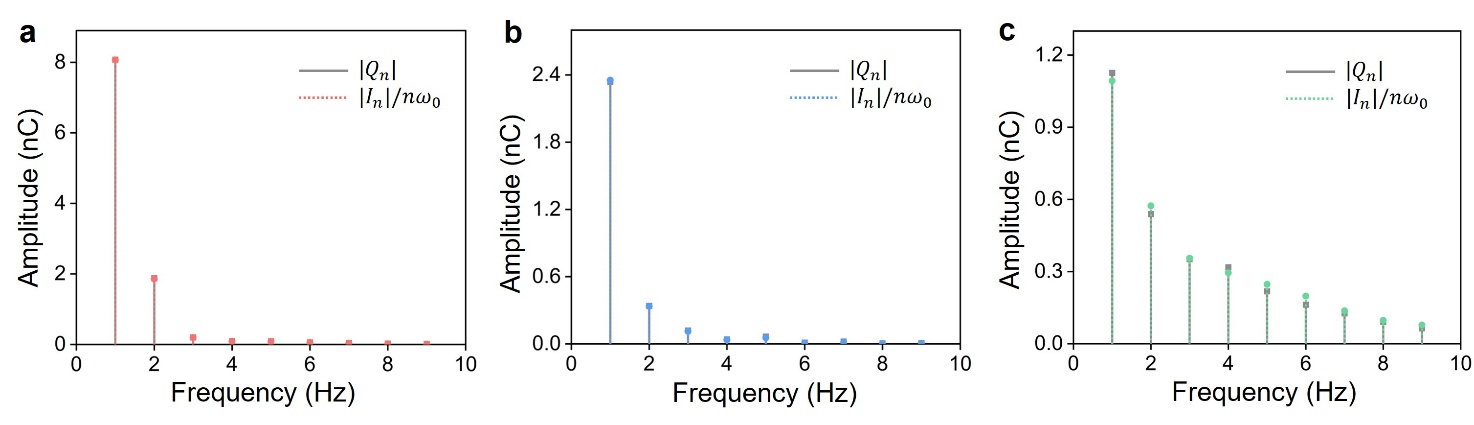}
\caption*{\textbf{Extended Data Fig. 5 | Harmonic-resolved charge-current correspondence across piezoelectret, ionoelectric-gel and solid-liquid generators.} a-c, Harmonic amplitudes of transferred charge \(\left| Q_{n} \right|\) and frequency-normalized current \(\left| I_{n} \right|/n\omega_{0}\) for piezoelectret (a), ionoelectric-gel (b) and solid-liquid (c) generators. The harmonic-by-harmonic correspondence follows the common downstream identity \(I = dQ/dt\) despite the different upstream source mechanisms. These systems are used as extension evidence for the shared charge-current organization rather than as complete tests of a common microscopic source model.}

\end{figure}

\begin{figure}[!htbp]
\centering
\includegraphics[width=0.99\linewidth,height=0.69\textheight,keepaspectratio]{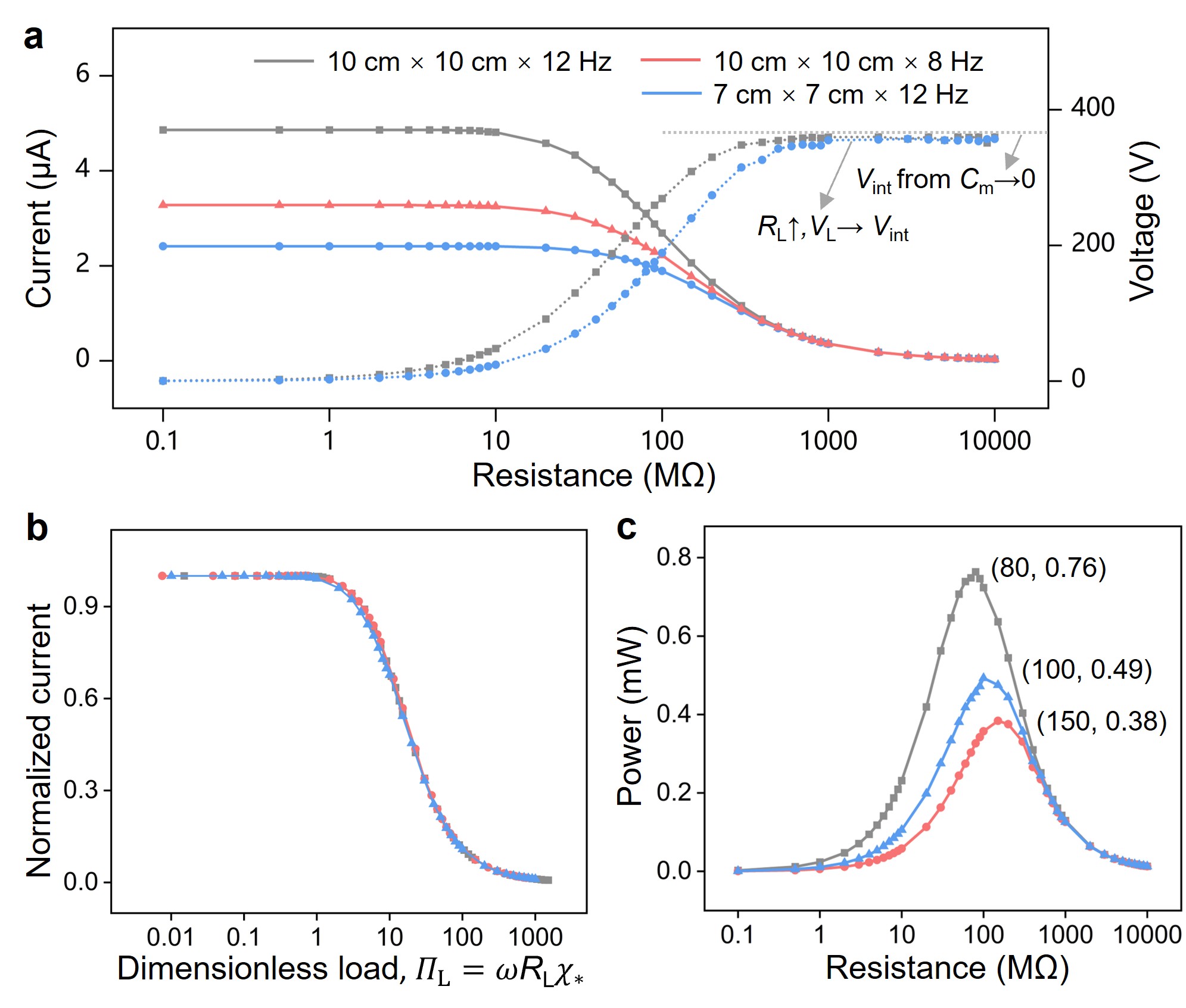}
\caption*{\textbf{Extended Data Fig. 6 | Electret load-constrained charge evolution, resistance scaling and power transfer.} a, Peak-to-peak current and the corresponding load voltage \({R_{L}I}_{pp}\) as functions of load resistance for 10 \(\times \ \)10 cm\textsuperscript{2} electret generators operated at 12 Hz and 8 Hz and a 7 \(\times \ \)7 cm\textsuperscript{2} generator operated at 12 Hz. Increasing \(R_{L}\) suppresses current while the load voltage approaches the independently measured small-capacitance voltage limit. Increasing active area or driving frequency shifts the load transition towards lower resistance, consistent with the TENG response. For \(R_{L}\) \(>\) 500 M\ensuremath{\Omega}, the current standard deviation is below 0.012 \ensuremath{\mu}A. b, Normalized current plotted against the dimensionless load parameter \(\Pi_{L} = \omega R_{L}\chi_{*}\). The measurements approximately collapse onto a common transition, consistent with competition between electrical charge relaxation and mechanical driving. c, Peak-to-peak power metric \(P_{pp} = I_{pp}^{2}R_{L}\) as a function of resistance for the same operating conditions. The optimal resistance shifts towards lower values with increasing active area or frequency, consistent with the load-scaling trend of the TENG.}

\end{figure}

\begin{figure}[!htbp]
\centering
\includegraphics[width=0.99\linewidth,height=0.69\textheight,keepaspectratio]{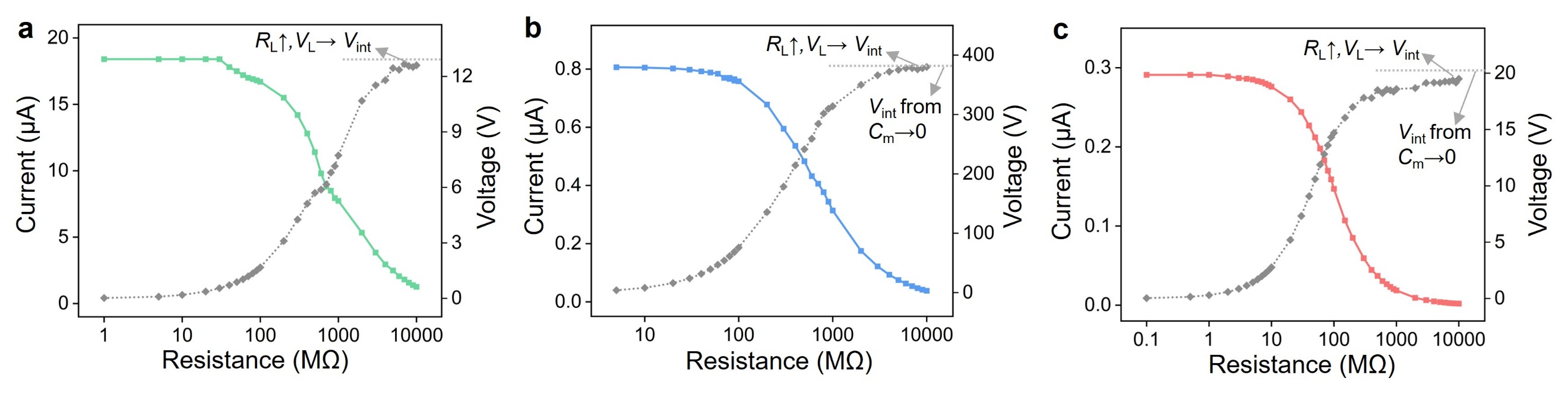}
\caption*{\textbf{Extended Data Fig. 7 | Convergence of small-capacitance and large-resistance limits to the intrinsic-voltage state across generator classes.} a-c, Peak-to-peak current and the corresponding load voltage \({R_{L}I}_{pp}\) as functions of load resistance for piezoelectret (a), ionoelectric-gel (b) and solid-liquid (c) generators. Increasing \(R_{L}\) progressively suppresses inter-conductor charge transfer and current, while the load voltage approaches the voltage independently obtained in the small-capacitance limit. The two electrical boundaries therefore converge towards the same zero-transfer, charge-frozen state under the examined conditions. For \(R_{L}\  >\) 1 G\ensuremath{\Omega}, the current standard deviation is below 0.081 \ensuremath{\mu}A, 0.008 \ensuremath{\mu}A and 0.002 \ensuremath{\mu}A in a-c, respectively.}

\end{figure}

\clearpage
\setcounter{equation}{0}
\section*{Supplementary Information}
\noindent\textit{Supplement to: Charge-state determination unifies electrostatic generator behaviour}
\par\medskip
\section*{Supplementary Note 1 | Definitions, electrical states and sign conventions}

\subsection*{1.1 Source state, geometry, boundary and history}

We denote the non-equilibrium electrical source state by

\begin{equation*}
\mathcal{S}(t) = \{\rho_{s},\sigma_{s},\mathbf{P}_{s},c_{i},\chi_{int},\ldots\}\tag{S1.1}\label{eq:S1.1}
\end{equation*}

The entries may represent prescribed volume or surface charge, polarization, mobile ionic populations or interfacial states. Different generator classes need not share the same microscopic variables.

The instantaneous geometry is denoted \(\Gamma(t)\), the electrical boundary \(\mathcal{B}(t)\), and the relevant charge history \(\mathcal{H}\). The complete charge distribution is

\begin{equation*}
\rho\left( \mathbf{r},t \right) = \rho_{s}\left( \mathbf{r},t \right) + \rho_{c}\left( \mathbf{r},t \right),\tag{S1.2}\label{eq:S1.2}
\end{equation*}

where \(\rho_{c}\) is the conductor-response charge selected by the electrostatic constraints.

The source charge in this bookkeeping may include the bound-charge representation of a prescribed polarization field. When that polarization is instead retained explicitly in the electric displacement, its equivalent bound charge is not counted again. Conductor-response charge denotes the free charge on metallic surfaces.

The general hierarchy is

\begin{equation*}
\left\lbrack \mathcal{S},\Gamma\mathcal{,B,H_{0}} \right\rbrack \rightarrow \rho \rightarrow \left\lbrack \mathbf{E},\phi \right\rbrack \rightarrow Q \rightarrow \lbrack I,V\rbrack\tag{S1.3}\label{eq:S1.3}
\end{equation*}

\subsection*{1.2 Conductor constraints}

For an ideal conductor in quasi-electrostatic equilibrium,

\begin{equation*}
\mathbf{E} = 0\quad\text{inside the conductor},\tag{S1.4}\label{eq:S1.4}
\end{equation*}

and each connected conductor is equipotential,

\begin{equation*}
\phi\left( \mathbf{r} \right) = V_{\alpha},\quad\quad\mathbf{r} \in S_{\alpha}\tag{S1.5}\label{eq:S1.5}
\end{equation*}

For an isolated conductor group \(g\), total free charge is conserved,

\begin{equation*}
\sum_{\alpha \in g}^{}{\int_{S_{\alpha}}^{}\sigma_{\alpha}}\left( \mathbf{r} \right)\, dS = Q_{g}^{tot}\tag{S1.6}\label{eq:S1.6}
\end{equation*}

For an ideal short between two conductor groups,

\begin{equation*}
V_{1} = V_{2},\tag{S1.7}\label{eq:S1.7}
\end{equation*}

while the total free charge of the connected group remains conserved.

\subsection*{1.3 Transferred charge and current}

For a specified electrode and transfer direction,

\begin{equation*}
Q(t) = Q_{e}(t) - Q_{e}\left( t_{0} \right)\tag{S1.8}\label{eq:S1.8}
\end{equation*}

The external conduction current is

\begin{equation*}
I(t) = \frac{dQ(t)}{dt}\tag{S1.9}\label{eq:S1.9}
\end{equation*}

This identity refers to charge and current in the same external transfer branch.

\subsection*{1.4 Intrinsic and measured voltage}

The intrinsic voltage \(V_{int}\) is the potential difference of a charge-frozen conductor state established by a specified source state, geometry and charge history. It is not a history-independent material constant.

A measured voltage \(V_{meas}\) is the potential difference of the charge state selected by the actual measurement boundary. Finite measurement capacitance can therefore alter the state being measured.

\subsection*{1.5 Reduced charge-response scale}

For a solved electrostatic state family, we define the local conductor charge-potential response coefficient as

\begin{equation*}
\chi_{Q} = - \left( \frac{\partial V}{\partial Q} \right)^{- 1}\tag{S1.10}\label{eq:S1.10}
\end{equation*}

For the linear one-dimensional CS-TENG considered here, this response coefficient reduces to the ratio of the independently solved short-circuit and charge-frozen states at the same final geometry,

\begin{equation*}
\chi_{Q} \equiv \frac{Q_{sc}}{V_{int}}\tag{S1.11}\label{eq:S1.11}
\end{equation*}

The reduced measurement and load parameters are then

\begin{equation*}
\Pi_{m} = \frac{C_{m}}{\chi_{*}},\quad\quad\Pi_{L} = \omega R_{L}\chi_{*},\tag{S1.12}\label{eq:S1.12}
\end{equation*}

when a scalar charge-response coefficient is an adequate representation of the solved electrostatic state family.

\section*{Supplementary Note 2 | A common charge-state determination route across electrostatic generator classes}

\subsection*{2.1 Generator-specific source states and common downstream closure}

The framework does not replace the microscopic physics that establishes a generator source state. TENGs and electrets can approximate fixed-source regimes after source establishment; piezoelectric generators possess deformation-dependent constitutive sources; piezoelectrets combine persistent and deformation-dependent source components; and ionoelectric or solid-liquid generators may require transport or electrochemical source evolution.

The source state \(\mathcal{S}\)denotes the material-specific non-equilibrium electrical variables established upstream of conductor-response charge; it does not imply a common microscopic source law. For a piezoelectric generator under a prescribed mechanical state, the conventional constitutive relation establishes a deformation-dependent polarization source, for example

\begin{equation*}
\mathbf{P}_{s}^{pz} = \mathbf{d}:\mathbf{T}\tag{S2.1}\label{eq:S2.1}
\end{equation*}

This polarization may be retained directly as a component of \(\mathcal{S}\), or equivalently represented through its associated bound volume and surface charges. In either representation, the deformation-induced polarization is upstream of the free charge subsequently selected on the electrodes by geometry and electrical boundary.

For transport-controlled systems, \(\mathcal{S}\) may instead contain mobile ionic populations, fixed or interfacial charge and other interfacial state variables. Their evolution is provided by the appropriate transport and electrochemical equations, as illustrated by the analytical correspondence in Supplementary Note 5; the dynamics of each experimental configuration are not separately simulated here. Solid-liquid generators can be organized analogously, with interfacial electrical variables entering the source state while droplet position, wetted footprint and interface geometry remain components of \(\Gamma\). These examples illustrate how generator-specific microscopic descriptions enter the common framework; they do not imply a universal source-evolution law.

The common physical statement begins downstream of these class-specific source mechanisms. Once the relevant source state, geometry, electrical boundary and preceding charge history are specified, the conductor-charge state must satisfy self-consistent electrostatic closure. Field, potential, transferred charge, current and voltage are then obtained from that state.

\subsection*{2.2 Reduced descriptors and charge-frozen limits}

For the analytical CS-TENG, the charge-response coefficient is obtained exactly from the solved short-circuit and charge-frozen state family. In other generator classes, an effective scalar response coefficient is useful only when the compared states refer to the same source state, geometry and history and when a scalar descriptor adequately represents the solved electrostatic response.

The same principle applies to the reduced measurement and load parameters. Their use across generator classes is a testable reduced consequence where justified, not a prerequisite of the source-state framework.

Small measurement capacitance and large load resistance are not dynamically identical boundaries. When the source state remains stable, however, both suppress inter-conductor charge transfer and approach the same zero-transfer, charge-frozen electrostatic state.

\subsection*{2.3 Evidence and scope in the present work}

The evidence hierarchy is summarized in Supplementary Table 1. The contact-separation TENG provides the full analytical and finite-geometry reference closure and the primary series of state-resolved experiments. Supplementary Notes 3-6 provide explicit source-to-conductor-charge calculations for electret, piezoelectric, ionoelectric and solid-liquid generators. Each case recovers an established theoretical description under its source, geometry and boundary assumptions, demonstrating how different source mechanisms enter the common charge-state determination route. The additional experimental configurations provide complementary tests of the measured electrical behaviours.

\par\medskip\noindent\textbf{Supplementary Table 1 \textbar{} Evidence hierarchy across generator classes.}\par\vspace{0.5em}

\begingroup\footnotesize
\setlength{\tabcolsep}{3.2pt}
\begin{longtable}{@{}>{\raggedright\arraybackslash}p{0.17\linewidth}>{\raggedright\arraybackslash}p{0.20\linewidth}>{\raggedright\arraybackslash}p{0.34\linewidth}>{\raggedright\arraybackslash}p{0.24\linewidth}@{}}
\toprule\noalign{}
\begin{minipage}[b]{\linewidth}\raggedright
\textbf{Generator class}
\end{minipage} & \begin{minipage}[b]{\linewidth}\raggedright
\textbf{Source-state regime}
\end{minipage} & \begin{minipage}[b]{\linewidth}\raggedright
\textbf{Evidence in this work}
\end{minipage} & \begin{minipage}[b]{\linewidth}\raggedright
\textbf{Closure level}
\end{minipage} \\
\midrule\noalign{}
\endhead
\bottomrule\noalign{}
\endlastfoot
Contact-separation TENG & Approximately fixed source after establishment
& Analytical and finite-geometry closure; boundary selection; history;
harmonics; load dynamics & Complete reference closure \\
Electret & Stored charge and persistent polarization & Derived electret
port relation; Boland rotating-generator recovery; secondary electrical
observations & Analytical source-to-port correspondence (Note 3) \\
Piezoelectric & Constitutive polarization and coupled mechanics &
Constitutive-source closure; Erturk-Inman bimorph electromechanical
equations; secondary electrical observations & Coupled analytical
correspondence (Note 4) \\
Piezoelectret & Mixed persistent and constitutive source & Experimental
boundary and spectral observations & Extension evidence \\
Ionoelectric gel & Transport-controlled ionic state & PNP-Darcy
source-to-electrode closure; Dobashi piezoionic-voltage and load
relations & Transport-coupled analytical correspondence (Note 5) \\
Solid-liquid generator & Contact-established interface charge with EDL
screening & Weighted electrode-charge response; droplet spreading/recoil
current correspondence & Interfacial-source analytical correspondence
(Note 6) \\
\end{longtable}
\endgroup

Taken together, the reference TENG solution and the four cross-mechanism analytical cases establish a common physical route from material-established electrical sources to self-consistent conductor-charge states and terminal observables. The device-specific source laws remain explicit and connect naturally to the common electrostatic and external-circuit closure described in Supplementary Notes 1 and 3-8.

\section*{Supplementary Note 3 | Electret generators within the source-driven framework}

Electret generators offer a direct test of the shared source-to-state calculation route. The established material charge or polarization defines an electrical source, whereas the free-charge distribution on the conducting electrodes is determined by electrostatic equilibrium, geometry, the external electrical boundary and charge history. We first derive the conventional electret port relation from this charge-state problem and then recover the analytical relations used for the rotational micro-electret generator of Boland et al\cite{siref1,siref2}.

\subsection*{3.1 Material source and conventional electret-port reduction}

Charge storage in electrets is commonly described in terms of trapped volume charge, surface charge and persistent or slowly relaxing polarization. The charging, transport, trapping and relaxation processes establish the material state; during a mechanical conversion cycle its source distribution can be treated as prescribed when its relaxation is slow compared with device motion\cite{siref3,siref4,siref5}. With the reversible dielectric response represented by the permittivity and persistent polarization retained explicitly, the governing electrostatic equations are

\begin{equation*}
\mathbf{E} = - \nabla\phi,\quad\mathbf{D} = \mathbf{\varepsilon E} + \mathbf{P}_{0},\quad\nabla \cdot \mathbf{D} = \rho_{t}\tag{S3.1}\label{eq:S3.1}
\end{equation*}

Any independent prescribed sheet charge enters through the normal-displacement jump at its interface. Polarization bound charge is counted once, either as an explicit polarization source or by its equivalent volume and surface charges. Electret transducer analyses traditionally reduce the material source to a bias potential and represent the moving transducer by a variable capacitance\cite{siref1,siref2,siref6,siref7,siref8}:

\begin{equation*}
V(t) = V_{e} - \frac{Q(t)}{C\left\lbrack x(t) \right\rbrack}\tag{S3.2}\label{eq:S3.2}
\end{equation*}

Here the effective bias and capacitance describe a solved terminal response; they are not used as primary source variables in the source-driven construction below.

\subsection*{3.2 Conductor-charge closure and recovery of the port equation}

Consider two metallic electrodes surrounding an electret dielectric. At a given mechanical coordinate x, the established trapped-charge and polarization distributions are prescribed, while the free surface charge on every metallic surface remains unknown. The electrostatic potential is decomposed into the source contribution and the conductor-response contribution:

\begin{equation*}
\phi\left( \mathbf{r} \right) = \phi_{s}\left( \mathbf{r}\mathcal{;S,}x \right) + \sum_{k}^{}{\int_{\Gamma_{k}(x)}^{}G_{x}}\left( \mathbf{r},\mathbf{r}' \right)\,\sigma_{c,k}\left( \mathbf{r}' \right)\, dA'\tag{S3.3}\label{eq:S3.3}
\end{equation*}

\begin{equation*}
q_{k} = \int_{\Gamma_{k}}^{}\sigma_{c,k}\, dA,\quad\quad q_{1} + q_{2} = q_{tot}^{(0)},\quad\quad\Delta q_{1} = - \Delta q_{2}\tag{S3.4}\label{eq:S3.4}
\end{equation*}

The potential is constant over each connected electrode, and the total free charge of an isolated electrode group obeys the stated conservation constraint. For the terminal equations below we define \(Q\) as the negative of the physical net free charge on electrode 2, with n pointing outward from the dielectric towards that metal. This port orientation is used consistently in Eqs. (S3.5)-(S3.13). Solving the short-circuit boundary-value problem with identical source, geometry and total-charge history establishes the reference port-charge coordinate:

\begin{equation*}
Q_{sc}(x) = - \int_{\Gamma_{2}(x)}^{}\sigma_{c,2}^{sc}\left( \mathbf{r};x \right)\, dA\tag{S3.5}\label{eq:S3.5}
\end{equation*}

To determine how terminal voltage changes the electrode-charge state, subtract two electrostatic solutions having the same established material source and instantaneous geometry. Their difference is source-free. An auxiliary unit-potential problem gives the incremental conductor-charge response and a strictly positive effective capacitance:

\begin{equation*}
\nabla \cdot \left( \mathbf{\varepsilon}\nabla\delta\phi \right) = 0,\quad\quad\delta\mathbf{D} = - \mathbf{\varepsilon}\nabla\delta\phi\tag{S3.6}\label{eq:S3.6}
\end{equation*}

\begin{equation*}
\nabla \cdot \left( \mathbf{\varepsilon}\nabla\psi_{x} \right) = 0,\quad\quad\psi_{x} = 0\ \left( \mathbf{r} \in \Gamma_{1} \right),\quad\psi_{x} = 1\ \left( \mathbf{r} \in \Gamma_{2} \right),\tag{S3.7}\label{eq:S3.7}
\end{equation*}

\begin{equation*}
\delta Q = \int_{\Gamma_{2}}^{}\delta\mathbf{D} \cdot \mathbf{n}\, dA = - \delta V\int_{\Gamma_{2}}^{}\mathbf{n} \cdot \mathbf{\varepsilon}\nabla\psi_{x}\, dA\tag{S3.8}\label{eq:S3.8}
\end{equation*}

\begin{equation*}
\int_{\Gamma_{2}}^{}\mathbf{n} \cdot \mathbf{\varepsilon}\nabla\psi_{x}\, dA = \int_{\Omega(x)}^{}\left( \nabla\psi_{x} \right)^{\mathsf{T}}\mathbf{\varepsilon}\nabla\psi_{x}\, d\Omega \equiv \chi(x) > 0\tag{S3.9}\label{eq:S3.9}
\end{equation*}

Because the incremental problem is linear at fixed source and geometry, integration about the independently solved short-circuit state yields the port-charge relation and its conventional bias-capacitance form:

\begin{equation*}
Q(x,V) = Q_{sc}(x) - \chi(x)V,\quad\quad\chi(x) = - \left. \ \frac{\partial Q}{\partial V} \right|_{\mathcal{S},x} > 0\tag{S3.10}\label{eq:S3.10}
\end{equation*}

\begin{equation*}
V = \frac{Q_{sc}(x) - Q}{\chi(x)} = \frac{Q_{sc}(x)}{\chi(x)} - \frac{Q}{\chi(x)}\tag{S3.11}\label{eq:S3.11}
\end{equation*}

\begin{equation*}
V = V_{e}(x) - \frac{Q}{C(x)}\tag{S3.12}\label{eq:S3.12}
\end{equation*}

Taking the full time derivative then gives the external-current law, including modulation of the short-circuit reference and of the charge-response coefficient:

\begin{equation*}
\dot{Q} = {\dot{Q}}_{sc}(x) - \chi(x)\dot{V} - \dot{\chi}(x)V\tag{S3.13}\label{eq:S3.13}
\end{equation*}

The conventional bias-source-variable-capacitance representation is therefore recovered from self-consistent electrode charge determination. The source-driven representation also retains the spatial charge distribution and the reference state from which the terminal reduction was obtained.

\subsection*{3.3 Recovery of the rotating electret-generator relations of Boland et al.}

Boland et al.\cite{siref1} considered a rotating electrode above a Teflon-AF-coated stator with a fixed air gap and periodically changing overlap area A(t). Their reported representative quantities include a four-pole structure, an effective radius of 4 mm, a 9-\ensuremath{\mu}m electret dielectric, relative permittivity 1.93, and an effective surface charge density of approximately \ensuremath{-}2.8 \ensuremath{\times} 10\ensuremath{^{-}}\textsuperscript{4} C m\ensuremath{^{-}}\textsuperscript{2}. We adopt the same local parallel-layer and negligible-edge-field idealization so that the equations can be compared directly.

The established electret sheet charge and the responding electrode charges satisfy local free-charge bookkeeping. Using the interfacial displacement jump and normal-field boundary condition gives the electric fields in the air gap and electret:

\begin{equation*}
Q_{implanted}(t) = \sigma A(t) = Q_{1}(t) + Q_{2}(t)\tag{S3.14}\label{eq:S3.14}
\end{equation*}

\begin{equation*}
D_{g}(t) = \frac{Q_{2}(t)}{A(t)}\tag{S3.15}\label{eq:S3.15}
\end{equation*}

\begin{equation*}
D_{g} - D_{d} = \sigma,\quad\quad \Rightarrow \quad\quad D_{d} = \frac{Q_{2}}{A} - \sigma\tag{S3.16}\label{eq:S3.16}
\end{equation*}

\begin{equation*}
E_{g} = \frac{Q_{2}}{\varepsilon_{0}A},\quad\quad E_{d} = \frac{Q_{2}/A - \sigma}{\varepsilon_{0}\varepsilon_{r}}\tag{S3.17}\label{eq:S3.17}
\end{equation*}

Integrating these fields across the two layers gives the terminal potential. The resulting expression has precisely the source term and geometry-weighted response-charge term of Boland et al., Eq. (3), with signs determined by the port orientation:

\begin{equation*}
V_{B}(t) = gE_{g} + dE_{d}\tag{S3.18}\label{eq:S3.18}
\end{equation*}

\begin{equation*}
V_{B}(t) = \frac{g}{\varepsilon_{0}}\frac{Q_{2}(t)}{A(t)} + \frac{d}{\varepsilon_{0}\varepsilon_{r}}\left\lbrack \frac{Q_{2}(t)}{A(t)} - \sigma \right\rbrack\tag{S3.19}\label{eq:S3.19}
\end{equation*}

\begin{equation*}
V_{B}(t) = - \frac{\sigma d}{\varepsilon_{0}\varepsilon_{r}} + \left( \frac{d}{\varepsilon_{0}\varepsilon_{r}} + \frac{g}{\varepsilon_{0}} \right)\frac{Q_{2}(t)}{A(t)}\tag{S3.20}\label{eq:S3.20}
\end{equation*}

The resistive branch closes the electrode-charge dynamics, leading to the same first-order charge evolution and load-current structure as the conventional model\cite{siref1}:

\begin{equation*}
V_{B}(t) = RI(t),\quad\quad I(t) = - \frac{dQ_{2}(t)}{dt}\tag{S3.21}\label{eq:S3.21}
\end{equation*}

\begin{equation*}
\frac{dQ_{2}}{dt} = \frac{\sigma d}{R\varepsilon_{0}\varepsilon_{r}} - \frac{1}{RA(t)}\left( \frac{d}{\varepsilon_{0}\varepsilon_{r}} + \frac{g}{\varepsilon_{0}} \right)Q_{2}(t)\tag{S3.22}\label{eq:S3.22}
\end{equation*}

\begin{equation*}
I(t) = - \frac{\sigma d}{R\varepsilon_{0}\varepsilon_{r}} + \frac{1}{RA(t)}\left( \frac{d}{\varepsilon_{0}\varepsilon_{r}} + \frac{g}{\varepsilon_{0}} \right)Q_{2}(t)\tag{S3.23}\label{eq:S3.23}
\end{equation*}

For the rotating-overlap branch we define an effective area rate k and write A(t) = kt. The load-law normalization is chosen to compare with Boland et al., Eqs. (6)-(9); the decreasing-overlap branch has the opposite slope and carries forward the previously established charge state.

\begin{equation*}
A(t) = kt,\quad\quad k = \frac{A_{\max}}{T_{p}/2} = n\pi r^{2}f,\quad 0 < t < \frac{1}{2nf}\tag{S3.24}\label{eq:S3.24}
\end{equation*}

A comparison of the printed 2003 equations reveals a factor-of-two difference between the overlap-area slope in its Eq. (5) and the effective rate used by its subsequent load-law Eqs. (6)-(9). We use the latter effective rate for the consistent port-law reduction below. With that convention, define the source and geometry combinations:

\begin{equation*}
\Lambda = \frac{d}{\varepsilon_{0}\varepsilon_{r}} + \frac{g}{\varepsilon_{0}},\quad\quad V_{s} = \frac{\sigma d}{\varepsilon_{0}\varepsilon_{r}},\quad\quad\alpha = \frac{\Lambda}{Rk},\quad\beta = \frac{V_{s}}{R}\tag{S3.25}\label{eq:S3.25}
\end{equation*}

\begin{equation*}
{\dot{Q}}_{2} + \frac{\alpha}{t}Q_{2} = \beta\tag{S3.26}\label{eq:S3.26}
\end{equation*}

\begin{equation*}
\frac{d}{dt}\left( t^{\alpha}Q_{2} \right) = \beta t^{\alpha}\quad \Rightarrow \quad Q_{2}(t) = \frac{\beta}{1 + \alpha}t + K_{0}t^{- \alpha}\tag{S3.27}\label{eq:S3.27}
\end{equation*}

For an overlap interval beginning at vanishing area, selecting the finite-charge solution removes the divergent integration branch. This yields the load current and its explicit resistance and rotation-speed dependence:

\begin{equation*}
Q_{2}(t) = \frac{V_{s}k}{Rk + \Lambda}\, t,\quad\quad I(t) = - {\dot{Q}}_{2}(t) = - \frac{V_{s}k}{Rk + \Lambda}\tag{S3.28}\label{eq:S3.28}
\end{equation*}

\begin{equation*}
I(t) = - \frac{\sigma d/\left( \varepsilon_{0}\varepsilon_{r} \right)}{R + \frac{1}{n\pi r^{2}f}\left( \frac{d}{\varepsilon_{0}\varepsilon_{r}} + \frac{g}{\varepsilon_{0}} \right)}\tag{S3.29}\label{eq:S3.29}
\end{equation*}

These equations recover the analytical structure of Boland et al., Eqs. (6) and (7), under the stated rotating-overlap and load-law convention. The source prefactor follows directly from their electrostatic Eq. (3) and circuit Eq. (4). Their printed Eq. (7) omits a relative-permittivity factor that appears in those preceding relations, so the comparison is made using the consistent electrostatic formulation. The corresponding resistive dissipation, stationary load condition and matched-load power scale are

\begin{equation*}
P_{R} = RI^{2} = \frac{RV_{s}^{2}k^{2}}{(Rk + \Lambda)^{2}}\tag{S3.30}\label{eq:S3.30}
\end{equation*}

\begin{equation*}
\frac{dP_{R}}{dR} = \frac{V_{s}^{2}k^{2}(\Lambda - Rk)}{(Rk + \Lambda)^{3}} = 0\quad \Rightarrow \quad R_{opt} = \frac{\Lambda}{k}\tag{S3.31}\label{eq:S3.31}
\end{equation*}

\begin{equation*}
R_{opt} = \frac{1}{n\pi r^{2}f}\left( \frac{d}{\varepsilon_{0}\varepsilon_{r}} + \frac{g}{\varepsilon_{0}} \right)\tag{S3.32}\label{eq:S3.32}
\end{equation*}

\begin{equation*}
P_{opt} = \frac{V_{s}^{2}k}{4\Lambda} = \frac{\sigma^{2}d^{2}n\pi r^{2}f}{4\varepsilon_{0}\varepsilon_{r}\left( d + \varepsilon_{r}g \right)}\tag{S3.33}\label{eq:S3.33}
\end{equation*}

\begin{equation*}
P_{opt} = \frac{\sigma^{2}n\pi r^{2}f\, d}{4\varepsilon_{0}\varepsilon_{r}\left( 1 + \varepsilon_{r}g/d \right)}\tag{S3.34}\label{eq:S3.34}
\end{equation*}

The source-charge conservation, electrostatic port-voltage relation and rotating-overlap load laws correspond to Boland et al., Eqs. (1), (3), (4) and (6)-(9), under the explicit normalization above\cite{siref1,siref2}. The recovered port and load relations illustrate how established electret charges determine the conductor response and terminal behaviour. In the experimental context of this model, Boland et al. reported power-speed measurements under 2, 4 and 7.6 M\ensuremath{\Omega} loads (their Fig. 5), including output above 25 \ensuremath{\mu}W at 4170 rpm\cite{siref1}. These measurements illustrate the resistance- and rotation-speed dependence addressed by the recovered relations.

\section*{Supplementary Note 4 | Piezoelectric generators within the source-driven framework}

Piezoelectricity provides a constitutive rather than a stored-charge source. In the unified calculation route, mechanical strain establishes polarization within the piezoelectric material, and the resulting field and conductor free-charge state are determined by electrostatic constraints. The mechanical and electrical subproblems remain bidirectionally coupled through the same piezoelectric constitutive law. Here we recover the conventional coupled port equations and relate them to the validated bimorph cantilever formulation of Erturk and Inman\cite{siref9,siref10,siref11,siref12}.

\subsection*{4.1 Constitutive polarization and the conventional coupled system}

For a linear piezoelectric body with displacement u, strain S, stress T, electric field E and electric displacement D, the strain-charge form of the constitutive law and the mechanical and electrical balance equations are

\begin{equation*}
\mathbf{S} = sym\left( \nabla\mathbf{u} \right),\quad\quad\mathbf{E} = - \nabla\phi\tag{S4.1}\label{eq:S4.1}
\end{equation*}

\begin{equation*}
\mathbf{T} = \mathbf{c}^{E}:\mathbf{S} - \mathbf{e}^{\mathsf{T}}\mathbf{E},\quad\quad\mathbf{D} = \mathbf{e}:\mathbf{S} + \mathbf{\varepsilon}^{S}\mathbf{E}\tag{S4.2}\label{eq:S4.2}
\end{equation*}

\begin{equation*}
\nabla \cdot \mathbf{T} + \mathbf{f} = \rho_{m}\ddot{\mathbf{u}},\quad\quad\nabla \cdot \mathbf{D} = \rho_{f}\tag{S4.3}\label{eq:S4.3}
\end{equation*}

The elastic tensor c\textsuperscript{E} is defined at constant electric field, and the dielectric tensor \ensuremath{\varepsilon}\textsuperscript{S} at constant strain. After spatial or modal reduction, a conventional linear port description separates mechanically induced source charge from the charge associated with applied electrode potential:

\begin{equation*}
\mathbf{Q}_{e} = \mathbf{Q}_{pz}\left( \mathbf{q} \right) + \mathbf{C}_{p}\mathbf{V}\tag{S4.4}\label{eq:S4.4}
\end{equation*}

\begin{equation*}
\begin{matrix}
\mathbf{Q}_{pz} = \mathbf{\Theta}^{\mathsf{T}}\mathbf{q},\quad\mathbf{I}_{e} = {\dot{\mathbf{Q}}}_{e} = \mathbf{\Theta}^{\mathsf{T}}\dot{\mathbf{q}} + \mathbf{C}_{p}\dot{\mathbf{V}} \\
\mathbf{M}\ddot{\mathbf{q}} + \mathbf{C}_{m}\dot{\mathbf{q}} + \mathbf{Kq} - \mathbf{\Theta V} = \mathbf{F}_{ext}
\end{matrix}\tag{S4.5}\label{eq:S4.5}
\end{equation*}

With a single resistive load, the port balance is a Norton-type current-source equation coupled to structural mass, damping, stiffness and inverse-piezoelectric feedback. This device-level representation follows from the field solution; it is not taken as a fundamental material source law.

\subsection*{4.2 Determination of electrode free charge from a polarization source}

At each instantaneous mechanical state, the direct piezoelectric constitutive term defines a polarization source. The corresponding bound volume and surface charges are distinct from the free charges induced on the metallic electrodes:

\begin{equation*}
\mathbf{P}_{pz} = \mathbf{e}:\mathbf{S},\quad\rho_{b} = - \nabla \cdot \mathbf{P}_{pz},\quad\sigma_{b} = \mathbf{P}_{pz} \cdot \mathbf{n}_{p}\tag{S4.6}\label{eq:S4.6}
\end{equation*}

\begin{equation*}
\mathbf{D} = \mathbf{\varepsilon}^{S}\mathbf{E} + \mathbf{P}_{pz},\quad\nabla \cdot \mathbf{D} = \rho_{f},\quad\phi = V_{k},\quad\int_{\Gamma_{k}}^{}\sigma_{c,k}\, dA = Q_{k}\tag{S4.7}\label{eq:S4.7}
\end{equation*}

\begin{equation*}
Q_{k} = - \int_{\Gamma_{k}}^{}\mathbf{D} \cdot \mathbf{n}\, dA,\quad\quad I_{k} = \frac{dQ_{k}}{dt}\tag{S4.8}\label{eq:S4.8}
\end{equation*}

\begin{equation*}
- \nabla \cdot \left( \mathbf{\varepsilon}^{S}\nabla\phi \right) = \rho_{f} - \nabla \cdot \left( \mathbf{e}:\mathbf{S} \right)\tag{S4.9}\label{eq:S4.9}
\end{equation*}

Electrodes are equipotential. Depending on the circuit, their potentials or integrated free charges provide the closure conditions. A unit-potential auxiliary field can be used to recover the source-induced electrode charge and the incremental potential response without assuming an equivalent circuit at the outset:

\begin{equation*}
Q_{src,k} = - \int_{\Omega_{p}}^{}\mathbf{P}_{pz} \cdot \nabla\psi_{k}\, d\Omega,\quad\quad C_{field,kj} = \int_{\Omega}^{}\left( \nabla\psi_{k} \right)^{\mathsf{T}}\mathbf{\varepsilon}^{S}\nabla\psi_{j}\, d\Omega\tag{S4.10}\label{eq:S4.10}
\end{equation*}

\begin{equation*}
Q_{e,k} = Q_{src,k} + \sum_{j}^{}C_{field,kj}V_{j},\quad\quad\mathbf{I}_{e} = {\dot{\mathbf{Q}}}_{src} + \mathbf{C}_{field}\dot{\mathbf{V}}\tag{S4.11}\label{eq:S4.11}
\end{equation*}

For the same instantaneous strain field, dielectric tensors, geometry and electrode conditions, the classical constitutive formulation and the source-driven formulation solve the same electrostatic boundary-value problem. Their potential difference is governed by a homogeneous equation and vanishes up to an irrelevant reference potential:

\begin{equation*}
\int_{\Omega}^{}(\nabla\psi)^{\mathsf{T}}\mathbf{\varepsilon}^{S}\nabla\psi\, d\Omega = 0\quad \Rightarrow \quad\nabla\psi = 0\tag{S4.12}\label{eq:S4.12}
\end{equation*}

The inverse-piezoelectric effect is retained by returning the solved field to the mechanical balance. The standard electric enthalpy then provides both the stress and electric displacement by differentiation:

\begin{equation*}
\mathcal{H}\left( \mathbf{S},\mathbf{E} \right) = \frac{1}{2}\mathbf{S}:\mathbf{c}^{E}:\mathbf{S} - \left( \mathbf{e}:\mathbf{S} \right) \cdot \mathbf{E} - \frac{1}{2}\mathbf{E} \cdot \mathbf{\varepsilon}^{S}\mathbf{E}\tag{S4.13}\label{eq:S4.13}
\end{equation*}

\begin{equation*}
\mathbf{T} = \frac{\partial\mathcal{H}}{\partial\mathbf{S}} = \mathbf{c}^{E}:\mathbf{S} - \mathbf{e}^{\mathsf{T}}\mathbf{E},\quad\quad\mathbf{D} = - \frac{\partial\mathcal{H}}{\partial\mathbf{E}} = \mathbf{\varepsilon}^{S}\mathbf{E} + \mathbf{P}_{pz}\tag{S4.14}\label{eq:S4.14}
\end{equation*}

Thus the common source-to-conductor-state step is nested within, rather than detached from, the complete electromechanical system. This is the basis for comparing the resulting device equations with established piezoelectric-generator theory.

\subsection*{4.3 Piezoelectric bimorph cantilever: recovery of the electrode-charge and coupling equations}

Erturk and Inman\cite{siref9} analyzed piezoelectric bimorph cantilevers with series and parallel electrical connections and experimentally validated a series-connected PZT-5A bimorph mounted on a brass substrate with a tip mass. We follow their Euler-Bernoulli beam assumptions, modal basis and series-port orientation. The representative geometry given in the source text is L = 50.8 mm, b = 31.8 mm, individual piezoelectric thickness 0.26 mm, substrate thickness 0.14 mm and tip mass 0.012 kg\cite{siref9}.

Let x be the beam coordinate and y the thickness coordinate measured from the composite neutral axis. The layer-centroid distance and the strain associated with bending are

\begin{equation*}
h_{c} = \frac{h_{s} + h_{p}}{2}\tag{S4.15}\label{eq:S4.15}
\end{equation*}

\begin{equation*}
S_{1}(x,y,t) = - y\frac{\partial^{2}w(x,t)}{\partial x^{2}}\tag{S4.16}\label{eq:S4.16}
\end{equation*}

For oppositely poled PZT layers, the bending strains also reverse sign across the neutral axis. Under the chosen series-port convention, their induced electrical source contributions add. Their local polarization, corresponding bound charge and thickness-averaged source polarization are

\begin{equation*}
P_{pz,3}(x,y,t) = {\bar{e}}_{31}S_{1}(x,y,t) = - {\bar{e}}_{31}y\, w_{xx}(x,t)\tag{S4.17}\label{eq:S4.17}
\end{equation*}

\begin{equation*}
\rho_{b} = - \nabla \cdot \mathbf{P}_{pz} = - \frac{\partial P_{pz,3}}{\partial y},\quad\quad\sigma_{b} = \mathbf{P}_{pz} \cdot \mathbf{n}\tag{S4.17a}\label{eq:S4.17a}
\end{equation*}

\begin{equation*}
{\bar{P}}_{pz,3}(x,t) = \frac{1}{h_{p}}\int_{h_{c} - h_{p}/2}^{h_{c} + h_{p}/2}P_{pz,3}(x,y,t)\, dy = - {\bar{e}}_{31}h_{c}w_{xx}(x,t)\tag{S4.18}\label{eq:S4.18}
\end{equation*}

Integration along the electrode area gives the modal source-charge contribution. The normal-displacement condition and equipotential electrode condition supply the actual free-charge response of one layer:

\begin{equation*}
Q_{pz,src}(t) = b\int_{0}^{L}{\bar{P}}_{pz,3}(x,t)\, dx = - {\bar{e}}_{31}h_{c}b\int_{0}^{L}w_{xx}(x,t)\, dx\tag{S4.18a}\label{eq:S4.18a}
\end{equation*}

\begin{equation*}
\frac{\partial D_{3}}{\partial y} \simeq 0,\quad\quad D_{3}(x,y,t) = D_{c}(x,t)\tag{S4.18b}\label{eq:S4.18b}
\end{equation*}

\begin{equation*}
v_{p}(t) = - \int_{h_{c} - h_{p}/2}^{h_{c} + h_{p}/2}E_{3}(x,y,t)\, dy = - \frac{h_{p}}{{\bar{\varepsilon}}_{33}^{S}}\left\lbrack D_{c}(x,t) - {\bar{P}}_{pz,3}(x,t) \right\rbrack\tag{S4.18c}\label{eq:S4.18c}
\end{equation*}

\begin{equation*}
D_{c}(x,t) = {\bar{P}}_{pz,3}(x,t) - \frac{{\bar{\varepsilon}}_{33}^{S}}{h_{p}}v_{p}(t)\tag{S4.18d}\label{eq:S4.18d}
\end{equation*}

\begin{equation*}
q_{p}(t) = - b\int_{0}^{L}D_{c}(x,t)\, dx = \frac{{\bar{\varepsilon}}_{33}^{S}bL}{h_{p}}v_{p}(t) - Q_{pz,src}(t)\tag{S4.19}\label{eq:S4.19}
\end{equation*}

For the two series-connected layers, the external port carries one series charge coordinate (rather than twice the single-layer charge). The resulting charge state and modal electromechanical coefficient are

\begin{equation*}
Q_{e}(t) = \frac{{\bar{\varepsilon}}_{33}^{S}bL}{2h_{p}}V(t) - \sum_{r \geq 1}^{}\kappa_{r}\eta_{r}(t)\tag{S4.20}\label{eq:S4.20}
\end{equation*}

\begin{equation*}
\kappa_{r} = - {\bar{e}}_{31}h_{c}b\int_{0}^{L}\varphi_{r}''(x)\, dx = - \frac{{\bar{e}}_{31}\left( h_{p} + h_{s} \right)b}{2}\,\varphi_{r}'(L)\tag{S4.21}\label{eq:S4.21}
\end{equation*}

The resistive-load constraint applied to this charge relation yields the familiar electrical governing equation and equivalent induced source current, corresponding to Eq. (28) of Erturk and Inman\cite{siref9}:

\begin{equation*}
\frac{{\bar{\varepsilon}}_{33}^{S}bL}{2h_{p}}\dot{V} + \frac{V}{R_{l}} = \sum_{r \geq 1}^{}\kappa_{r}{\dot{\eta}}_{r}\tag{S4.22}\label{eq:S4.22}
\end{equation*}

\begin{equation*}
i_{src}(t) = \sum_{r \geq 1}^{}\kappa_{r}{\dot{\eta}}_{r}(t)\tag{S4.23}\label{eq:S4.23}
\end{equation*}

The electric field also generates inverse-piezoelectric bending moment. Projecting the moment onto the beam modes recovers the reciprocal coupling coefficient and the mechanical mode equation:

\begin{equation*}
M_{pz} = - \frac{{\bar{e}}_{31}bV}{2h_{p}}\left\lbrack h_{p}h_{c} - \left( - h_{p}h_{c} \right) \right\rbrack = - {\bar{e}}_{31}bh_{c}V\tag{S4.24}\label{eq:S4.24}
\end{equation*}

\begin{equation*}
\vartheta_{s} = \frac{{\bar{e}}_{31}b}{2h_{p}}\left\lbrack \frac{h_{s}^{2}}{4} - \left( h_{p} + \frac{h_{s}}{2} \right)^{2} \right\rbrack = - {\bar{e}}_{31}h_{c}b,\quad\chi_{r}^{s} = \vartheta_{s}\varphi_{r}'(L)\tag{S4.25}\label{eq:S4.25}
\end{equation*}

\begin{equation*}
{\ddot{\eta}}_{r} + 2\zeta_{r}\omega_{r}{\dot{\eta}}_{r} + \omega_{r}^{2}\eta_{r} + \chi_{r}^{s}V = f_{r}(t)\tag{S4.26}\label{eq:S4.26}
\end{equation*}

Under transverse translational base excitation, the modal participation factor is

\begin{equation*}
\sigma_{r} = - \left\lbrack m\int_{0}^{L}\varphi_{r}(x)\, dx + M_{t}\varphi_{r}(L) \right\rbrack\tag{S4.27}\label{eq:S4.27}
\end{equation*}

Equations (S4.22) and (S4.26) form a bidirectionally coupled state and circuit system. They connect the constitutive polarization source to the actual electrode charge and retain the same mechanical feedback as the classical bimorph formulation.

\subsection*{4.4 Recovery of classical frequency-response functions}

For steady harmonic base acceleration, define the dynamic stiffness of a single beam mode and write the phasor mechanical and electrical equations:

\begin{equation*}
Z_{r}(\omega) = \omega_{r}^{2} - \omega^{2} + j2\zeta_{r}\omega_{r}\omega\tag{S4.28}\label{eq:S4.28}
\end{equation*}

\begin{equation*}
Z_{r}{\widehat{\eta}}_{r} + \chi_{r}^{s}\widehat{V} = \sigma_{r}{\widehat{a}}_{b}\tag{S4.29}\label{eq:S4.29}
\end{equation*}

\begin{equation*}
\left\lbrack \frac{1}{R_{l}} + j\omega\frac{{\bar{\varepsilon}}_{33}^{S}bL}{2h_{p}} \right\rbrack\widehat{V} = j\omega\kappa_{r}{\widehat{\eta}}_{r}\tag{S4.30}\label{eq:S4.30}
\end{equation*}

Eliminating the modal coordinate yields the voltage response to base acceleration, while back-substitution determines the relative tip displacement:

\begin{equation*}
\left\{ \left\lbrack \frac{1}{R_{l}} + j\omega\frac{{\bar{\varepsilon}}_{33}^{S}bL}{2h_{p}} \right\rbrack Z_{r} + j\omega\kappa_{r}\chi_{r}^{s} \right\}\widehat{V} = j\omega\kappa_{r}\sigma_{r}{\widehat{a}}_{b}\tag{S4.31}\label{eq:S4.31}
\end{equation*}

\begin{equation*}
\frac{\widehat{V}}{{\widehat{a}}_{b}} = \frac{j2\omega R_{l}\kappa_{r}\sigma_{r}}{\left( 2 + j\omega R_{l}{\bar{\varepsilon}}_{33}^{S}bL/h_{p} \right)\left( \omega_{r}^{2} - \omega^{2} + j2\zeta_{r}\omega_{r}\omega \right) + j2\omega R_{l}\kappa_{r}\chi_{r}^{s}}\tag{S4.32}\label{eq:S4.32}
\end{equation*}

\begin{equation*}
{\widehat{\beta}}_{s}(\omega,x) = \frac{{\widehat{w}}_{rel}(x)}{{\widehat{a}}_{b}} = \frac{\left( 2 + j\omega R_{l}{\bar{\varepsilon}}_{33}^{S}bL/h_{p} \right)\sigma_{r}\varphi_{r}(x)}{\left( 2 + j\omega R_{l}{\bar{\varepsilon}}_{33}^{S}bL/h_{p} \right)Z_{r} + j2\omega R_{l}\kappa_{r}\chi_{r}^{s}}\tag{S4.33}\label{eq:S4.33}
\end{equation*}

Equation (S4.32) corresponds to the single-mode coupled voltage frequency-response function of Erturk and Inman, Eq. (68), when expressed with their series-connected electrical parameters. Their laser vibrometer measured absolute rather than base-relative tip velocity, so the base-motion contribution is also retained:

\begin{equation*}
\frac{{\widehat{\dot{w}}}_{abs}(L)}{{\widehat{a}}_{b}} = \frac{1}{j\omega} + j\omega{\widehat{\beta}}_{s}(\omega,L)\tag{S4.34}\label{eq:S4.34}
\end{equation*}

This is the same measurement relation used in their Eq. (79). The electrical-current and power definitions used in comparing load responses are

\begin{equation*}
{\widehat{I}}_{l} = \frac{\widehat{V}}{R_{l}},\quad\quad P_{avg} = \frac{\left| \widehat{V} \right|^{2}}{2R_{l}},\quad\quad P_{peak\ metric} = \frac{\left| \widehat{V} \right|^{2}}{R_{l}}\tag{S4.35}\label{eq:S4.35}
\end{equation*}

Erturk and Inman\cite{siref9} compared measured and calculated voltage and tip-velocity frequency responses at 1, 33 and 470 k\ensuremath{\Omega} (their Fig. 7), extended the load comparison across eight resistances (Fig. 8), and reported current and power frequency responses (Fig. 9). The peak-amplitude power convention used in Fig. 9 is twice the cycle-averaged power for a sinusoidal voltage. These experiments provide the validation context for the coupled electrical and mechanical response relations recovered above.

\section*{Supplementary Note 5 | Ionoelectric generators within the source-driven framework}

Ionoelectric conversion tests the common framework with a moving, transport-controlled source. Pressure-driven solvent flow, diffusion and electrical migration establish evolving ion concentrations and fluxes; the conductor-charge response is then determined by the coupled electrostatic and electrode boundary problem. We retain the Poisson-Nernst-Planck (PNP) and Darcy transport physics and use the piezoionic mechanoreceptor of Dobashi et al.\cite{siref13} to demonstrate recovery of its characteristic voltage and load relations.

\subsection*{5.1 Ion transport and pressure-driven source establishment}

For a gel containing monovalent cations and anions, the one-dimensional Nernst-Planck fluxes include diffusion, electromigration and convection. Ionic species conservation, the charge-density relation, Poisson electrostatics and Darcy flow are expressed as

\begin{equation*}
J_{+} = - D_{+}\frac{\partial c_{+}}{\partial x} - \frac{eD_{+}c_{+}}{k_{B}T}\frac{\partial\phi}{\partial x} + \beta_{+}c_{+}v\tag{S5.1}\label{eq:S5.1}
\end{equation*}

\begin{equation*}
J_{-} = - D_{-}\frac{\partial c_{-}}{\partial x} + \frac{eD_{-}c_{-}}{k_{B}T}\frac{\partial\phi}{\partial x} + \beta_{-}c_{-}v\tag{S5.2}\label{eq:S5.2}
\end{equation*}

\begin{equation*}
\frac{\partial c_{\pm}}{\partial t} + \nabla \cdot \mathbf{J}_{\pm} = 0\tag{S5.3}\label{eq:S5.3}
\end{equation*}

\begin{equation*}
\rho_{ion} = ec_{+} - ec_{-} + \rho_{f},\tag{S5.4}\label{eq:S5.4}
\end{equation*}

\begin{equation*}
- \nabla \cdot \left( \mathbf{\varepsilon}\nabla\phi \right) = \rho_{ion}\tag{S5.5}\label{eq:S5.5}
\end{equation*}

\begin{equation*}
\mathbf{v} = - \frac{\kappa}{\eta}\nabla p,\tag{S5.6}\label{eq:S5.6}
\end{equation*}

The diffusivities and species-dependent convection factors describe coupling between the ions and polymer/solvent network. These equations determine an evolving ionic source, including field-driven feedback on ionic migration. The source state therefore contains more information than instantaneous net space charge:

\begin{equation*}
\mathcal{S}(t) = \left\{ c_{+}\left( \mathbf{r},t \right),c_{-}\left( \mathbf{r},t \right),\rho_{f}\left( \mathbf{r} \right),\mathbf{v}\left( \mathbf{r},t \right),\xi_{int}(t) \right\}\tag{S5.7}\label{eq:S5.7}
\end{equation*}

\begin{equation*}
\rho_{s}\left( \mathbf{r},t \right) = ec_{+}\left( \mathbf{r},t \right) - ec_{-}\left( \mathbf{r},t \right) + \rho_{f}\left( \mathbf{r} \right)\tag{S5.8}\label{eq:S5.8}
\end{equation*}

Even where local electroneutrality makes net bulk charge small, different ionic fluxes can generate nonzero ionic current. Species-resolved concentrations and fluxes are thus retained in the upstream state rather than replaced with a single fixed surface charge.

\subsection*{5.2 From ionic source to electrode free-charge state}

At an instantaneous ionic state, the electrode free-charge distribution follows from Poisson electrostatics, conductor equipotentiality and integrated charge or circuit constraints. A boundary-integral representation and the electrode charge definition are

\begin{equation*}
- \nabla \cdot \left( \mathbf{\varepsilon}\nabla\phi \right) = \rho_{s},\quad\quad\phi|_{\Gamma_{k}} = V_{k}(t),\tag{S5.9}\label{eq:S5.9}
\end{equation*}

\begin{equation*}
\phi_{s}\left( \mathbf{r},t \right) + \sum_{k}^{}{\int_{\Gamma_{k}}^{}G_{el}}\left( \mathbf{r},\mathbf{r}' \right)\sigma_{c,k}\left( \mathbf{r}',t \right)\, dA' = V_{i}(t),\quad\quad\mathbf{r} \in \Gamma_{i}\tag{S5.10}\label{eq:S5.10}
\end{equation*}

\begin{equation*}
Q_{k}(t) = \int_{\Gamma_{k}}^{}\sigma_{c,k}\left( \mathbf{r},t \right)\, dA\tag{S5.11}\label{eq:S5.11}
\end{equation*}

Introduce a weighting-potential solution for each electrode and apply electrostatic reciprocity. The resulting source-induced electrode charge and terminal-potential response follow from the same boundary-value problem:

\begin{equation*}
\nabla \cdot \left( \mathbf{\varepsilon}\nabla\psi_{k} \right) = 0,\quad\quad\psi_{k}\left( \mathbf{r} \right) = \delta_{kj},\quad\quad\mathbf{r} \in \Gamma_{j}\tag{S5.12a}\label{eq:S5.12a}
\end{equation*}

\begin{equation*}
Q_{k}^{src}(t) = - \int_{\Omega}^{}\psi_{k}\rho_{s}d\Omega,Q_{k}(t) = Q_{k}^{src}(t) + \sum_{j}^{}C_{kj}V_{j}(t)\tag{S5.12b}\label{eq:S5.12b}
\end{equation*}

For the derivative formula in Eq. (S5.12d), the weighting potential and electrode geometry are held fixed during the differentiation, as in the Dobashi representative configuration. When geometry varies, the time derivative of the weighting field and the evolving integration domain contribute additional terms. At fixed geometry, the ionic charge continuity equation relates source changes to species fluxes and any reaction terms, giving the weighted source-charge rate:

\begin{equation*}
\mathbf{j}_{ion} = e\sum_{i}^{}z_{i}\mathbf{J}_{i},\frac{\partial\rho_{s}}{\partial t} = - \nabla \cdot \mathbf{j}_{ion} + e\sum_{i}^{}z_{i}\mathcal{R}_{i}\tag{S5.12c}\label{eq:S5.12c}
\end{equation*}

\begin{equation*}
\frac{dQ_{k}^{src}}{dt} = - \int_{\Omega}^{}\mathbf{j}_{ion} \cdot \nabla\psi_{k}d\Omega + \int_{\partial\Omega}^{}\psi_{k}\mathbf{j}_{ion} \cdot \mathbf{n}dA - e\sum_{i}^{}z_{i}\int_{\Omega}^{}\psi_{k}\mathcal{R}_{i}d\Omega\tag{S5.12d}\label{eq:S5.12d}
\end{equation*}

External transfer through the circuit is imposed on the integrated conductor charges rather than prescribed independently from the ionic source. The two-electrode resistive closure is

\begin{equation*}
I_{ext}(t) = \frac{dQ_{1}}{dt} = - \frac{dQ_{2}}{dt},\quad\quad V_{1}(t) - V_{2}(t) = RI_{ext}(t)\tag{S5.13}\label{eq:S5.13}
\end{equation*}

For strongly coupled ionic motion, the transport and electrostatic/electrode problems are solved together or iterated to self-consistency. This preserves the same state-determination sequence while retaining the material-specific transport physics.

\subsection*{5.3 Equivalence to the classical PNP-Darcy electrostatic solution}

Substituting the source-density expression into the electrostatic closure restores the standard ionic Poisson equation. If the two formulations use identical instantaneous ionic concentrations, dielectric properties, electrode conditions and outer boundaries, their difference solves a homogeneous electrostatic problem:

\begin{equation*}
- \nabla \cdot \left( \mathbf{\varepsilon}\nabla\phi \right) = e\left( c_{+} - c_{-} \right) + \rho_{fixed}\tag{S5.14}\label{eq:S5.14}
\end{equation*}

\begin{equation*}
\int_{\Omega}^{}(\nabla\psi)^{\mathsf{T}}\mathbf{\varepsilon}\nabla\psi\, d\Omega = 0\quad \Rightarrow \quad\nabla\psi = 0\tag{S5.15}\label{eq:S5.15}
\end{equation*}

\begin{equation*}
\phi_{cl} = \phi_{sd},\quad\quad\mathbf{E}_{cl} = \mathbf{E}_{sd},\quad\quad Q_{k,cl} = Q_{k,sd}\tag{S5.16}\label{eq:S5.16}
\end{equation*}

The uniqueness identity establishes equivalence of the potential, electric field and electrode free charges for a specified ionic state. Using the same coupled transport, initial-state and interface equations then gives the same complete ionoelectric evolution. The common source-driven contribution is to organize this established coupled problem around the explicit conductor-charge state linking material transport to terminal observations.

\subsection*{5.4 Dobashi piezoionic mechanoreceptor: voltage and load correspondence}

Dobashi et al.\cite{siref13} studied a mechanically compressed gel containing positive and negative mobile ions. Pressure-driven solvent flow carries cations and anions differently, establishing a transient ionic source between a sensing electrode beneath the compressed region and a reference electrode in the undeformed region. For their mechanoreceptor setting, the species-resolved source charge and ionic current are

\begin{equation*}
\rho_{s}\left( \mathbf{r},t \right) = e\left\lbrack c_{cat}\left( \mathbf{r},t \right) - c_{an}\left( \mathbf{r},t \right) \right\rbrack + \rho_{f}\left( \mathbf{r} \right)\tag{S5.17}\label{eq:S5.17}
\end{equation*}

\begin{equation*}
\mathbf{j}_{ion}\left( \mathbf{r},t \right) = e\left\lbrack \mathbf{J}_{cat}\left( \mathbf{r},t \right) - \mathbf{J}_{an}\left( \mathbf{r},t \right) \right\rbrack\tag{S5.18}\label{eq:S5.18}
\end{equation*}

For a uniform one-dimensional region of length L between the electrodes, the weighting field is linear. In this reduced description, lateral ionic flux and explicit Faradaic interfacial charge transfer are neglected; where such effects are important, corresponding boundary and reaction terms must be retained. The weighting-field projection then connects the source current to the electrode-charge state:

\begin{equation*}
\psi_{s}(x) = \frac{x}{L},0 \leq x \leq L\tag{S5.19}\label{eq:S5.19}
\end{equation*}

\begin{equation*}
\frac{dQ_{s}^{src}}{dt} = - \frac{A}{L}\int_{0}^{L}j_{ion}(x,t)dx\tag{S5.20}\label{eq:S5.20}
\end{equation*}

Under open-circuit, stationary-geometry conditions the conductor total charge is fixed. In the uniform steady-state limit this electrode charge constraint reduces to the familiar local condition of vanishing total ionic current. The ionic convection factors, Darcy velocity, source current and field-driven conduction contribution can be written as

\begin{equation*}
\beta_{cat} = \frac{D_{cat}}{D_{0,cat}},\beta_{an} = \frac{D_{an}}{D_{0,an}},v = - \frac{\kappa}{\eta}\frac{\partial p}{\partial x}\tag{S5.21}\label{eq:S5.21}
\end{equation*}

\begin{equation*}
j_{s} = eN\left( \beta_{cat} - \beta_{an} \right)v\tag{S5.22}\label{eq:S5.22}
\end{equation*}

\begin{equation*}
j_{ion} = j_{s} + \sigma_{ion}E,E = - \frac{\partial\phi}{\partial x}\tag{S5.23}\label{eq:S5.23}
\end{equation*}

\begin{equation*}
E = - \frac{eN}{\sigma_{ion}}\left( \beta_{cat} - \beta_{an} \right)v\tag{S5.24}\label{eq:S5.24}
\end{equation*}

Integration of the field along the device leads to the simplified pressure-induced open-circuit voltage relation:

\begin{equation*}
\Delta V = - \frac{eN\kappa}{\sigma_{ion}\eta}\left( \frac{D_{cat}}{D_{0,cat}} - \frac{D_{an}}{D_{0,an}} \right)\Delta P\tag{S5.25}\label{eq:S5.25}
\end{equation*}

This corresponds to the steady, one-dimensional piezoionic-voltage scaling discussed by Dobashi et al.\cite{siref13}. It relates the pressure difference, cation-anion mobility contrast and ionic conductivity to the observed polarity and voltage. The role of the source-driven description is to show how the voltage follows from source transport, the weighted electrode response and the open-circuit charge constraint.

For a finite resistive load, the ionic source current and field-driven return current determine the source-induced electrode-charge rate. We denote the source current by \(I_{s}\) = A \(j_{s}\), the ionic conductance by \(G_{ion}\) = \(\sigma_{ion}A/L\), and the corresponding ionic resistance by \(R_{ion}\) = 1/\(G_{ion}\). \(C_{geom}\) is the fixed-geometry electrostatic terminal response coefficient; combining the induced-charge rate with the terminal charge conservation gives

\begin{equation*}
\frac{dQ_{s}^{src}}{dt} = - Aj_{s} + \frac{\sigma_{ion}A}{L}V\tag{S5.26}\label{eq:S5.26}
\end{equation*}

\begin{equation*}
C_{geom}\frac{dV}{dt} + \left( G_{ion} + \frac{1}{R} \right)V = I_{s}\tag{S5.27}\label{eq:S5.27}
\end{equation*}

In the steady linear limit the terminal relation reduces to the usual source-with-internal-resistance expression. Its load current, output power and matched-load condition are

\begin{equation*}
I_{R} = \frac{V_{oc}}{R + R_{ion}},P_{R} = \frac{V_{oc}^{2}R}{\left( R + R_{ion} \right)^{2}}\tag{S5.28}\label{eq:S5.28}
\end{equation*}

\begin{equation*}
R_{opt} = R_{ion},P_{\max} = \frac{V_{oc}^{2}}{4R_{ion}}\tag{S5.29}\label{eq:S5.29}
\end{equation*}

Dobashi et al.\cite{siref13} reported maximum load-dependent power near an external resistance of 500 \ensuremath{\Omega} (their Fig. 2K), providing an experimental reference for the internal-resistance matching relation in Eq. (S5.29), rather than a universal matching value. The polarity reversal accompanying the crossover in effective cation and anion diffusivities (Figs. 2G-H) corresponds to a sign change in the ionic source contrast entering Eq. (S5.25). The increase in voltage with ion concentration (Fig. 2E) and the step-compression transients (Figs. 2B, I and J) are likewise linked to source transport and its time evolution; reproducing the transient response requires updating the ionic state before the conductor-charge and circuit closure.

\section*{Supplementary Note 6 | Solid-liquid contact-electrification generators within the source-driven framework}

Solid-liquid triboelectric generators provide an interfacial-source case. Contact electrification establishes interfacial charge, while the liquid electrical double layer (EDL) screens and modifies its electrostatic action. Subsequent liquid spreading, recoil and separation change the geometric coupling between that source and a metallic electrode. The following derivation follows the contact-electrification/EDL model of Wang et al.\cite{siref14} and relates it directly to the common conductor-charge determination process.

\subsection*{6.1 Established interfacial source and EDL screening}

In a Gouy-Chapman-Stern-type effective description, the interfacial source entering the downstream electrostatic problem is parameterized by the contact-established charge density and an EDL screening factor\cite{siref14,siref15,siref16}:

\begin{equation*}
\sigma_{eff} = \beta_{EDL}\sigma_{sat},\quad\quad\beta_{EDL} = \frac{C_{s}}{C_{s} + C_{EDL}}.\tag{S6.1}\label{eq:S6.1}
\end{equation*}

The screening factor is a reduced description of the specified interface conditions. At each instantaneous wetted area, this effective surface source is treated as a prescribed input to the conductor-response problem.

\subsection*{6.2 Interfacial source mapped to the metallic electrode}

For a given electrode arrangement and dielectric geometry, introduce a unit electrode weighting potential in the corresponding source-free auxiliary electrostatic problem:

\begin{equation*}
\nabla \cdot \left( \mathbf{\varepsilon}\nabla\psi_{k} \right) = 0,\quad\quad\psi_{k}\left( \mathbf{r};t \right) = \delta_{kj}\quad\left( \mathbf{r} \in \Gamma_{j} \right).\tag{S6.2}\label{eq:S6.2}
\end{equation*}

Let \ensuremath{\alpha}(t) denote the instantaneous wetted interfacial region. Electrostatic reciprocity gives the source-induced electrode charge obtained by integrating over this region:

\begin{equation*}
Q_{k}^{src}(t) = - \int_{\alpha(t)}^{}\sigma_{eff}\left( \mathbf{r},t \right)\,\psi_{k}\left( \mathbf{r};t \right)\, dA.\tag{S6.3}\label{eq:S6.3}
\end{equation*}

Thus the material source strength and wetting-dependent spatial coupling determine the induced free charge. For multiple electrodes, the same field solution also gives the terminal potential-response matrix:

\begin{equation*}
Q_{k}(t) = Q_{k}^{src}(t) + \sum_{j}^{}C_{kj}(t)V_{j}(t),\quad\quad C_{kj} = \int_{\Omega}^{}\left( \nabla\psi_{k} \right)^{\mathsf{T}}\mathbf{\varepsilon}\nabla\psi_{j}\, d\Omega.\tag{S6.4}\label{eq:S6.4}
\end{equation*}

The downstream calculation consequently follows the sequence of established contact source and EDL screening, solved conductor-charge state, and circuit-selected voltage/current. It is this mapping, rather than an assumed current waveform, that connects the liquid-solid device to the common framework.

\subsection*{6.3 Recovery of the contacting-droplet current relation}

Wang et al.\cite{siref14} considered a droplet impacting and spreading over a dielectric film backed by a metallic electrode, followed by recoil and separation. Let \ensuremath{\alpha}(t) denote the instantaneous wetted region, r(t) the contact radius, A(t) the wetted area and \ensuremath{\Gamma}w(t) the area-averaged electrode weighting potential. These geometric factors are

\begin{equation*}
A(t) = \pi r^{2}(t),\quad\quad\Gamma_{w}(t) = \frac{1}{A(t)}\int_{\alpha(t)}^{}\psi_{w}\left( \mathbf{r};t \right)\, dA.\tag{S6.5}\label{eq:S6.5}
\end{equation*}

The induced charge then takes the compact source-times-geometric-coupling form

\begin{equation*}
Q_{ind}(t) = - \beta_{EDL}\sigma_{sat}A(t)\Gamma_{w}(t).\tag{S6.6}\label{eq:S6.6}
\end{equation*}

For a shorted external circuit, differentiating the induced conductor charge gives the transferred current. If the averaged weighting factor changes slowly over the event, the simplified radius-dependent current is

\begin{equation*}
I_{sc}(t) = \frac{dQ_{ind}}{dt} = - \beta_{EDL}\sigma_{sat}\left( \Gamma_{w}(t)\frac{dA}{dt} + A(t)\frac{d\Gamma_{w}}{dt} \right).\tag{S6.7}\label{eq:S6.7}
\end{equation*}

\begin{equation*}
I_{sc}(t) \simeq - 2\pi\beta_{EDL}\sigma_{sat}\Gamma_{w}\, r(t)\dot{r}(t).\tag{S6.8}\label{eq:S6.8}
\end{equation*}

The spreading, recoiling and detachment trajectory can be represented by the piecewise smooth radius function adopted in the source model\cite{siref14}:

\begin{equation*}
r(t) = \left\{ \begin{matrix}
r_{\max}\sin\left( \frac{\pi t}{2t_{s}} \right), & 0 \leq t \leq t_{s}, \\
R + \frac{r_{\max} - R}{2}\left\lbrack 1 + \cos\left( \frac{\pi\left( t - t_{s} \right)}{t_{r}} \right) \right\rbrack, & t_{s} < t \leq t_{s} + t_{r}, \\
\frac{R}{2}\left\lbrack 1 + \cos\left( \frac{\pi\left( t - t_{s} - t_{r} \right)}{t_{d}} \right) \right\rbrack, & t_{s} + t_{r} < t \leq t_{s} + t_{r} + t_{d}.
\end{matrix} \right.\ \ \ \ \ \ \ \ \ \ \ \\tag{S6.9}\label{eq:S6.9}
\end{equation*}

Because the radius increases during spreading and decreases during recoil and separation, the geometry-driven charge rate reverses sign, generating oppositely directed current pulses from the same established source. The current scales reflect the contact area rate and the EDL-screened charge density. These analytical relations correspond to the induced-charge and pulse-generation mechanism reported by Wang et al.\cite{siref14}. In the approximately constant-weighting limit, the characteristic spreading and detachment peak amplitudes scale with r\textsubscript{max}\textsuperscript{2}/t\textsubscript{s} and R\textsuperscript{2}/t\textsubscript{d}, respectively, and are both proportional to \ensuremath{\beta}\textsubscript{EDL}\ensuremath{\sigma}\textsubscript{sat}\ensuremath{\Gamma}\textsubscript{w}\cite{siref14}. This makes the dependence on the established interfacial source and the droplet dynamics explicit.

\textbf{\hfill\break }

\section*{Supplementary Note 7 | Charge-distribution closure of the large-area contact-separation TENG}

\subsection*{7.1 Source establishment and unknown conductor charges}

Consider a large-area contact-separation TENG comprising a dielectric of thickness \(d_{0}\) and relative permittivity \(\varepsilon_{r}\), a backing electrode, and a counter-electrode separated by an air gap \(z\).

After source establishment, the dielectric carries

\begin{equation*}
\sigma_{s} = \sigma_{tp} + \sigma_{te},\tag{S7.1}\label{eq:S7.1}
\end{equation*}

The two terms denote the pre-existing and contact-generated source contributions. The complete initial conductor-charge bookkeeping may additionally contain the compensation charge associated with source establishment. Retaining this term throughout the conductor constraints shows that it enters only the geometry-independent outer-surface charge and cancels from the geometry-driven transferred-charge relations derived below.

Using the sign convention of the analytical derivation, the physical surface-charge sequence is written

\begin{equation*}
- \sigma_{1},\quad + \sigma_{2},\quad - \sigma_{s},\quad + \sigma_{3},\quad - \sigma_{4}\tag{S7.2}\label{eq:S7.2}
\end{equation*}

The quantities \(\sigma_{1}\)-\(\sigma_{4}\) are unknown conductor-response charges.

\subsection*{7.2 Short-circuit electrostatic constraints}

Total free charge of the connected conductor group is conserved,

\begin{equation*}
- \sigma_{1} + \sigma_{2} + \sigma_{3} - \sigma_{4} = \sigma_{te} + \sigma_{c}\tag{S7.3}\label{eq:S7.3}
\end{equation*}

Vanishing electric field inside the two conductors gives

\begin{equation*}
- \sigma_{s} + \sigma_{1} + \sigma_{2} + \sigma_{3} - \sigma_{4} = 0,\tag{S7.4}\label{eq:S7.4}
\end{equation*}

and

\begin{equation*}
\sigma_{s} + \sigma_{1} - \sigma_{2} - \sigma_{3} - \sigma_{4} = 0\tag{S7.5}\label{eq:S7.5}
\end{equation*}

Short-circuit equipotentiality reduces to

\begin{equation*}
\frac{\sigma_{3}d_{0}}{\varepsilon_{0}\varepsilon_{r}} = \frac{\sigma_{2}z}{\varepsilon_{0}}\tag{S7.6}\label{eq:S7.6}
\end{equation*}

These equations solve the complete conductor-charge state before field or output quantities are evaluated.

\subsection*{7.3 Complete short-circuit charge distribution}

Adding and subtracting Eqs. (S7.4) and (S7.5) gives

\begin{equation*}
\sigma_{1} = \sigma_{4},\quad\quad\sigma_{2} + \sigma_{3} = \sigma_{s}\tag{S7.7}\label{eq:S7.7}
\end{equation*}

Combining with Eqs. (S7.3) and (S7.6),

\begin{equation*}
\sigma_{1} = \sigma_{4} = \frac{\sigma_{tp} - \sigma_{c}}{2},\tag{S7.8}\label{eq:S7.8}
\end{equation*}

\begin{equation*}
\sigma_{2}(z) = \frac{\sigma_{s}d_{0}}{\varepsilon_{r}z + d_{0}},\tag{S7.9}\label{eq:S7.9}
\end{equation*}

and

\begin{equation*}
\sigma_{3}(z) = \frac{\sigma_{s}\varepsilon_{r}z}{\varepsilon_{r}z + d_{0}}\tag{S7.10}\label{eq:S7.10}
\end{equation*}

The geometry-dependent internal redistribution is therefore controlled by

\begin{equation*}
\sigma_{2}(z) + \sigma_{3}(z) = \sigma_{s}\tag{S7.11}\label{eq:S7.11}
\end{equation*}

The outer common-mode charge is independent of separation in this ideal large-area solution. Accordingly, the compensation-charge contribution remains in the absolute outer-surface bookkeeping but cancels from the geometry-dependent internal redistribution, transferred charge, intrinsic voltage and load evolution.

\subsection*{7.4 Field derived from the solved charge distribution}

After Eqs. (S7.8)-(S7.10) have been obtained, the fields follow by superposition. In the dielectric,

\begin{equation*}
\mathbf{E}_{d}(z) = \frac{\sigma_{s}z}{\varepsilon_{0}\left( \varepsilon_{r}z + d_{0} \right)}\,\mathbf{e}_{z},\tag{S7.12}\label{eq:S7.12}
\end{equation*}

and in the air gap,

\begin{equation*}
\mathbf{E}_{g}(z) = - \frac{\sigma_{s}d_{0}}{\varepsilon_{0}\left( \varepsilon_{r}z + d_{0} \right)}\,\mathbf{e}_{z}\tag{S7.13}\label{eq:S7.13}
\end{equation*}

These satisfy the short-circuit potential condition

\begin{equation*}
E_{d}d_{0} + E_{g}z = 0\tag{S7.14}\label{eq:S7.14}
\end{equation*}

\subsection*{7.5 Transferred charge from two solved states}

If the device is short-circuit equilibrated at the reference separation and then moved to a new separation, the short-circuit transferred charge is the difference between the two solved conductor-charge states,

\begin{equation*}
Q_{sc}\left( z;z_{0} \right) = S\left\lbrack \sigma_{2}\left( z_{0} \right) - \sigma_{2}(z) \right\rbrack\tag{S7.15}\label{eq:S7.15}
\end{equation*}

Therefore

\begin{equation*}
Q_{sc}\left( z;z_{0} \right) = \frac{\sigma_{s}\varepsilon_{r}Sd_{0}\left( z - z_{0} \right)}{\left( \varepsilon_{r}z + d_{0} \right)\left( \varepsilon_{r}z_{0} + d_{0} \right)}\tag{S7.16}\label{eq:S7.16}
\end{equation*}

For a prescribed trajectory \(z(t)\),

\begin{equation*}
I_{sc}(t) = \frac{dQ_{sc}}{dt} = \frac{\sigma_{s}\varepsilon_{r}Sd_{0}}{\left\lbrack \varepsilon_{r}z(t) + d_{0} \right\rbrack^{2}}\frac{dz}{dt}\tag{S7.17}\label{eq:S7.17}
\end{equation*}

The geometrical charge sensitivity is

\begin{equation*}
\frac{\partial Q_{sc}}{\partial z} = \frac{\sigma_{s}\varepsilon_{r}Sd_{0}}{\left( \varepsilon_{r}z + d_{0} \right)^{2}}\tag{S7.18}\label{eq:S7.18}
\end{equation*}

\subsection*{7.6 Charge isolation and intrinsic voltage}

Suppose the generator is short-circuit equilibrated at \(z_{0}\) and then isolated. The total free charge of each conductor group becomes fixed. In the one-dimensional infinite-plate limit, the layer-integrated conductor charges remain those of the state established at \(z_{0}\).

When the separation changes to \(z_{1}\), the terminal potential follows from the fields of this charge-frozen state. Using the short-circuit reference condition at \(z_{0}\),

\begin{equation*}
V_{int}\left( z_{1};z_{0} \right) = \frac{\sigma_{s}d_{0}\left( z_{1} - z_{0} \right)}{\varepsilon_{0}\left( \varepsilon_{r}z_{0} + d_{0} \right)}\tag{S7.19}\label{eq:S7.19}
\end{equation*}

The dependence on \(z_{0}\) makes the charge history explicit.

\subsection*{7.7 Reduced response scale after charge closure}

At the same final geometry,

\begin{equation*}
Q_{sc} = \frac{\sigma_{s}\varepsilon_{r}Sd_{0}\left( z_{1} - z_{0} \right)}{\left( \varepsilon_{r}z_{1} + d_{0} \right)\left( \varepsilon_{r}z_{0} + d_{0} \right)}\tag{S7.20}\label{eq:S7.20}
\end{equation*}

The ratio of the independently solved short-circuit and charge-frozen states is

\begin{equation*}
\chi_{Q}\left( z_{1} \right) \equiv \frac{Q_{sc}}{V_{int}} = \frac{\varepsilon_{0}\varepsilon_{r}S}{\varepsilon_{r}z_{1} + d_{0}}\tag{S7.21}\label{eq:S7.21}
\end{equation*}

For this analytical state family,

\begin{equation*}
\chi_{Q} = - \left( \frac{\partial V}{\partial Q} \right)^{- 1}\tag{S7.22}\label{eq:S7.22}
\end{equation*}

is an equivalent mathematical identity, but not the starting point of the derivation.

\subsection*{7.8 Analytical scaling}

For fixed geometry and source density,

\begin{equation*}
Q_{sc} \propto S,\quad\quad I_{sc} \propto S,\quad\quad V_{int} \propto S^{0},\quad\quad\chi_{Q} \propto S\tag{S7.23}\label{eq:S7.23}
\end{equation*}

For a geometrically identical trajectory traversed at different frequency,

\begin{equation*}
Q_{sc} \propto \omega_{0}^{0},\quad\quad I_{sc} \propto \omega_{0}\tag{S7.24}\label{eq:S7.24}
\end{equation*}

\section*{Supplementary Note 8 | Self-consistent charge-distribution closure in finite geometry}

\subsection*{8.1 Continuous finite-geometry electrostatic problem}

The large-area analytical solution treats the conductor response as laterally uniform. In a finite device, this approximation is relaxed and the conductor response is represented by unknown spatial surface-charge densities. For the two-electrode contact-separation geometry, we denote the unknown conductor-response charge densities by \(\sigma_{1}\left( \mathbf{r} \right)\) and \(\sigma_{2}\left( \mathbf{r} \right)\) on conductor surfaces \(S_{1}\) and \(S_{2}\), respectively. The prescribed material or interfacial source state is represented by \(\sigma_{s}\left( \mathbf{r} \right)\) on the source surface \(S_{s}\); if polarization-bound surface charge is represented explicitly, it is included among the prescribed source terms rather than treated as an unknown conductor response.

For a field point \(\mathbf{r}\) on either conductor, the electrostatic potential is obtained by superposition,

\begin{equation*}
\phi\left( \mathbf{r} \right) = \frac{1}{4\pi\varepsilon}\left\lbrack \int_{S_{1}}^{}\frac{\sigma_{1}\left( \mathbf{r}' \right)}{\left| \mathbf{r} - \mathbf{r}' \right|}\, dS' + \int_{S_{2}}^{}\frac{\sigma_{2}\left( \mathbf{r}' \right)}{\left| \mathbf{r} - \mathbf{r}' \right|}\, dS' + \int_{S_{s}}^{}\frac{\sigma_{s}^{eff}\left( \mathbf{r}' \right)}{\left| \mathbf{r} - \mathbf{r}' \right|}\, dS' \right\rbrack + \phi_{ref},\tag{S8.1}\label{eq:S8.1}
\end{equation*}

where \(\sigma_{s}^{eff}\) denotes the prescribed source contribution used in the finite-geometry representation. The additive potential reference \(\phi_{ref}\) is arbitrary and was set to zero in the numerical implementation. The physical solution is determined by potential differences, conductor equipotentiality and charge constraints.

The continuum potential representation uses a reference-medium kernel. The coefficients in Eqs. (S8.8)-(S8.10) explicitly use the free-space Coulomb kernel. In a layered dielectric, the material-interface and induced-polarization contributions must be treated consistently through an equivalent source representation or the appropriate dielectric Green operator. This dielectric treatment is distinct from, and complementary to, the equipotential and total-charge constraints imposed on the conductors.

For each ideal conductor,

\begin{equation*}
\phi\left( \mathbf{r} \right) = V_{k},\quad\quad\mathbf{r} \in S_{k},\tag{S8.2}\label{eq:S8.2}
\end{equation*}

where \(V_{k}\) is spatially constant over conductor \(k\). The integrated free charge represented on each conductor surface is

\begin{equation*}
Q_{k} = \int_{S_{k}}^{}\sigma_{k}\left( \mathbf{r} \right)\, dS\tag{S8.3}\label{eq:S8.3}
\end{equation*}

Equations (S8.1)-(S8.3), together with the electrical boundary condition, define the finite-geometry conductor-charge problem. Potential is used here as the Maxwell constraint that determines the unknown conductor charge; spatial potential and electric-field maps are evaluated after the charge distribution has been solved.

For an ideal short circuit between the two conductor groups,

\begin{equation*}
V_{1} = V_{2} \equiv V_{sc},\quad\quad Q_{1} + Q_{2} = Q_{tot},\tag{S8.4}\label{eq:S8.4}
\end{equation*}

where \(Q_{tot}\) is fixed by the preceding conductor-charge state. For electrical isolation, the two conductor totals are individually fixed,

\begin{equation*}
Q_{1} = Q_{1}^{(0)},\quad\quad Q_{2} = Q_{2}^{(0)},\tag{S8.5}\label{eq:S8.5}
\end{equation*}

while their local surface-charge distributions remain free to rearrange as geometry changes.

The formulation above is written for two unknown conductor surfaces because this is the minimal finite-plate problem used for the present contact-separation closure. Additional conductor surfaces can be included by adding the corresponding unknown surface-charge vectors and equipotential constraints without changing the structure of the method.

\subsection*{8.2 Method-of-moments discretization}

For a square conductor of side length \(L\), each solved surface is divided into an \(N \times N\) grid of square elements. The number of elements on each surface is

\begin{equation*}
n = N^{2},\quad\quad b = \frac{L}{N},\quad\quad\Delta S = b^{2},\tag{S8.6}\label{eq:S8.6}
\end{equation*}

where \(b\) is the element side length. A piecewise-constant pulse basis is used,

\begin{equation*}
\sigma_{k}\left( \mathbf{r} \right) \approx \sum_{m = 1}^{n}\sigma_{k,m}f_{m}\left( \mathbf{r} \right),\quad\quad f_{m}\left( \mathbf{r} \right) = \left\{ \begin{matrix}
1, & \mathbf{r} \in S_{m}, \\
0, & \mathbf{r} \notin S_{m}.
\end{matrix} \right.\\tag{S8.7}\label{eq:S8.7}
\end{equation*}

Point matching is applied at the center \(\mathbf{r}_{i}\) of each element. The contribution of element \(m\) on conductor \(b\) to the potential at collocation point \(i\) on conductor \(a\) is written

\begin{equation*}
G_{im}^{ab} = \frac{1}{4\pi\varepsilon_{0}}\int_{S_{m}^{(b)}}^{}\frac{dS'}{\left| \mathbf{r}_{i}^{(a)} - \mathbf{r}' \right|}\tag{S8.8}\label{eq:S8.8}
\end{equation*}

For distinct sufficiently small elements, the centroid approximation gives

\begin{equation*}
G_{im}^{ab} \approx \frac{\Delta S}{4\pi\varepsilon_{0}R_{im}^{ab}},\quad\quad R_{im}^{ab} = \left| \mathbf{r}_{i}^{(a)} - \mathbf{r}_{m}^{(b)} \right|,\quad\quad(a,i) \neq (b,m)\tag{S8.9}\label{eq:S8.9}
\end{equation*}

Equation (S8.9) is used only for far-field interactions in which the element side length is smaller than both characteristic normal separations of the layered structure, that is, smaller than the dielectric spacing \emph{d} and the air-gap separation \emph{z}. Whenever the panel dimension is comparable to or larger than either separation, the finite rectangular-panel integral in Eq. (S8.8) is evaluated directly rather than replaced by a point charge at the centroid. In the BEM implementation, these finite-panel interaction coefficients are evaluated using 32nd-order numerical quadrature. The same direct finite-panel treatment is used for other near-field off-diagonal interactions. This hybrid far-field/finite-panel evaluation retains the efficiency of the centroid approximation while preserving numerical accuracy in the small-thickness and small-gap regimes, including cases in which the panel dimension greatly exceeds \emph{d} or \emph{z}.

The prescribed source contribution at the collocation point is collected into the known vector

\begin{equation*}
s_{i}^{(a)} = \frac{1}{4\pi\varepsilon_{0}}\int_{S_{s}}^{}\frac{\sigma_{s}^{eff}\left( \mathbf{r}' \right)}{\left| \mathbf{r}_{i}^{(a)} - \mathbf{r}' \right|}\, dS'\tag{S8.10}\label{eq:S8.10}
\end{equation*}

The equipotential equations on the two conductors then become

\begin{equation*}
\sum_{m = 1}^{n}G_{im}^{11}\sigma_{1,m} + \sum_{m = 1}^{n}G_{im}^{12}\sigma_{2,m} + s_{i}^{(1)} = V_{1},\quad\quad i = 1,\ldots,n,\tag{S8.11}\label{eq:S8.11}
\end{equation*}

and

\begin{equation*}
\sum_{m = 1}^{n}G_{im}^{21}\sigma_{1,m} + \sum_{m = 1}^{n}G_{im}^{22}\sigma_{2,m} + s_{i}^{(2)} = V_{2},\quad\quad i = 1,\ldots,n\tag{S8.12}\label{eq:S8.12}
\end{equation*}

These discrete equations are the direct finite-dimensional form of electrostatic potential superposition and conductor equipotentiality.

\subsection*{8.3 Self-interaction of a square element}

The centroid point-charge approximation is not used when a square element acts on its own collocation point because \(R_{ii} = 0\). The diagonal term is evaluated from the finite square-element integral. For an element of side \(b\) centered on the collocation point,

\begin{equation*}
G_{ii}^{self} = \frac{1}{4\pi\varepsilon_{0}}\int_{- b/2}^{b/2}{\int_{- b/2}^{b/2}\frac{dx\, dy}{\sqrt{x^{2} + y^{2}}}}\tag{S8.13}\label{eq:S8.13}
\end{equation*}

The integral has the closed form

\begin{equation*}
G_{ii}^{self} = \frac{b\, arcosh(17)}{4\pi\varepsilon_{0}} = \frac{b\ln\left( 1 + \sqrt{2} \right)}{\pi\varepsilon_{0}}\tag{S8.14}\label{eq:S8.14}
\end{equation*}

This separate treatment removes the artificial singularity of the centroid approximation while retaining the same pulse-basis discretization for the complete matrix system.

\subsection*{8.4 Matrix form and electrical boundary conditions}

Define the surface-charge vectors

\begin{equation*}
\mathbf{\sigma}_{1} = \left( \sigma_{1,1},\ldots,\sigma_{1,n} \right)^{T},\quad\quad\mathbf{\sigma}_{2} = \left( \sigma_{2,1},\ldots,\sigma_{2,n} \right)^{T},\tag{S8.15}\label{eq:S8.15}
\end{equation*}

and the interaction matrices \(\mathbf{G}_{ab}\) with entries \(G_{im}^{ab}\). Let \(\mathbf{1}\) denote the \(n\)-component vector of ones and \(\mathbf{a} = \Delta S\,\mathbf{1}\). The prescribed source contributions are \(\mathbf{s}_{1}\) and \(\mathbf{s}_{2}\).

For the short-circuit state, \(V_{1} = V_{2} = V_{sc}\) and the two conductor surfaces share one total-charge constraint. The resulting \(2n + 1\) system is

\begin{equation*}
\begin{bmatrix}
\mathbf{G}_{11} & \mathbf{G}_{12} & - \mathbf{1} \\
\mathbf{G}_{21} & \mathbf{G}_{22} & - \mathbf{1} \\
\mathbf{a}^{T} & \mathbf{a}^{T} & 0
\end{bmatrix}\begin{bmatrix}
\mathbf{\sigma}_{1} \\
\mathbf{\sigma}_{2} \\
V_{sc}
\end{bmatrix} = \begin{bmatrix}
 - \mathbf{s}_{1} \\
 - \mathbf{s}_{2} \\
Q_{tot}
\end{bmatrix}\tag{S8.16}\label{eq:S8.16}
\end{equation*}

The primary numerical output of Eq. (S8.16) is the self-consistent conductor-charge distribution. The common conductor potential is an auxiliary unknown required by equipotentiality rather than an independently imposed source variable.

For a charge-isolated state, the total free charge of each conductor group is frozen at the value established in the reference state. The two conductor potentials are then independent unknowns,

\begin{equation*}
\begin{bmatrix}
\mathbf{G}_{11} & \mathbf{G}_{12} & - \mathbf{1} & \mathbf{0} \\
\mathbf{G}_{21} & \mathbf{G}_{22} & \mathbf{0} & - \mathbf{1} \\
\mathbf{a}^{T} & \mathbf{0}^{T} & 0 & 0 \\
\mathbf{0}^{T} & \mathbf{a}^{T} & 0 & 0
\end{bmatrix}\begin{bmatrix}
\mathbf{\sigma}_{1} \\
\mathbf{\sigma}_{2} \\
V_{1} \\
V_{2}
\end{bmatrix} = \begin{bmatrix}
 - \mathbf{s}_{1} \\
 - \mathbf{s}_{2} \\
Q_{1}^{(0)} \\
Q_{2}^{(0)}
\end{bmatrix}\tag{S8.17}\label{eq:S8.17}
\end{equation*}

The intrinsic potential difference of this solved finite-geometry state is

\begin{equation*}
V_{int} = V_{1} - V_{2}\tag{S8.18}\label{eq:S8.18}
\end{equation*}

Other electrical boundaries are implemented by replacing the last constraint rows with the corresponding conductor-charge and potential relations. Thus the numerical method changes the electrical constraint, not the underlying electrostatic interaction matrix.

\subsection*{8.5 Integrated charge, transferred charge, potential and electric field}

After the spatial conductor charge has been solved, the integrated charge represented on conductor \(k\) is

\begin{equation*}
Q_{k} = \Delta S\sum_{m = 1}^{n}\sigma_{k,m}\tag{S8.19}\label{eq:S8.19}
\end{equation*}

For two independently solved short-circuit geometries \(z_{0}\) and \(z\), the transferred charge is the difference between the corresponding integrated conductor states,

\begin{equation*}
Q_{sc}\left( z;z_{0} \right) = Q_{1}(z) - Q_{1}\left( z_{0} \right) = - \left\lbrack Q_{2}(z) - Q_{2}\left( z_{0} \right) \right\rbrack,\tag{S8.20}\label{eq:S8.20}
\end{equation*}

with the overall sign determined by the adopted transfer direction. For a quasi-static prescribed trajectory \(z(t)\),

\begin{equation*}
I_{sc}(t) = \frac{dQ_{sc}}{dt}\tag{S8.21}\label{eq:S8.21}
\end{equation*}

Spatial potential is evaluated only after the conductor-charge solution has been obtained,

\begin{equation*}
\phi\left( \mathbf{r} \right) = \sum_{k = 1}^{2}{\sum_{m = 1}^{n}\frac{\sigma_{k,m}}{4\pi\varepsilon_{0}}}\int_{S_{m}^{(k)}}^{}\frac{dS'}{\left| \mathbf{r} - \mathbf{r}' \right|} + \phi_{src}\left( \mathbf{r} \right) + \phi_{ref},\tag{S8.22}\label{eq:S8.22}
\end{equation*}

where \(\phi_{src}\) is the known contribution of the prescribed source charges. The electric field then follows from

\begin{equation*}
\mathbf{E}\left( \mathbf{r} \right) = - \nabla\phi\left( \mathbf{r} \right)\tag{S8.23}\label{eq:S8.23}
\end{equation*}

The finite-geometry calculation therefore preserves the same physical hierarchy as the analytical formulation: the source state and electrical constraints determine the conductor-charge distribution first; integrated charge, potential and electric field are then evaluated from that solved state.

\subsection*{8.6 Numerical implementation}

The finite-geometry calculations used in this work were implemented in Python. Square conductor surfaces were discretized using \(N =\) 100 in the calculations reported here, corresponding to 100 \(\times\) 100 = 10000 square elements per solved surface. The discretization density is a controllable numerical parameter rather than a physical model parameter and can be increased directly when finer spatial resolution is required. The additive potential reference was set to zero.

The calculations shown in the main text use the experimental contact-separation geometry and the corresponding prescribed source state. Representative charge, potential and electric-field maps were evaluated at the internal separations reported in Fig. 2. The finite solution develops edge redistribution and fringing automatically through the nonlocal interaction matrix; no lateral conductor-charge profile is prescribed in advance.

\subsection*{8.7 Convergence towards the large-area analytical limit}

The finite-geometry and large-area analytical formulations describe the same charge-distribution closure at different spatial resolutions. As the lateral dimension becomes large compared with the internal separation, the central finite-plate solution approaches the laterally uniform analytical state, while the residual difference is confined increasingly to the edge region.

To quantify the practical large-area regime, representative TENG geometries were scanned over lateral dimensions

\begin{equation*}
L = 1\text{-}15\ cm\tag{S8.24}\label{eq:S8.24}
\end{equation*}

and dielectric thicknesses

\begin{equation*}
d_{0} = 10\text{-}200\ \mu m\tag{S8.25}\label{eq:S8.25}
\end{equation*}

The internal separation was varied over the normalized range reported in Fig. 2f. For a compared output quantity \(X\), the relative deviation between the finite-plate and infinite-plate predictions was defined as

\begin{equation*}
\eta_{X} = \frac{\left| X_{finite} - X_{\infty} \right|}{\left| X_{\infty} \right|} \times 100\%\tag{S8.26}\label{eq:S8.26}
\end{equation*}

The normalized separation \(z/L\) is the natural geometrical parameter for this comparison: in the limit \(z/L \rightarrow 0\), edge contributions occupy a progressively smaller fraction of the electrostatically active region and the finite-plate solution converges to the infinite-plate result. The parameter sweep in Fig. 2f therefore characterizes the accuracy of the analytical large-area approximation over representative TENG dimensions rather than defining a device-specific validity condition.

\section*{Supplementary Note 9 | Boundary-selected charge distributions and finite-capacitance voltage measurement}

\subsection*{9.1 Initial state and new unknowns}

The TENG is first short-circuit equilibrated at \(z_{0}\), giving

\begin{equation*}
\{\sigma_{1}^{(0)},\sigma_{2}^{(0)},\sigma_{3}^{(0)},\sigma_{4}^{(0)}\}\tag{S9.1}\label{eq:S9.1}
\end{equation*}

The short is replaced by a measurement capacitance \(C_{m}\). After motion to \(z_{1}\), the new unknown state is

\begin{equation*}
\{\sigma'_{1},\sigma'_{2},\sigma'_{3},\sigma'_{4},q_{m}\}\tag{S9.2}\label{eq:S9.2}
\end{equation*}

The measurement boundary is part of the electrostatic problem; \(q_{m}\) is not prescribed independently.

\subsection*{9.2 Charge conservation and conductor equilibrium}

If positive \(q_{m}\) transfers in the chosen direction,

\begin{equation*}
- \sigma'_{1} + \sigma'_{2} = - \sigma_{1}^{(0)} + \sigma_{2}^{(0)} - \frac{q_{m}}{S},\tag{S9.3}\label{eq:S9.3}
\end{equation*}

\begin{equation*}
\sigma'_{3} - \sigma'_{4} = \sigma_{3}^{(0)} - \sigma_{4}^{(0)} + \frac{q_{m}}{S}\tag{S9.4}\label{eq:S9.4}
\end{equation*}

Vanishing internal conductor fields gives

\begin{equation*}
- \sigma_{s} + \sigma'_{1} + \sigma'_{2} + \sigma'_{3} - \sigma'_{4} = 0,\tag{S9.5}\label{eq:S9.5}
\end{equation*}

\begin{equation*}
\sigma_{s} + \sigma'_{1} - \sigma'_{2} - \sigma'_{3} - \sigma'_{4} = 0\tag{S9.6}\label{eq:S9.6}
\end{equation*}

Therefore

\begin{equation*}
\sigma'_{1} = \sigma_{1}^{(0)},\quad\quad\sigma'_{4} = \sigma_{4}^{(0)},\tag{S9.7}\label{eq:S9.7}
\end{equation*}

\begin{equation*}
\sigma'_{2} = \sigma_{2}^{(0)} - \frac{q_{m}}{S},\quad\quad\sigma'_{3} = \sigma_{3}^{(0)} + \frac{q_{m}}{S}\tag{S9.8}\label{eq:S9.8}
\end{equation*}

\subsection*{9.3 Measurement boundary}

The generator terminal potential derived from the new charge state is

\begin{equation*}
V_{p} = \frac{\sigma'_{2}z_{1}}{\varepsilon_{0}} - \frac{\sigma'_{3}d_{0}}{\varepsilon_{0}\varepsilon_{r}}\tag{S9.9}\label{eq:S9.9}
\end{equation*}

The measurement capacitance carries the same transferred charge,

\begin{equation*}
V_{m} = \frac{q_{m}}{C_{m}},\tag{S9.10}\label{eq:S9.10}
\end{equation*}

and equilibrium requires

\begin{equation*}
V_{p} = V_{m}\tag{S9.11}\label{eq:S9.11}
\end{equation*}

Solving the full system gives

\begin{equation*}
q_{m} = \frac{\sigma_{s}\varepsilon_{r}Sd_{0}\left( z_{1} - z_{0} \right)C_{m}}{\left( \varepsilon_{r}z_{0} + d_{0} \right)\left\lbrack C_{m}\left( \varepsilon_{r}z_{1} + d_{0} \right) + \varepsilon_{0}\varepsilon_{r}S \right\rbrack}\tag{S9.12}\label{eq:S9.12}
\end{equation*}

The measured voltage is

\begin{equation*}
V_{meas} = \frac{q_{m}}{C_{m}} = \frac{\sigma_{s}\varepsilon_{r}Sd_{0}\left( z_{1} - z_{0} \right)}{\left( \varepsilon_{r}z_{0} + d_{0} \right)\left\lbrack C_{m}\left( \varepsilon_{r}z_{1} + d_{0} \right) + \varepsilon_{0}\varepsilon_{r}S \right\rbrack}\tag{S9.13}\label{eq:S9.13}
\end{equation*}

\subsection*{9.4 Complete post-measurement charge distribution}

Define

\begin{equation*}
D_{m} = C_{m}\left( \varepsilon_{r}z_{1} + d_{0} \right) + \varepsilon_{0}\varepsilon_{r}S\tag{S9.14}\label{eq:S9.14}
\end{equation*}

Then

\begin{equation*}
\sigma'_{1} = \sigma'_{4} = \frac{\sigma_{tp}}{2},\tag{S9.15}\label{eq:S9.15}
\end{equation*}

\begin{equation*}
\sigma'_{2} = \frac{\sigma_{s}d_{0}\left\lbrack C_{m}\left( \varepsilon_{r}z_{0} + d_{0} \right) + \varepsilon_{0}\varepsilon_{r}S \right\rbrack}{\left( \varepsilon_{r}z_{0} + d_{0} \right)D_{m}},\tag{S9.16}\label{eq:S9.16}
\end{equation*}

and

\begin{equation*}
\sigma'_{3} = \frac{\sigma_{s}\varepsilon_{r}\left\lbrack C_{m}z_{1}\left( \varepsilon_{r}z_{0} + d_{0} \right) + \varepsilon_{0}\varepsilon_{r}Sz_{0} \right\rbrack}{\left( \varepsilon_{r}z_{0} + d_{0} \right)D_{m}}\tag{S9.17}\label{eq:S9.17}
\end{equation*}

As required,

\begin{equation*}
\sigma'_{2} + \sigma'_{3} = \sigma_{s}\tag{S9.18}\label{eq:S9.18}
\end{equation*}

The physically primary result of the finite-\(C_{m}\) calculation is therefore the family of boundary-selected conductor-charge states \(\{\sigma'_{i}\left( C_{m} \right)\}\).

\subsection*{9.5 Limiting states}

As \(C_{m} \rightarrow 0\),

\begin{equation*}
q_{m} \rightarrow 0,\quad\quad V_{meas} \rightarrow V_{int},\tag{S9.19}\label{eq:S9.19}
\end{equation*}

and the conductor charges approach the initially isolated state.

As \(C_{m} \rightarrow \infty\),

\begin{equation*}
q_{m} \rightarrow Q_{sc},\quad\quad V_{meas} \rightarrow 0,\tag{S9.20}\label{eq:S9.20}
\end{equation*}

and the conductor charges approach the short-circuit state at \(z_{1}\).

\subsection*{9.6 Reduced measurement law}

Using

\begin{equation*}
\chi_{Q} = \frac{Q_{sc}}{V_{int}},\tag{S9.21}\label{eq:S9.21}
\end{equation*}

Eq. (S9.12) reduces to

\begin{equation*}
\frac{q_{m}}{Q_{sc}} = \frac{C_{m}}{C_{m} + \chi_{Q}},\tag{S9.22}\label{eq:S9.22}
\end{equation*}

and

\begin{equation*}
\frac{V_{meas}}{V_{int}} = \frac{\chi_{Q}}{C_{m} + \chi_{Q}}\tag{S9.23}\label{eq:S9.23}
\end{equation*}

Define

\begin{equation*}
\Pi_{m} = \frac{C_{m}}{\chi_{Q}}\tag{S9.24}\label{eq:S9.24}
\end{equation*}

Then

\begin{equation*}
\frac{q_{m}}{Q_{sc}} = \frac{\Pi_{m}}{1 + \Pi_{m}},\quad\quad\frac{V_{meas}}{V_{int}} = \frac{1}{1 + \Pi_{m}}\tag{S9.25}\label{eq:S9.25}
\end{equation*}

At \(C_{m} = \chi_{Q}\), both normalized observables equal 1/2.

Because \(\chi_{Q} \propto S\), finite measurement capacitance produces an area-dependent measured state even though \(V_{int} \propto S^{0}\) in the analytical large-area limit.

\section*{Supplementary Note 10 | Nonlinear charge-distribution trajectories and harmonic generation}

\subsection*{10.1 Harmonic generation begins with the solved charge trajectory}

For a fixed-source TENG,

\begin{equation*}
\mathcal{S}(t) \simeq \mathcal{S}_{0},\tag{S10.1}\label{eq:S10.1}
\end{equation*}

and the time dependence enters primarily through the mechanical geometry. The causal sequence is

\begin{equation*}
\mathcal{S}_{0},\Gamma(t) \rightarrow \rho\left( \mathbf{r},t \right) \rightarrow Q(t)\tag{S10.2}\label{eq:S10.2}
\end{equation*}

For the analytical CS-TENG,

\begin{equation*}
Q\left( z;z_{0} \right) = \frac{\sigma_{s}Sd_{0}}{\varepsilon_{r}z_{0} + d_{0}} - \frac{\sigma_{s}Sd_{0}}{\varepsilon_{r}z + d_{0}}\tag{S10.3}\label{eq:S10.3}
\end{equation*}

All waveform nonlinearity is therefore contained in the solved geometry-dependent conductor charge.

\subsection*{10.2 Exact sinusoidal-motion harmonic series}

Let

\begin{equation*}
z(t) = z_{c} + A\cos\left( \omega_{0}t \right)\tag{S10.4}\label{eq:S10.4}
\end{equation*}

Define

\begin{equation*}
D_{c} = \varepsilon_{r}z_{c} + d_{0},\quad\quad\beta = \frac{\varepsilon_{r}A}{D_{c}}\tag{S10.5}\label{eq:S10.5}
\end{equation*}

For \(|\beta| < 1\),

\begin{equation*}
Q(t) = Q_{const} - \frac{\sigma_{s}Sd_{0}}{D_{c}}\frac{1}{1 + \beta\cos\left( \omega_{0}t \right)}\tag{S10.6}\label{eq:S10.6}
\end{equation*}

Using

\begin{equation*}
\frac{1}{1 + \beta\cos\theta} = \frac{1}{\sqrt{1 - \beta^{2}}}\left\lbrack 1 + 2\sum_{n = 1}^{\infty}( - \kappa)^{n}\cos(n\theta) \right\rbrack,\tag{S10.7}\label{eq:S10.7}
\end{equation*}

where

\begin{equation*}
\kappa = \frac{\beta}{1 + \sqrt{1 - \beta^{2}}},\tag{S10.8}\label{eq:S10.8}
\end{equation*}

the oscillatory charge is

\begin{equation*}
Q_{ac}(t) = - \frac{2\sigma_{s}Sd_{0}}{D_{c}\sqrt{1 - \beta^{2}}}\sum_{n = 1}^{\infty}( - \kappa)^{n}\cos\left( n\omega_{0}t \right)\tag{S10.9}\label{eq:S10.9}
\end{equation*}

Thus

\begin{equation*}
A_{Q,n} = \frac{2\left| \sigma_{s} \right|Sd_{0}}{D_{c}\sqrt{1 - \beta^{2}}}\kappa^{n},\tag{S10.10}\label{eq:S10.10}
\end{equation*}

and

\begin{equation*}
\frac{A_{Q,n}}{A_{Q,1}} = \kappa^{n - 1}\tag{S10.11}\label{eq:S10.11}
\end{equation*}

Increasing amplitude \(A\) or decreasing mean gap \(z_{c}\) increases \(\beta\), increases \(\kappa\), and strengthens normalized higher harmonics.

\subsection*{10.3 Local nonlinear interpretation}

For a general charge-geometry relation,

\begin{equation*}
Q(z) = Q_{c} + Q'_{c}\Delta z + \frac{1}{2}Q''_{c}\Delta z^{2} + \frac{1}{6}Q'''_{c}\Delta z^{3} + \cdots\tag{S10.12}\label{eq:S10.12}
\end{equation*}

For the analytical CS-TENG,

\begin{equation*}
Q'(z) = \frac{\sigma_{s}\varepsilon_{r}Sd_{0}}{\left( \varepsilon_{r}z + d_{0} \right)^{2}},\tag{S10.13}\label{eq:S10.13}
\end{equation*}

\begin{equation*}
Q''(z) = - \frac{2\sigma_{s}\varepsilon_{r}^{2}Sd_{0}}{\left( \varepsilon_{r}z + d_{0} \right)^{3}},\tag{S10.14}\label{eq:S10.14}
\end{equation*}

\begin{equation*}
Q'''(z) = \frac{6\sigma_{s}\varepsilon_{r}^{3}Sd_{0}}{\left( \varepsilon_{r}z + d_{0} \right)^{4}}\tag{S10.15}\label{eq:S10.15}
\end{equation*}

The increasing curvature at smaller separation explains why stronger nonlinearity is sampled there.

\subsection*{10.4 Area and frequency}

All analytical charge-harmonic amplitudes scale as

\begin{equation*}
A_{Q,n} \propto S,\tag{S10.16}\label{eq:S10.16}
\end{equation*}

whereas their normalized ratios in Eq. (S10.11) are independent of \(S\). Thus area changes the overall output scale more strongly than the normalized spectral shape in the ideal large-area model.

For the same geometrical path traversed at a different frequency, the quasi-static \(Q\)-geometry trajectory is unchanged. Frequency enters when this trajectory is differentiated to obtain current.

\section*{Supplementary Note 11 | Exact spectral correspondence between transferred charge and current}

\subsection*{11.1 Harmonic identity}

For a periodic transferred-charge trajectory,

\begin{equation*}
Q(t) = \sum_{n}^{}Q_{n}e^{in\omega_{0}t}\tag{S11.1}\label{eq:S11.1}
\end{equation*}

Since

\begin{equation*}
I(t) = \frac{dQ}{dt},\tag{S11.2}\label{eq:S11.2}
\end{equation*}

it follows exactly that

\begin{equation*}
I_{n} = i{n\omega}_{0}Q_{n}\tag{S11.3}\label{eq:S11.3}
\end{equation*}

Therefore

\begin{equation*}
\left| I_{n} \right| = {n\omega}_{0}\left| Q_{n} \right|,\quad\quad\frac{\left| I_{n} \right|}{{n\omega}_{0}} = \left| Q_{n} \right|\tag{S11.4}\label{eq:S11.4}
\end{equation*}

For a consistent polarity convention,

\begin{equation*}
\phi_{I,n} - \phi_{Q,n} = \frac{\pi}{2}\quad(mod\mspace{6mu} 2\pi)\tag{S11.5}\label{eq:S11.5}
\end{equation*}

\subsection*{11.2 Why current appears more harmonic-rich}

Relative to the fundamental,

\begin{equation*}
\frac{\left| I_{n} \right|}{\left| I_{1} \right|} = n\frac{\left| Q_{n} \right|}{\left| Q_{1} \right|}\tag{S11.6}\label{eq:S11.6}
\end{equation*}

Thus current does not contain a separate set of harmonics; differentiation weights each harmonic already present in the charge trajectory by its own frequency.

For the exact analytical CS-TENG series,

\begin{equation*}
\frac{\left| Q_{n} \right|}{\left| Q_{1} \right|} = \kappa^{n - 1},\tag{S11.7}\label{eq:S11.7}
\end{equation*}

so

\begin{equation*}
\frac{\left| I_{n} \right|}{\left| I_{1} \right|} = n\kappa^{n - 1}\tag{S11.8}\label{eq:S11.8}
\end{equation*}

\subsection*{11.3 Experimental spectral metrics}

For harmonics above the established noise floor, define

\begin{equation*}
\eta_{n} = \frac{\left| I_{n} \right|}{\omega_{n}\left| Q_{n} \right|}\tag{S11.9}\label{eq:S11.9}
\end{equation*}

The prediction is

\begin{equation*}
\eta_{n} = 1\tag{S11.10}\label{eq:S11.10}
\end{equation*}

The phase residual is

\begin{equation*}
\Delta\phi_{n} = \phi_{I,n} - \phi_{Q,n} - \frac{\pi}{2},\tag{S11.11}\label{eq:S11.11}
\end{equation*}

with prediction

\begin{equation*}
\Delta\phi_{n} = 0\quad(mod\mspace{6mu} 2\pi)\tag{S11.12}\label{eq:S11.12}
\end{equation*}

after accounting for polarity.

A valid comparison requires the same mechanical condition, the same physical transfer branch or sufficiently equivalent electrical boundaries, adequate bandwidth, and consistent time referencing.

\section*{Supplementary Note 12 | Load-constrained evolution of the self-consistent charge distribution}

\subsection*{12.1 Physical update loop}

A finite resistance constrains inter-conductor charge exchange. During a small interval,

\begin{equation*}
dQ = I(t)\, dt = \frac{V_{p}(t)}{R_{L}}\, dt\tag{S12.1}\label{eq:S12.1}
\end{equation*}

This transfer changes the total free charge of the conductor groups, after which the spatial charge distribution is re-established at the updated geometry. The physical update is

\begin{equation*}
\rho(t) \rightarrow V_{p}(t) \rightarrow dQ \rightarrow \{ Q_{\alpha}^{tot}(t + dt)\} \rightarrow \rho(t + dt)\tag{S12.2}\label{eq:S12.2}
\end{equation*}

\subsection*{12.2 Analytical CS-TENG charge state under finite transfer}

Starting from the short-circuit reference state at \(z_{0}\), let \(Q(t)\) be the accumulated load-mediated transfer. Then

\begin{equation*}
\sigma_{2}(t) = \sigma_{2}^{(0)} - \frac{Q(t)}{S},\tag{S12.3}\label{eq:S12.3}
\end{equation*}

\begin{equation*}
\sigma_{3}(t) = \sigma_{3}^{(0)} + \frac{Q(t)}{S},\tag{S12.4}\label{eq:S12.4}
\end{equation*}

while the outer-surface charges remain fixed in the analytical model.

At the same instantaneous geometry \(z(t)\), the unrestricted short-circuit state requires

\begin{equation*}
\sigma_{2}^{sc}\left\lbrack z(t) \right\rbrack = \frac{\sigma_{s}d_{0}}{\varepsilon_{r}z(t) + d_{0}}\tag{S12.5}\label{eq:S12.5}
\end{equation*}

The transferred charge needed to reach that state is

\begin{equation*}
Q_{eq}(t) = S\left\lbrack \sigma_{2}^{(0)} - \sigma_{2}^{sc}\left( z(t) \right) \right\rbrack = \frac{\sigma_{s}\varepsilon_{r}Sd_{0}\left\lbrack z(t) - z_{0} \right\rbrack}{\left\lbrack \varepsilon_{r}z(t) + d_{0} \right\rbrack\left( \varepsilon_{r}z_{0} + d_{0} \right)}\tag{S12.6}\label{eq:S12.6}
\end{equation*}

\subsection*{12.3 Exact reduced evolution equation}

After the instantaneous charge distribution is known, the terminal potential is obtained from its fields. Eliminating the explicit surface-charge variables gives

\begin{equation*}
V_{p}(t) = \frac{Q_{eq}(t) - Q(t)}{\chi_{Q}(t)},\tag{S12.7}\label{eq:S12.7}
\end{equation*}

with

\begin{equation*}
\chi_{Q}(t) = \frac{\varepsilon_{0}\varepsilon_{r}S}{\varepsilon_{r}z(t) + d_{0}}\tag{S12.8}\label{eq:S12.8}
\end{equation*}

Equation (S12.7) is an exact reduced expression of the one-dimensional analytical charge solution, not the assumed starting law.

Because \(I = V_{p}/R_{L} = \dot{Q}\),

\begin{equation*}
\frac{dQ}{dt} = \frac{Q_{eq}(t) - Q(t)}{R_{L}\chi_{Q}(t)}\tag{S12.9}\label{eq:S12.9}
\end{equation*}

Define

\begin{equation*}
\tau_{L}(t) = R_{L}\chi_{Q}(t)\tag{S12.10}\label{eq:S12.10}
\end{equation*}

\subsection*{12.4 Step response}

At fixed final geometry, \(Q_{eq}\) and \(\chi_{Q}\) are constant. Then

\begin{equation*}
Q(t) = Q_{eq} + \left\lbrack Q(0) - Q_{eq} \right\rbrack e^{- t/\left( R_{L}\chi_{Q} \right)},\tag{S12.11}\label{eq:S12.11}
\end{equation*}

\begin{equation*}
V_{p}(t) = V_{p}(0)e^{- t/\left( R_{L}\chi_{Q} \right)},\tag{S12.12}\label{eq:S12.12}
\end{equation*}

and

\begin{equation*}
I(t) = \frac{V_{p}(0)}{R_{L}}e^{- t/\left( R_{L}\chi_{Q} \right)}\tag{S12.13}\label{eq:S12.13}
\end{equation*}

Thus relaxation is exponential in time. At large \(R_{L}\) under finite-time driving,

\begin{equation*}
Q(t) \rightarrow Q\left( t_{0} \right),\quad\quad V_{p}(t) \rightarrow V_{int}(t),\tag{S12.14}\label{eq:S12.14}
\end{equation*}

and

\begin{equation*}
I(t) \sim \frac{V_{int}(t)}{R_{L}}\tag{S12.15}\label{eq:S12.15}
\end{equation*}

The high-resistance current asymptote is therefore inverse in resistance.

\subsection*{12.5 Exact arbitrary-motion solution}

For a general prescribed trajectory,

\begin{equation*}
\dot{Q} + a(t)Q = a(t)Q_{eq}(t),\quad\quad a(t) = \frac{1}{R_{L}\chi_{Q}(t)}\tag{S12.16}\label{eq:S12.16}
\end{equation*}

With

\begin{equation*}
A(t) = \int_{t_{0}}^{t}a(s)\, ds,\tag{S12.17}\label{eq:S12.17}
\end{equation*}

the exact solution is

\begin{equation*}
Q(t) = e^{- A(t)}\left\lbrack Q\left( t_{0} \right) + \int_{t_{0}}^{t}e^{A(u)}a(u)Q_{eq}(u)\, du \right\rbrack\tag{S12.18}\label{eq:S12.18}
\end{equation*}

No constant-\(\chi_{Q}\) approximation is required for the analytical CS-TENG trajectory.

\subsection*{12.6 Exact area/frequency scaling by nondimensionalization}

Let

\begin{equation*}
\theta = \omega_{0}t,\tag{S12.19}\label{eq:S12.19}
\end{equation*}

and choose a fixed reference geometry at which the characteristic charge-response coefficient is denoted \(\chi_{*}\). This \(\chi_{*}\) is the reference value of the main-text response coefficient \(\chi_{Q}\) used to nondimensionalize the complete trajectory.

\begin{equation*}
\chi_{*} = \frac{\varepsilon_{0}\varepsilon_{r}S}{\varepsilon_{r}z_{*} + d_{0}}\tag{S12.20}\label{eq:S12.20}
\end{equation*}

Write

\begin{equation*}
\chi_{Q}(t) = \chi_{*}g(\theta),\quad\quad Q_{eq}(t) = Q_{*}q_{eq}(\theta),\quad\quad Q(t) = Q_{*}q(\theta),\tag{S12.21}\label{eq:S12.21}
\end{equation*}

where \(Q_{*} \propto S\). Equation (S12.9) becomes

\begin{equation*}
\frac{dq}{d\theta} = \frac{q_{eq}(\theta) - q(\theta)}{\Pi_{L}g(\theta)},\tag{S12.22}\label{eq:S12.22}
\end{equation*}

with

\begin{equation*}
\Pi_{L} = \omega_{0}R_{L}\chi_{*}\tag{S12.23}\label{eq:S12.23}
\end{equation*}

For a fixed normalized mechanical trajectory, the normalized nonlinear load response is therefore controlled by the single dimensionless parameter \ensuremath{\Pi}\_L. Any characteristic resistance defined by a fixed normalized response criterion satisfies

\begin{equation*}
R^{*} = \frac{\Pi_{L}^{*}}{\omega_{0}\chi_{*}} \propto \frac{1}{\omega_{0}S}\tag{S12.24}\label{eq:S12.24}
\end{equation*}

The scaling is exact within the analytical model; the numerical value of \(\Pi_{L}^{*}\) depends on the operational definition of the transition.

\subsection*{12.7 Harmonic reduced limit}

When one effective \(\chi_{Q}\) is adequate,

\begin{equation*}
Q_{n} = \frac{Q_{eq,n}}{1 + i\omega_{n}R_{L}\chi_{Q}},\tag{S12.25}\label{eq:S12.25}
\end{equation*}

\begin{equation*}
I_{n} = \frac{i\omega_{n}Q_{eq,n}}{1 + i\omega_{n}R_{L}\chi_{Q}}\tag{S12.26}\label{eq:S12.26}
\end{equation*}

Each harmonic experiences

\begin{equation*}
\Pi_{L,n} = \omega_{n}R_{L}\chi_{Q}\tag{S12.27}\label{eq:S12.27}
\end{equation*}

Higher harmonics therefore enter the load-constrained regime at lower resistance. These independent-harmonic expressions are reduced approximations when \(\chi_{Q}(t)\) varies strongly during a nonlinear cycle.

\subsection*{12.8 Peak-to-peak power metric and optimum resistance}

To match the experimental reporting convention used for the load curves, current is characterized by its peak-to-peak amplitude

\begin{equation*}
I_{pp} = I_{\max} - I_{\min},\tag{S12.28}\label{eq:S12.28}
\end{equation*}

and the corresponding load-power metric is defined as

\begin{equation*}
P_{pp} = I_{pp}^{2}R_{L}\tag{S12.29}\label{eq:S12.29}
\end{equation*}

For the nondimensional trajectory \(Q(t) = Q_{*}q(\theta)\) with \(\theta = \omega_{0}t\), define

\begin{equation*}
\mathcal{J}_{pp}\left( \Pi_{L} \right) = \max_{\theta}\left( \frac{dq}{d\theta} \right) - \min_{\theta}\left( \frac{dq}{d\theta} \right)\tag{S12.30}\label{eq:S12.30}
\end{equation*}

Then

\begin{equation*}
I_{pp} = \omega_{0}Q_{*}\mathcal{J}_{pp}\left( \Pi_{L} \right),\tag{S12.31}\label{eq:S12.31}
\end{equation*}

and therefore

\begin{equation*}
P_{pp} = \frac{\omega_{0}Q_{*}^{2}}{\chi_{*}}\Pi_{L}\mathcal{J}_{pp}^{2}\left( \Pi_{L} \right)\tag{S12.32}\label{eq:S12.32}
\end{equation*}

For a fixed normalized trajectory, \(\mathcal{J}_{pp}\) depends only on the dimensionless load parameter \(\Pi_{L}\). The peak-to-peak power metric therefore reaches its maximum at a dimensionless value \(\Pi_{L,opt}\) independent of \(S\) and \(\omega_{0}\), giving

\begin{equation*}
R_{L,opt} = \frac{\Pi_{L,opt}}{\omega_{0}\chi_{*}} \propto \frac{1}{\omega_{0}S}\tag{S12.33}\label{eq:S12.33}
\end{equation*}

In the single-frequency constant-\(\chi_{Q}\) limit, the current peak amplitude is

\begin{equation*}
\left| \widehat{I} \right| = \frac{\omega\left| {\widehat{Q}}_{eq} \right|}{\sqrt{1 + \left( \omega R_{L}\chi_{Q} \right)^{2}}},\tag{S12.34}\label{eq:S12.34}
\end{equation*}

so that \(I_{pp} = 2\left| \widehat{I} \right|\) and

\begin{equation*}
P_{pp} = 4R_{L}\frac{\omega^{2}\left| {\widehat{Q}}_{eq} \right|^{2}}{1 + \left( \omega R_{L}\chi_{Q} \right)^{2}}\tag{S12.35}\label{eq:S12.35}
\end{equation*}

The multiplicative factor does not affect the optimum resistance, which remains

\begin{equation*}
R_{L,opt} = \frac{1}{\omega\chi_{Q}}\tag{S12.36}\label{eq:S12.36}
\end{equation*}

Equation (S12.36) is the single-frequency reduced limit, whereas Eq. (S12.33) gives the more general nonlinear scaling for the peak-to-peak power metric used experimentally.

\subsection*{12.9 Finite-geometry resistive update}

For the finite-geometry numerical model, after solving \(\mathbf{\sigma}_{k}\) and terminal potential \(V_{k}\) at time step \(k\),

\begin{equation*}
\Delta Q_{k} = \frac{V_{k}}{R_{L}}\Delta t\tag{S12.37}\label{eq:S12.37}
\end{equation*}

The conductor totals are updated,

\begin{equation*}
Q_{1}^{k + 1} = Q_{1}^{k} - \Delta Q_{k},\quad\quad Q_{2}^{k + 1} = Q_{2}^{k} + \Delta Q_{k},\tag{S12.38}\label{eq:S12.38}
\end{equation*}

and the spatial electrostatic problem is re-solved at the new geometry. This gives the discrete charge-distribution evolution

\begin{equation*}
\mathbf{\sigma}_{k} \rightarrow V_{k} \rightarrow \Delta Q_{k} \rightarrow \mathbf{Q}_{c,k + 1} \rightarrow \mathbf{\sigma}_{k + 1}\tag{S12.39}\label{eq:S12.39}
\end{equation*}

\clearpage
\section*{Supplementary Methods}

\subsection*{Supplementary Method 1 | Contact-separation TENG fabrication and operation}

The contact-separation TENG consisted of a 100-\ensuremath{\mu}m FEP dielectric with a conductive-ink back electrode and a 100-\(\mu\)m aluminium independent electrode. Both layers were fixed on acrylic supports. Devices with 5 \(\times\) 5, 7 \(\times\) 7 and 10 \(\times\) 10 cm\textsuperscript{2} active areas were prepared using the same material stack. The devices were mounted on a LinMot linear motor (model PS01-37x120-C). Before measurements, the TENG was operated repeatedly in contact-separation mode until the initial charging process had reached stable output.

The contact point defined the zero-separation reference. Retraction of the motor by a prescribed distance established the internal gap, and the subsequent programmed motor trajectory defined \(z(t)\). Sinusoidal motor programs were used for spectral measurements, with controller compensation adjusted when necessary to reduce forward/reverse motion asymmetry.

\subsection*{Supplementary Method 2 | Secondary generator configurations}

The electret generator used a commercial PTFE sheet with an area of 10 \(\times\) 10 cm\textsuperscript{2} and a thickness of 100 \ensuremath{\mu}m. The material was used as received without additional charging. A finite non-contact gap was maintained between the PTFE and the independent electrode.

The piezoelectric generator used a commercial 1-mm-thick PZT ceramic with supplier-reported \(d_{33} =\) 650 pC N\ensuremath{^{-}}\textsuperscript{1}. The piezoelectret was a commercial 6 \(\times\) 6 cm\textsuperscript{2}, 100-\ensuremath{\mu}m-thick sheet with a supplier-reported effective coefficient of 200 pC N\ensuremath{^{-}}\textsuperscript{1}. Both devices were compressed by an acrylic loading plate connected to the linear motor through a force sensor (model SBT730).

The solid-liquid generator consisted of an FEP-covered pair of aluminium electrodes, each with the active area used in the corresponding experiment, with a 4-cm separation between the electrodes. Water droplets were delivered from a reservoir/dropper structure. The ionoelectric-gel sample was prepared from EMIMTFSI and PDMS at a 1:10 mixing ratio, thermally cured at 80 \textdegree{}C for 12 h and subsequently UV-cured for 1 h. These two experimental systems provide complementary electrical observations. The analytical source-to-electrode demonstrations in Supplementary Notes 5 and 12 concern the specific representative literature geometries described in those notes, rather than separate numerical models of the present experimental configurations.

\subsection*{Supplementary Method 3 | Electrical measurement chain}

A Keithley 6514 electrometer was used for charge, current and voltage acquisition. The electrometer output was acquired through an NI-9223 data-acquisition module and recorded in LabVIEW. The acquisition program displayed the time-domain waveform and was also used to obtain peak-to-peak signal amplitudes.

\subsection*{Supplementary Method 4 | Finite-capacitance measurement boundary}

The generator, external high-voltage capacitor and electrometer were connected in series. The electrometer was operated in charge mode and recorded the charge transferred through the measurement branch. The external capacitance was varied to span the boundary-dependent transition of the generator under test. The corresponding measured voltage was obtained from \(V_{meas} = q_{m}/C_{m}\).

\subsection*{Supplementary Method 5 | Harmonic analysis}

Each harmonic dataset contained 2 min of data sampled at 1000 Hz. Fourier coefficients were extracted at the mechanical fundamental and integer harmonics. The fundamental angular frequency is denoted \(\omega_{0}\), and the \(n\)th harmonic frequency is \(\omega_{n} = n\omega_{0}\). Charge and current spectra were compared using \(I_{n} = i{n\omega}_{0}Q_{n}\).

\subsection*{Supplementary Method 6 | Resistive-load scans}

Commercial resistors were inserted in the external charge-transfer path. The full available range was approximately 1 k\(\Omega\)-10 G\(\Omega\), with the relevant transition region depending on generator class. Peak-to-peak current \(I_{pp}\) was recorded from the LabVIEW waveform. The corresponding curves were calculated as \(V_{pp} = I_{pp}R\) and \(P_{pp} = I_{pp}^{2}R\), using the same peak-to-peak convention throughout each series. The high-resistance \(V_{pp}\) limit was compared with the small-capacitance voltage limit.

\subsection*{Supplementary Method 7 | Finite-geometry Python implementation}

The finite-geometry model used the experimental TENG geometry and representative internal separations of 0.5 and 5 mm. The conductor surfaces were divided into square elements, with N = 100 divisions along each lateral direction unless otherwise stated. The complete conductor free-charge distribution was solved in Python using the method-of-moments formulation described in Supplementary Note 8. The self-interaction term was evaluated separately, and the additive potential reference was set to zero. Potential and electric field were calculated after the conductor-charge solution had been obtained. The analytical-limit parameter sweep covered lateral dimensions of 1-15 cm and dielectric thicknesses of 10-200 \ensuremath{\mu}m, with relative error evaluated against the corresponding large-area analytical result.

\section*{Supplementary References}
\begingroup\small\raggedright\renewcommand{\refname}{}

\endgroup

\end{document}